\documentclass[final,times,10pt,authoryear]{elsarticle}

\usepackage{amssymb}
\usepackage{lipsum}
\usepackage[T1]{fontenc}
\usepackage{stfloats}
\usepackage{graphicx}
\usepackage[normalem]{ulem}

\usepackage{txfonts}
\usepackage[table,xcdraw]{xcolor}
\usepackage{amssymb,amsmath,latexsym,mathrsfs}
\usepackage{amsmath}
\usepackage[hidelinks]{hyperref}
\usepackage{xcolor}
\usepackage{float}
\usepackage{epsfig}
\usepackage{soul}
\usepackage{makecell}
\usepackage{colortbl} 
\usepackage{xcolor}   
\usepackage{makecell} 
\usepackage{colortbl} 
\usepackage{xcolor}   
\usepackage{longtable}
\usepackage{adjustbox}
\usepackage{multirow}
\usepackage{hyperref}

\journal{High Energy Astrophysics}

\begin{document}

\begin{frontmatter}



\title{A New Master Supernovae Ia sample and the investigation of the Hubble tension}

\author[1,2,3,4]{M. G. Dainotti}{}
\cortext[cor1]{Corresponding author}
\ead{maria.dainotti@nao.ac.jp}
\affiliation[1]{organization={Division of Science,National Astronomical Observatory of Japan},
            addressline={2 Chome-21-1 Osawa, Mitaka}, 
            city={Tokyo},
            postcode={181-8588}, 
            country={Japan}}
            
\affiliation[2]{organization={The Graduate University for Advanced Studies, SOKENDAI},
            addressline={Shonankokusaimura, Hayama, Miura District}, 
            city={Kanagawa},
            postcode={240-0115}, 
            country={Japan}}
\affiliation[3]{organization={Space Science Institutes },
            addressline={4765 Walnut St Ste B}, 
            city={Boulder},
            postcode={ 80301}, 
            state={CO},
            country={USA}}
\affiliation[4]{organization={Nevada Center for Astrophysics, University of Nevada,89154},
            addressline={4505 Maryland Parkway}, 
            city={Las Vegas},
            postcode={ 80301}, 
            state={NV},
            country={USA}}

\author[5,6]{B. De Simone}
\affiliation[5]{organization={Dipartimento di Fisica ``E.R. Caianiello”, Università di Salerno},
        addressline={Via Giovanni Paolo II, 132}, 
        city={Fisciano, Salerno},
        postcode={84084},
        country={Italy}}

\author[23]{A. Garg}
\affiliation[23]{organization = Department of Astronomy, Astrophysics and Space Engineering, Indian Institute of Technology, 
city= {Indore},
state = {Madhya Pradesh},
postcode = {403552}, 
country={India}}

\affiliation[6]{organization={INFN Gruppo Collegato di Salerno - Sezione di Napoli. c/o Dipartimento di Fisica ”E.R. Caianiello”, Università di Salerno},
            addressline={ Via Giovanni Paolo II, 132}, 
            city={Fisciano, Salerno},
            postcode={84084}, 
            country={Italy}}

\author[1,2,8,9]{K. Kohri}

\affiliation[8]{organization={Theory Center, IPNS and QUP (WPI), KEK},
            addressline={1-1, Oho}, 
            city={Tsukuba},
            postcode={305-0801}, 
            country={Japan}}
            
\affiliation[9]{organization={Kavli IPMU (WPI), UTIAS, The University of Tokyo},
            addressline={Kashiwa}, 
            city={Chiba},
            postcode={277-8583},
            country={Japan}}

\author[11,12]{A. Bashyal}
\affiliation[11]{organization ={Central Department of Physics, Tribhuvan University},
            addressline={Kirtipur}, 
            city={Kathmandu},
            postcode={44618},
            country={Nepal}}
\affiliation[12]{organization ={Pokhara Astronomical Society},
            city={Pokhara},
            postcode={33700},
            country={Nepal}}

\author[31]{A. Aich}
\affiliation[31]{organization={Prayoga Institute of Education Research},
     addressline={Kanakapura Rd, off Ravgodlu, Post},
     city={Bengaluru},
     postcode={560082},
     state={Karnataka},
     country={India}}

\author[7]{A. Mondal}
\affiliation[7]{organization={Indian Institute of Science Education and Research},
addressline={Homi Bhabha Road}, city={Pashan, Pune, Maharashtra}, postcode={411008}, country={India}}

\author[15,16,17]{S. Nagataki}

\affiliation[15]{
organization={Astrophysical Big Bang Laboratory (ABBL), RIKEN Cluster for Pioneering Research},
            addressline={2-1 Hirosawa, Wako}, 
            city={Saitama},
            postcode={351-0198}, 
            country={Japan}}

\affiliation[16]{
organization={RIKEN Interdisciplinary Theoretical \& Mathematical Science Program (iTHEMS)},
            addressline={2-1 Hirosawa, Wako}, 
            city={Saitama},
            postcode={351-0198}, 
            country={Japan}}

\affiliation[17]{
organization={Astrophysical Big Bang Group (ABBG), Okinawa Institute of Science and Technology (OIST)},
            addressline={1919-1 Tancha, Onna-son, Kunigami-gun}, 
            city={Okinawa},
            postcode={904-0495}, 
            country={Japan}}

\author[18,19]{G. Montani}
\affiliation[18]{organization={Physics Department, “Sapienza” University of Rome},
        addressline={P.le Aldo Moro 5}, 
        city={Rome},
        postcode={00185},
        country={Italy}}

\author[27]{T. Jareen}
\affiliation[27]{organization={The Thanu Padmanabhan Centre for Cosmology and Science Popularization},
    addressline={Gurugram},
    postcode={122505},
    state={Haryana},
    country= {India}}

\author[30]{V. M. Jabir
}
\affiliation[30]{organization={Indira Gandhi National Tribal University (IGNTU)},
    addressline={Lal Pur, Amarkantak},
    postcode={484886},
    state={Madhya Pradesh},
    country= {India}}

\author[29]{S. Khanjani}
\affiliation[29]{organization={Bahá'í Institute for Higher Education (BIHE), Iran}
   }

\affiliation[19]{organization={ENEA, Fusion and Nuclear Safety Department},
        addressline={ C.R. Frascati, Via E. Fermi 45}, 
        city={ Frascati},
        postcode={00044},
        country={Italy}}

\author[32]{M. Bogdan}
\affiliation[32]{organization={Department of Mathematics, University of Wroclaw},
        city={Wroclaw},
        postcode={50-384},
        country={Poland}}

\author[10]{N. Fraija}
\affiliation[10]{
organization={Instituto de Astronomía, Universidad Nacional Autonóma de México Circuito Exterior},
            addressline={C.U., A. Postal 70-264}, 
            city={México D.F.},
            postcode={04510}, 
            country={Mexico}}
            
\author[10]{A. C. C. do E. S. Pedreira}

\author[14]{R. H. Dejrah}
\affiliation[14]{
organization={Department of Physics, Faculty of Sciences, Ankara University},
            city={Ankara},
            postcode={06100}, 
            country={Türkiye}}

\author[13]{A. Singh}
\affiliation[13]{
organization={National Institute of Science Education and Research},
            city={Bhubaneswar},
            postcode={752050}, 
            state={Odisha},
            country={India}}            
          
\author[24]{M. Parakh}
\affiliation[24]{
organization={Indian Institute of Science Education and Research},
            addressline={Bhopal Bypass Road}, 
            city={Bhauri, Bhopal},
            postcode={462066}, 
            state={Madhya Pradesh},
            country={India}}
            
\author[25]{R. Mandal}
\affiliation[25]{organization = {Department of Earth and Space Sciences, Indian Institute of Space Science and Technology},
    city={Thiruvananthapuram 695547},
    state = {Kerala},
    country = {India}
}

\author[28]{K. Jarial}
\affiliation[28]{organization={Department of Physics, Sri Venkateswara College, University of Delhi},
    addressline={H5Q9+96J, Moti Bagh II, Dhaula Kuan Enclave I, Dhaula Kuan, New Delhi},
    postcode={110021},
    state={Delhi},
    country= {India}}
    
\author[5,6]{G. Lambiase}

\author[26]{H. Sarkar}
\affiliation[26]{organization={Indian Institute of Science Education and Research},
    addressline={Knowledge City, Sector 81, SAS Nagar},
    city={Manauli},
    postcode={140306},
    state={Punjab},
    country= {India}}

\begin{abstract}
Modern cosmological research still thoroughly debates the discrepancy between local probes and the Cosmic Microwave Background observations in the Hubble constant (\texorpdfstring{$H_0$}{H0}) measurements, ranging from 4 to 6$\sigma$.
In the current study, we examine this tension using the Supernovae Ia (SNe Ia) data from the Pantheon, Pantheon+ (P+), Joint Lightcurve Analysis (JLA), and Dark Energy Survey, (DES) catalogs combined together into the so-called Master Sample. The sample contains 3714 SNe Ia, and is divided all of them into redshift-ordered bins. Three binning techniques are presented: the equi-population, the moving window (MW), and the equi-spacing in the \texorpdfstring{$\log-z$}{log-z}. We perform a Markov-Chain Monte Carlo analysis (MCMC) for each bin to determine the $H_0$ value, estimating it within the standard flat \texorpdfstring{$\Lambda$CDM}{LCDM} and the \texorpdfstring{$w_{0}w_{a}$CDM}{w0waCDM} models. These \texorpdfstring{$H_0$}{H0} values are then fitted with the following phenomenological function: \texorpdfstring{$\mathcal{H}_0(z) = \tilde{H}_0 / (1 + z)^\alpha$}{H0(z) = H0tilde / (1 + z)^alpha}, where \texorpdfstring{$\tilde{H}_0$}{H0tilde} is a free parameter representing \texorpdfstring{$\mathcal{H}_0(z)$}{H0(z)} fitted in \texorpdfstring{$z=0$}{z=0}, and \texorpdfstring{$\alpha$}{alpha} is the evolutionary parameter. Our results indicate a decreasing trend characterized by \texorpdfstring{$\alpha \sim 0.01$}{alpha ~ 0.01}, whose consistency with zero ranges from $1 \sigma$ in 5 cases to 1 case at 3 $\sigma$ and 11 cases at $> 3 \sigma$ in several samples and configurations. Such a trend in the SNe Ia catalogs could be due to evolution with redshift for the astrophysical variables or unveiled selection biases. Alternatively, intrinsic physics, possibly the \texorpdfstring{$f(R)$}{f(R)} theory of gravity, could be responsible for this trend.
\end{abstract}


\end{frontmatter}

\section{Introduction} \label{sec:intro}

\begin{figure}[ht]
    \centering
    \includegraphics[width=1.00\linewidth]{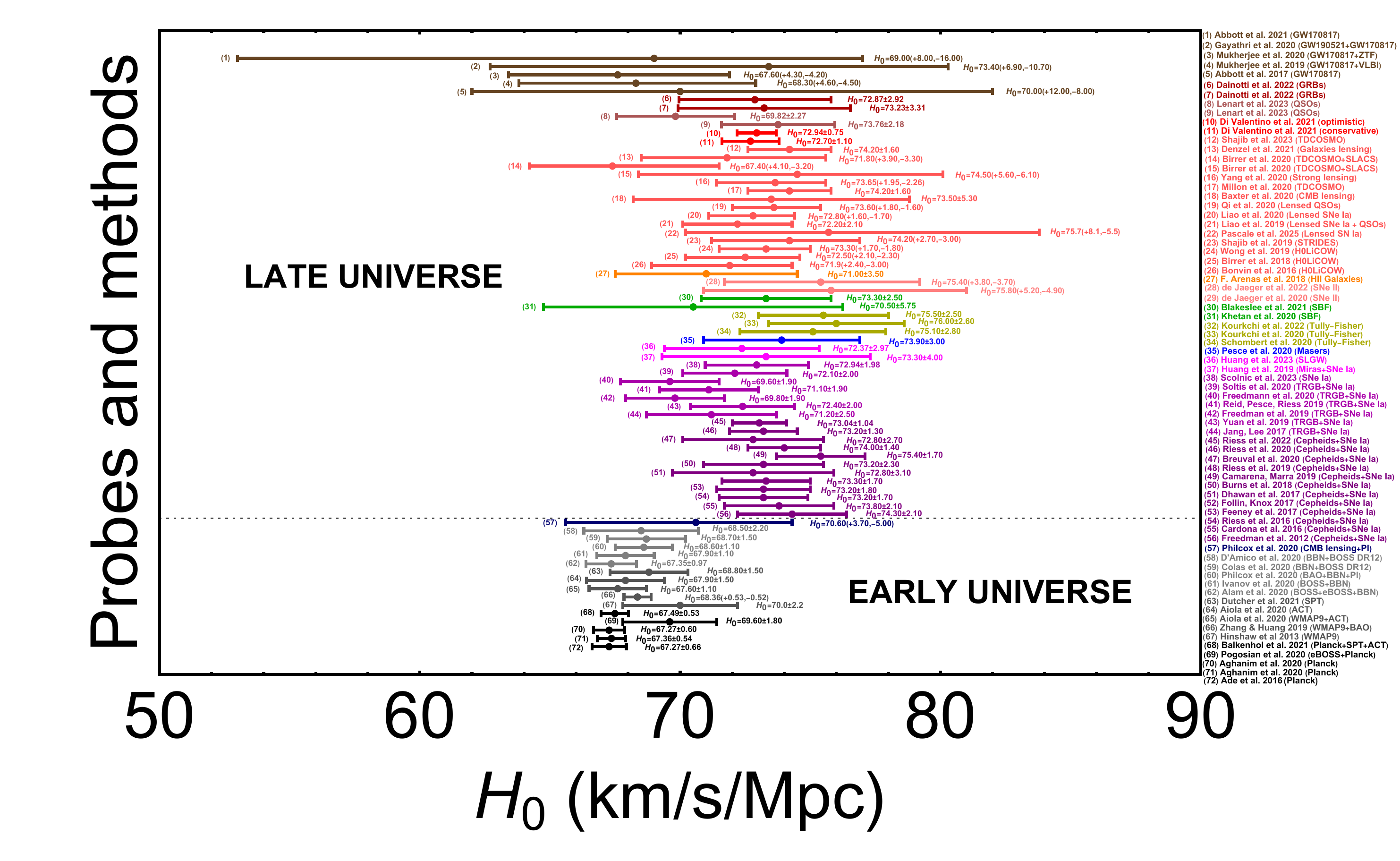}
    \caption{The value of $H_0$ measured through different probes in the literature. References: 
    (1) \citealt{2021ApJ...909..218A},
    (2) \citealt{2020ApJ...890L..20G},
    (3) \citealt{2020MNRAS.494.1956M},
    (4) \citealt{2021A&A...646A..65M},
    (5) \citealt{2017ApJ...851L..16A},
    (6) \citealt{10.1093/mnras/stac2752},
    (7) \citealt{10.1093/mnras/stac2752},
    (8) \citealt{2023ApJS..264...46L},
    (9) \citealt{2023ApJS..264...46L},
    (10) \citealt{2021MNRAS.502.2065D},
    (11) \citealt{2021MNRAS.502.2065D},
    (12) \citealt{2023A&A...673A...9S},
    (13) \citealt{2021MNRAS.501..784D},
    (14) \citealt{2020A&A...643A.165B},
    (15) \citealt{2020A&A...643A.165B},
    (16) \citealt{2020MNRAS.497L..56Y},
    (17) \citealt{2020A&A...642A.193M},
    (18) \citealt{Baxter_2020},
    (19) \citealt{2020ChPhC..44e5101Q},
    (20) \citealt{2020ApJ...895L..29L},
    (21) \citealt{2019ApJ...886L..23L},
    (22) \citealt{2024arXiv240318902P}
    (23) \citealt{2019MNRAS.483.5649S},
    (24) \citealt{Wong_2019},
    (25) \citealt{2019MNRAS.484.4726B},
    (26) \citealt{2017MNRAS.465.4914B},
    (27) \citealt{Fern_ndez_Arenas_2017},
    (28) \citealt{2022MNRAS.514.4620D},
    (29) \citealt{2020MNRAS.496.3402D},
    (30) \citealt{2021ApJ...911...65B},
    (31) \citealt{2021A&A...647A..72K},
    (32) \citealt{2022MNRAS.511.6160K},
    (33) \citealt{2020ApJ...896....3K},
    (34) \citealt{2020AJ....160...71S},
    (35) \citealt{2019BAAS...51g.176P},
    (36) \citealt{2023JCAP...08..003H},
    (37) \citealt{2019hst..prop15867H},
    (38) \citealt{2023arXiv231116830S},
    (39) \citealt{2021ApJ...908L...5S},
    (40) \citealt{Freedman_2020},
    (41) \citealt{2019ApJ...886L..27R},
    (42) \citealt{2019ApJ...882...34F},
    (43) \citealt{2019ApJ...886...61Y},
    (44) \citealt{2017arXiv170201118J},
    (45) \citealt{2022ApJ...938...36R},
    (46) \citealt{2020ApJ...896L..43R},
    (47) \citealt{Breuval_2020},
    (48) \citealt{2019ApJ...876...85R},
    (49) \citealt{2020PhRvR...2a3028C},
    (50) \citealt{2018ApJ...869...56B},
    (51) \citealt{2018A&A...609A..72D},
    (52) \citealt{2018MNRAS.477.4534F},
    (53) \citealt{2018MNRAS.476.3861F},
    (54) \citealt{2016ApJ...826...56R},
    (55) \citealt{2017JCAP...03..056C},
    (56) \citealt{2012ApJ...758...24F},
    (57) \citealt{2020JCAP...05..032P},
    (58) \citealt{2020JCAP...05..005D},
    (59) \citealt{2020JCAP...06..001C},
    (60) \citealt{2020JCAP...05..032P},
    (61) \citealt{2020JCAP...05..042I},
    (62) \citealt{2017MNRAS.470.2617A},
    (63) \citealt{2021PhRvD.104b2003D},
    (64) \citealt{2020JCAP...12..047A},
    (65) \citealt{2020JCAP...12..047A},
    (66) \citealt{2019CoTPh..71..826Z},
    (67) \citealt{2013ApJS..208...19H},
    (68) \citealt{2021PhRvD.104h3509B},
    (69) \citealt{2020ApJ...904L..17P},
    (70) \citealt{Planck2018},
    (71) \citealt{Planck2018}, and
    (72) \citealt{2016A&A...594A..13P}.}
    \label{fig:H0probes}
\end{figure}

The most relevant discussion in modern cosmology is the so-called $H_0$ tension. This is the difference observed, roughly in $\sim 4 - 6$ $\sigma$, between the local estimation of $H_{\rm 0}$ obtained with the observation of Cepheids together with Supernovae Type Ia (SNe Ia) and the cosmological value of $H_0$ measured with the Cosmic Microwave Background (CMB) power spectrum \citep{Planck2018}.
The Planck CMB measurement of $H_0$ provides $H_{0,CMB}=67.4 \pm 0.5 \, km\,s^{-1}\, Mpc^{-1}$, while the SH0ES project \citep{2022ApJ...934L...7R} reports a value of $H_{0,\text{SH0ES}}=73.04 \pm 1.04 \, km\,s^{-1}\, Mpc^{-1}$, achieved through the calibration of 42 local SNe Ia with Cepheid variables and their host galaxies. 
Figure \ref{fig:H0probes} provides a comprehensive overview of the $H_0$ estimations obtained from various probes. This figure depicts the distribution of $H_0$ values recorded over the years and clearly illustrates the evident tension through the direct comparison of all values.
The full discussion of the $H_0$ tension obtained with several probes is summarized in Appendix \ref{sec:H0literature} and we refer to \citealt{2025arXiv250401669D} for a more complete review of the main observational tensions in modern cosmology.
Ideally, the value of $H_0$ should be independent of the sources used for its measurement; however, the significant discrepancies in its estimations have sparked extensive discussions in the literature. Evidence for a redshift-decreasing trend for $H_0$, labeled $\mathcal{H}_0(z)$ \footnote{In what follows, we define the redshift variable $z(t)=(1/a(t)) - 1$, having set to unity the present-day value of the scale factor $a(t)$.}, emerges from a binned analysis of the SNe Ia samples \citep{Dainotti2021hubble}. This approach analyzes each bin within the $\Lambda$CDM model. The running of $\mathcal{H}_0(z)$ has a natural theoretical interpretation, including the introduction of $\mathcal{H}_0(z)$ as an effective $H_{0}$ for any cosmological model described by the Hubble parameter $H(z)$. In the $\Lambda$CDM framework, it is defined as: $\mathcal{H}_0(z)=H(z)/E_{\Lambda \text{CDM}}(z)$, where $E_{\Lambda \text{CDM}}(z)$ represents the expansion rate associated with the model. Indeed, in this framework, $\mathcal{H}_0(z) \equiv H(z=0)$, while for cosmological models other than $\Lambda$CDM, $\mathcal{H}_0(z)$ varies with the redshift $z$. 

It is worth stressing how the definition above of $\mathcal{H}_0(z)$ can be generalized also in other cases, having the generic expansion rate $E_x$ derived from an $x$ cosmological model. For instance, in the present data analysis, we will consider a $w_{0}w_{a}$CDM model in addition to the $\Lambda$CDM. The $w_{0}w_{a}$CDM is described by the Chevallier-Polarski-Linder parametrization (CPL, \citealt{2001IJMPD..10..213C, 2003PhRvL..90i1301L}). This consideration highlights the diagnostic capability of the quantity $\mathcal{H}_0(z)$ in discriminating the reliability of a given dynamical proposal compared to the actual dynamics of the universe. Any time we use $\mathcal{H}_0(z)$ (as referred to a given redshift binning of the sources) and we find that it does not significantly vary along the bins, we provide a reliable representation of the universe properties in terms of the $x$ model. 

The concept of $\mathcal{H}_0(z)$ as a function of the redshift $z$ was introduced and tested in \citet{Dainotti2021hubble,Dainotti2022hubble}, with additional insights provided by \citet{2020arXiv201102858K,2022arXiv220113384K}.
It was argued that $\mathcal{H}_0(z)$ can be induced by a rescaling of the Einstein constant through a mechanism that occurs naturally in a metric $f(R)$ modified gravity formulation in the Jordan frame. This idea has been further explored in subsequent works. In particular, the power law behavior of the running $H_0$ with $z$, originally proposed in \citet{Dainotti2022hubble}, was exactly reproduced in \citet{Schiavone2023}. This phenomenological ansatz provides a simple yet effective interpolation of the binned analysis, explaining the astrophysical origins (redshift evolution of sources) and physical ones (new dynamical effects) of the observed decrease in $H_0$ values.
These dynamical effects can be explained with the metric $f(R)$ gravity in the Jordan frame  \citep{montani2024runninghubbleconstantevolutionary}. In these theories, the non-minimally coupled scalar field can naturally account for the decrease in $H_0$ values across increasing redshift bins. This decline is explained by the rolling down of the scalar field, which reflects the most significant dynamical contribution from the modified gravity framework.

The emergence of a decreasing trend in $\mathcal{H}_0(z)$ has been investigated in such dynamic DE models \citep{Montani.2023xpd, Montani2024xys}. In \citealt{MONTANI2024101486}, the quintessence scalar-field potential is reformulated as a running cosmological constant with redshift. In a subsequent study \citep{montani2024runninghubbleconstantevolutionary}, a standard DE component with a quintessence Equation of State (EoS) parameter, was shown to be influenced by non-equilibrium physics (e.g., bulk viscosity), leading to an effective phantom energy density. As the redshift increases, this non-equilibrium effect further supports the running $\mathcal{H}_0(z)$.

In the present study, the methodologies proposed by \citet{Dainotti2021hubble,Dainotti2022hubble} and \citet{DESIMONE2024} are examined and improved through the incorporation of additional binning techniques, along with a more comprehensive statistical analysis of the model's residuals.
Indeed, \cite{lovick2023nongaussianlikelihoodstypeia, 2024JHEAp..41...30D}, almost at the same time and completely independently discovered that the statistical assumption on the residual of the distance moduli of the SNe Ia normalized by the covariance matrix, do not fullfill the Gaussianity assumption. When the best-likelihoods are used see \cite{2023MNRAS.521.3909B, 2023ApJ...951...63D}, then the uncertainties on cosmological parameters is reduced up to 43\%. This is the reason why, in this analysis, we use these statistical assumptions. This choice is justified by the importance of investigating potential evolutionary trends in local observations. The redshift binning approach for SNe Ia samples is advantageous because it allows for a detailed examination of SNe Ia properties and highlights possible trends with redshift in the astrophysical parameters: this represents crucial support for the development of unbiased techniques in SN observations. If this is not the case, the emergence of local evolutions in the $H_0$ may be an observed effect of a more complex cosmological scenario, and the binning approach would provide an invaluable benchmark for alternative theoretical frameworks.
This work follows the same approach as \citet{Dainotti2021hubble,Dainotti2022hubble} to highlight in which samples and bins we could still recover the $H_0$ decreasing trend. The decreasing of the $H_0$ with the redshift in SNe Ia has been discussed in these papers:
\citet{2020PhRvD.102b3520K,
PhysRevD.106.L041301, 
Dainotti2021hubble, 
Colgain2022,
Dainotti2022hubble,2023A&A...674A..45J,DESIMONE2024,lopezhenandez2024dynamicaltendencyh0late,2024EPJC...84..317M,2024MNRAS.530.5091X}. 
It is worth stressing that the \emph{running} of a cosmological parameter with redshift is a more general result, given that the decrease is a special case of running. 
Some works are discussing a redshift-dependent or time-dependent $H_0$, and some (including the analysis here and in \citealt{Dainotti2021hubble,Dainotti2022hubble}) are discussing a more specific trend in $H_0$, i.e., decreasing. Since experienced groups have carefully selected each of the samples, the combination of these samples presented here brings the advantage of a larger sample, and we have here provided a combination of the likelihoods which takes into account the difference in calibration of all the adopted samples. We here focus on the variation of $H_0$ and other cosmological parameters, but in other works, including some of us, have also used BAOs \citep{2020PhRvD.102j3525K,2024arXiv240606389C,2024arXiv240519953C,Dainotti2022hubble}.
Although many probes can be used together to tackle the $H_0$ tension, we start with analyzing the SNe Ia samples because it is the first step of the distance ladder at cosmological redshifts.
A further advantageous aspect of the current analysis is creating a uniquely large sample of SNe Ia derived from previously published catalogs. This sample is free of duplicated SNe and includes a relatively high number of available events (3714 SNe Ia). It is suitable not only for a redshift binning approach but also for any other cosmological analysis that utilizes SNe Ia as standard candles.

The paper is structured as follows. In Section \ref{sec:cosmologicalmodels}, the flat $\Lambda$CDM and $w_{0}w_{a}$CDM models are summarized. Section \ref{sec:data} introduces the SNe Ia data. Section \ref{sec:methodology} describes the binning approaches, the residuals analysis, and the fitting procedures applied to the $H_0$ values of the different bins. The results of the current analysis are reported in Section \ref{sec:results}. The general discussion of the results is presented in Section \ref{sec:discussion}, while the summary and conclusions are drawn in the last Section \ref{sec:conclusions}.
In Appendix \ref{sec:H0literature}, an overview of the methods and proposals for solving the $H_0$ tension is reported. In Appendix \ref{sec:SH0ES}, we test the behavior of the $H_0$ through a combination of binned SNe Ia and the SH0ES constraints.
Here we stress that the values of $H_0, \tilde{H}_0,$ and $\mathcal{H}_0(z)$ are expressed in units of $km\,s^{-1}\,Mpc^{-1}$ and the measurement units will be omitted except for the plots.


\section{The flat \texorpdfstring{$\Lambda$CDM}{LambdaCDM} and \texorpdfstring{$w_0w_a$CDM}{w0waCDM} cosmological models}
\label{sec:cosmologicalmodels}
In the case of a homogeneous and isotropic universe, the Hubble function $H(z)$ can be written in the following form \citep{2003RvMP...75..559P}:

\begin{equation}
    H(z)=H_0 \sqrt{\Omega_M\,\left(1+z\right)^{3} + \Omega_{\Lambda} +\Omega_{k}\,\left(1+z\right)^{2} + \Omega_{r}\,\left(1+z\right)^{4}},
    \label{eq:H(z)_LCDM}
\end{equation}
where the dimensionless parameters $\Omega_M$ and $\Omega_{\Lambda}$ are associated with the matter contribution and the constant cosmological term ($\Lambda$ being the cosmological constant), respectively. We stress that all the $\Omega_i$ parameters are estimated today (at $t=t_{0}$). To have that $H_0=H(z=0)$,  $\Omega_M$, and $\Omega_{\Lambda}$ obey the normalization condition $\Omega_M+\Omega_{\Lambda}=1$, it is important to stress that the curvature term $\Omega_{k}$ is null for a flat model, and the radiation contribution $\Omega_{r}$ is negligible in the late-universe dynamics.\\
For what it concerns the CPL parameterization in the $w_{0}w_{a}$CDM model, the EoS for the DE is in the form:

\begin{equation}
    w_{DE}(z)=w_0+w_a\frac{z}{1+z}.
    \label{eq:w0waCDM}
\end{equation}

The Hubble function $H(z)$ through the CPL parameterization for the flat $w_{0}w_{a}$CDM model becomes:

\begin{equation}
H(z)=H_0\,\sqrt{\Omega_M\,\left(1+z\right)^{3}+\Omega_{DE}\,\left(1+z\right)^{3\,\left(1+w_{0}+w_{a}\right)}\,e^{-3\,w_{a}\,z/(1+z)}}
\label{eq:H(z)-Linder-w(z)}.
\end{equation}

Choosing $w_{0}=-1$ and $w_{a}=0$, the $\Lambda$CDM is retrieved and $\Omega_{DE}$ becomes $\Omega_{\Lambda}$. The different cosmological models allow us to define the luminosity distance, denoted with $d_l$, that, in the context of SNe Ia, assumes the following form:

\begin{equation}
d_{L}(z_{hel},z_{HD})=c\,(1+z_{hel})\int_{0}^{z_{HD}} \frac{dz'}{H(z)},
\label{eq:dl-LCDM_zHD}
\end{equation}
where $z_{hel}$ is the heliocentric redshift and $z_{HD}$ is the Hubble-diagram redshift, corrected according to the peculiar velocity of the SNe host galaxy and is considered in the CMB frame \citep{2004PASA...21...97D, Scolnic2018P,2020ApJ...902...14S}. To use SNe Ia as cosmological probes, we have to consider their observed distance modulus, $\mu_{\text{obs}}$, and compare it with their theoretical distance modulus $\mu_{\textrm{th}}$, defined as follows:
\begin{equation}
\mu_{\textrm{th}}=5\hspace{0.5ex}log_{10}\ d_L(z,\Omega_M, H_0, ...) +25,
\label{eq:mu_theory}
\end{equation}
where $d_L$ is expressed in Mpc.

\section{The Supernovae Ia data (SNe Ia)} \label{sec:data}
In this Section, details about the main SNe Ia catalogs are presented. The focus is on the following ones: \emph{Pantheon}, \emph{P+}, \emph{JLA} and \emph{DES}.
The Pantheon sample \citep{Scolnic2018P} is a collection of 1048 SNe Ia confirmed spectroscopically with data collected from various surveys and compiled into a single data set. In this sample, the redshift range is $0.01<z<2.26$. The observed $\mu_{\textrm{obs}}$ can be obtained through the modified Tripp formula \citep{Tripp1998}:
\begin{equation}
\label{eq:mu_obs}
\mu_{\textrm{obs}}=m_{B}-M+ a \cdot x_{1} - b \cdot c + \Delta M + \Delta B,
\end{equation}
where $x_{1}$ and $c$ are the stretch and color parameters respectively, $m_{B}$ is the SN apparent magnitude in $B$-band, $M$ is the $B$-band absolute magnitude for a reference SN with $x_{1}=0$ and $c=0$, $\Delta M$ is the correction factor based on the SN host galaxy mass and $\Delta B$ is a bias correction based on the simulations of \citet{Scolnic2018P}.
This study on the Pantheon sample requires the implementation of the BEAMS with Bias Correction method \citep[BBC;][]{2016ApJ...822L..35S} to create a Hubble diagram corrected for selection biases.
As explained in \citet{Tripp1998} and \citet{Scolnic2018P}, there is a degeneracy between $H_0$ and $M$.
It should be emphasized that in the Pantheon release, the absolute magnitude is fixed to $M=-19.35$ so that $H_0=70$.
The value of $M=-19.35$ can be derived from \citet{Scolnic2018P}, computing $M$ through Equation \ref{eq:mu_obs}.
In this paper, $H_0$ is not derived using the BBC method. However, it is obtained by the cosmological analysis fixing and varying the value of $\Omega_M$, directly comparing the $\mu_{\textrm{obs}}$ tabulated in \citet{Scolnic2018P} with $\mu_{\textrm{th}}$ for each SN.

In addition, different models for the stretch and color of the SNe Ia population can be applied to a given sample: C11 \citep{2011A&A...529L...4C} and G10 \citep{2010A&A...523A...7G} are the most suitable models.

Since, in principle, there are no reasons to prefer one model over the other, we average them following the same approach as \citealt{Scolnic2018P}: the bias corrections of G10 and C11 are taken, and this constitutes the systematic part of the covariance matrix, denoted $C_{sys}$.

For the Pantheon data, in the present analysis, for the $\Lambda$CDM, the total matter density is fixed at $\Omega_M=0.298$, while for the $w_{0}w_{a}$CDM, the values of $\Omega_M$, $w_{0}$ and $w_{a}$ are fixed at $0.308$, $-1.009$, and $-0.129$, respectively, see table \ref{tab:Omega}.

We also utilize the P+ data set \citep{Scolnic2022PP}. P+ is an extension of the Pantheon data, with an increased sample size of up to $1701$ SNe Ia. The redshift range for P+ is $0.001<z<2.3$. 
The $\mu_{obs}$ in P+ are the same as in Equation \ref{eq:mu_obs}.
For the fitting of the SN light curves in P+, the authors applied the SALT$2$ method. The calibration of the P+ photometric system, called SuperCal-fragilistic Cross Calibration \citep{brout2021pantheon}, is based on an expansion of the available filters: the specific filters in P+ include Pan-STARRS1 (PS1), Sloan Digital Sky Survey (SDSS), The Supernova Legacy Survey (SNLS) and The Hubble Space Telescope (HST). 
In P+, the calibration of the parameter $M$ is such that the value of $H_0$ is the one inferred locally from the SH0ES Collaboration \citep{2022ApJ...934L...7R}.
The $1701$ light curves are drawn from $1550$ different SNe Ia. Indeed, the P+ catalog contains a non-negligible number of \emph{duplicated SNe} (151), namely, data of the same SN Ia observed in different surveys and by different telescopes. Since some SNe are present more than twice, we found 158 duplicates inside the P+.
From a statistical point of view, it is acceptable to count the same SNe Ia more than once if we consider different telescopes and wavelengths; the question here is more ontological since we aim to consider one entity as a singular SNe Ia.
Here, we remove duplicates added to the SNe Ia P+ sample and prioritize the most populated survey in the P+. For example, if two surveys observe an SN in P+, we exclude the duplicate SN from the survey, contributing to fewer SNe Ia in the total P+. 
 
However, understanding how this data duplication can influence MCMC is essential. In Bayesian analysis, MCMC methods are employed to sample posterior distributions. When data points are duplicated, the likelihood function tends to be unbalanced towards a given region of the data space from which the posteriors are drawn.
Duplicated data can lead to over-confident posterior estimates. This overconfidence can cause the MCMC algorithm to underestimate the true variability in the parameter values, potentially resulting in misleading inferences. To mitigate these concerns, it is crucial to pre-process the data to identify duplicates and address them before applying the MCMC methods.
For this reason, the analysis of the P+ in the current paper considers removing these duplicates. For the P+ data, $\Omega_M=0.334$ for $\Lambda$CDM and $w_{0}=-0.93$, $w_{a}=-0.1$, and $\Omega_M=0.403$ for $w_{0}w_{a}$CDM, see table \ref{tab:Omega}.

The JLA catalog \citep{Betoule2014} includes combined data from multiple surveys, including SDSS-II, SNLS, and HST. The catalog constitutes a data set of $740$ SNe Ia in the redshift range $0.01<z<1.2$. Similarly to the Pantheon sample, JLA obtains $\mu_{\text{obs}}$ from the Tripp formula, similar to Equation \ref{eq:mu_obs}, but without the bias correction. The systematic uncertainty for the SDSS and SNLS surveys is reduced through a joint photometric recalibration of the surveys by \citet{Betoule_2013}, and the SALT$2$ method is introduced to retrieve the input distances between low-$z$ and high-$z$ \citep{Betoule2014}. $374$ spectroscopically verified SNe Ia within the SALT$2$ parameter range (from the SDSS-II survey) are included in the catalog. The remaining JLA data set is added following the approach of \citet{Conley2011}.
The JLA sample is analyzed using $\Omega_M=0.295$ for the $\Lambda$CDM model, whereas for the $w_{0}w_{a}$CDM model, the parameters are $\Omega_M=0.304$, $w_{0} = -0.957$ and $w_a = -0.336$, see table \ref{tab:Omega}. In addition, in the JLA catalog, $M$ is computed such that $H_0=70$.

DES data set consists of 1635 confirmed SNe Ia in a range of $0.10 < z < 1.13$ obtained by the collaboration of four primary probes \citep{2024ApJ...973L..14A}. The DES-SN5YR sample, an extension of the DES data set with the addition of 194 low-$z$ external SNe Ia, consists of 1829 SNe Ia.  
The $\mu_{\textrm{obs}}$ can be obtained by adding the term $\gamma G_{\textrm{host}}$, where $\gamma G_{\textrm{host}}$ is a correction for the observed correlations between the host properties and the SN luminosity. In the analysis of this data set, we use $\Omega_M=0.352$ for the $\Lambda$CDM model, and $\Omega_M=0.495$, $w_{0} = -0.36$, and $w_a = -8.8$ for the $w_{0}w_{a}$CDM model. Similar to the majority of samples investigated here, the DES calibration of $M$ implies that $H_0=70$, see table \ref{tab:Omega}.

\subsection{Combination of different samples: the Master Sample}

In our analysis of the SNe Ia data, we use the original calibrations to ensure consistency with their data release. Specifically, we adopt a calibration value of $H_0=73.04$ for P+ and $H_0=70$ for the other samples.
We decided to keep the original calibrations to avoid changing the $M$ values, which could imply additive statistical fluctuations. However, we observe that the value $\alpha$ remains compatible in 1 $\sigma$ between the no-recalibration and recalibration approaches.
We aim to demonstrate that the trend, as mentioned earlier, exists regardless of the calibration approach. 
When combining P+ with the other samples, we account for the different calibration values in  $H_0$. In such cases, we change the value of $M$ by evaluating the quantity $\mu_{\text{obs}}-M$, see Equation \ref{eq:mu_obs}, for both samples and adding again a new value of $M$, pertinent to the sample, as a free-to-vary parameter in the MCMC analysis of the full sample. The inferred $M$ value is then adopted in all bins regardless of their redshift range. This procedure of recalibrating $M$ in the full range of each SN Ia sample differs from the approach adopted in \citealt{Dainotti2021hubble}, where $M$ is estimated only in the first bin: the new procedure ensures a uniform recalibration for all SNe Ia from the same data release and, as observed previously in \citealt{Dainotti2021hubble}, it does not alter the evolutionary trend of $H_0$, if present in the results.
To achieve the $H_0=70$ calibration in the P+ component of the Master Sample, the value $M=-19.253$ \citep{2022ApJ...934L...7R} has been removed from the $\mu_{obs}$ and it has been replaced with the re-computed value $M=-19.351$ to re-calibrate the observed distance moduli in the context of $\Lambda$CDM framework.
When combining these samples, we account for systematic differences, such as variations in calibration, bias corrections, and host galaxy dependencies. The covariance matrices of each sample, encompassing both statistical and systematic uncertainties, are combined to maintain consistency between the data sets.
Given that the original data release for Pantheon, P+, JLA, and DES are calibrated on $\Lambda$CDM cosmology, in the case of $w_{0}w_{a}$CDM model the process of removing the default $M$ and re-calibrating it is applied to all the samples.

\subsection{The Master Sample}
To create a comprehensive and statistically significant sample of SNe Ia, we combine data from four main catalogs: DES, P+, Pantheon, and JLA. This combined dataset, called the Master Sample, consists of 3714 SNe Ia. During the combination process, duplicate entries were removed by assigning priority to catalogs in the reversed chronological order: DES, P+, Pantheon, and JLA. The Master Sample is composed as follows:

\begin{enumerate}
 \item \textbf{DES}: includes 1829 SNe Ia, making it the largest contributor, and we remove no entries from this sample.  
    \item \textbf{P+}: This sample has 1701 SNe Ia, with 158 duplicated SNe within itself that are removed for successive analysis. Concerning the cross-check of the different samples, the duplicated entries between DES and P+ without duplicates are 335. Also, the duplicates with DES are removed, leaving a sample of 1208 SNe Ia.
    \item \textbf{Pantheon}: Originally composed of 1048 SNe Ia from this data set, has 867 duplicates with P+ and DES together. Thus, the Pantheon sample contributes 181 SNe Ia.
    \item \textbf{JLA}: It initially has 740 SNe Ia and has 244 duplicates with DES, P+, and Pantheon combination. JLA contributes then with 496 SNe Ia.
 
\end{enumerate}

\begin{figure}[t!]
    \centering
    \includegraphics[width=0.45\linewidth]{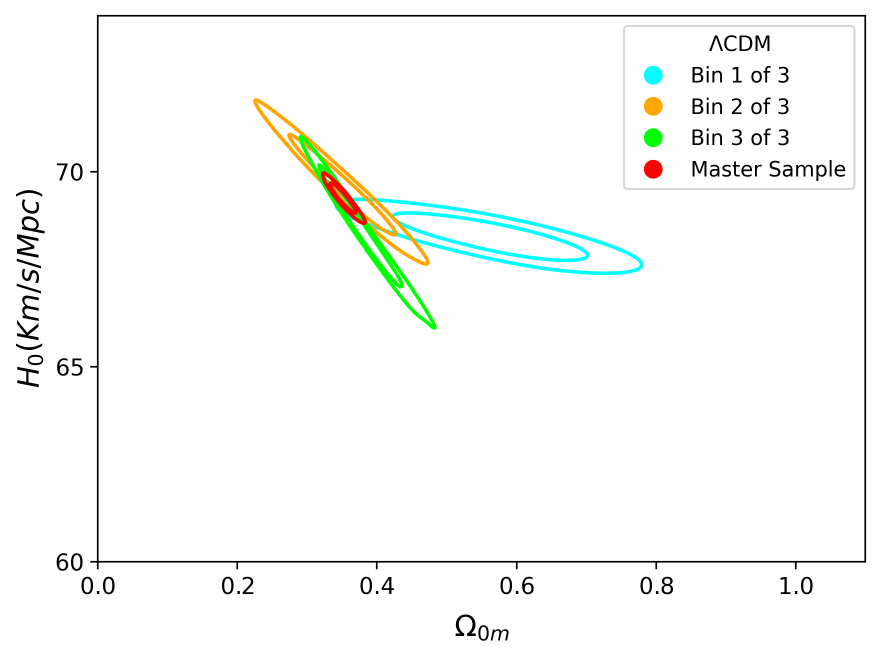}  
    \includegraphics[width=0.45\linewidth]{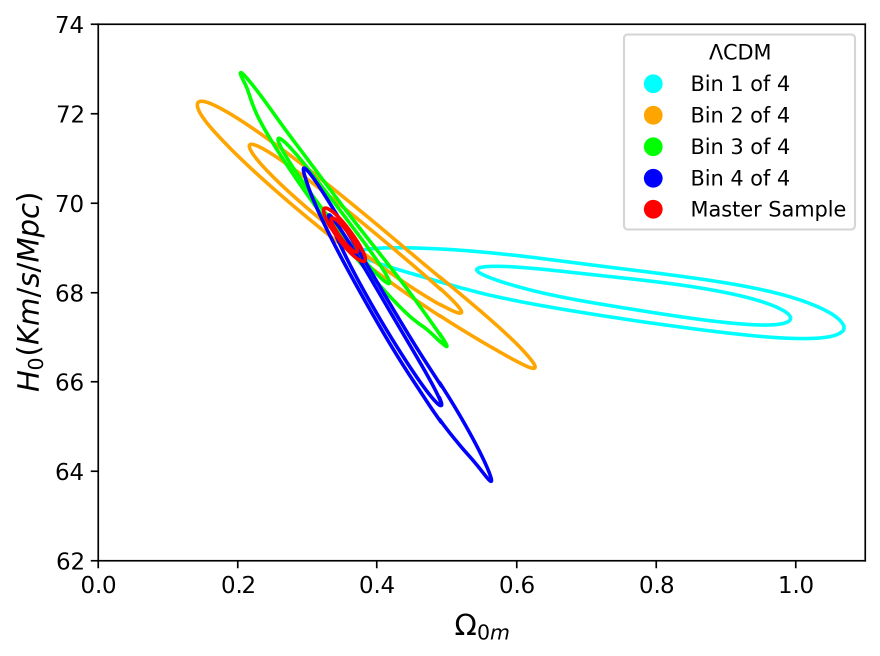}
    \includegraphics[width=0.45\linewidth]{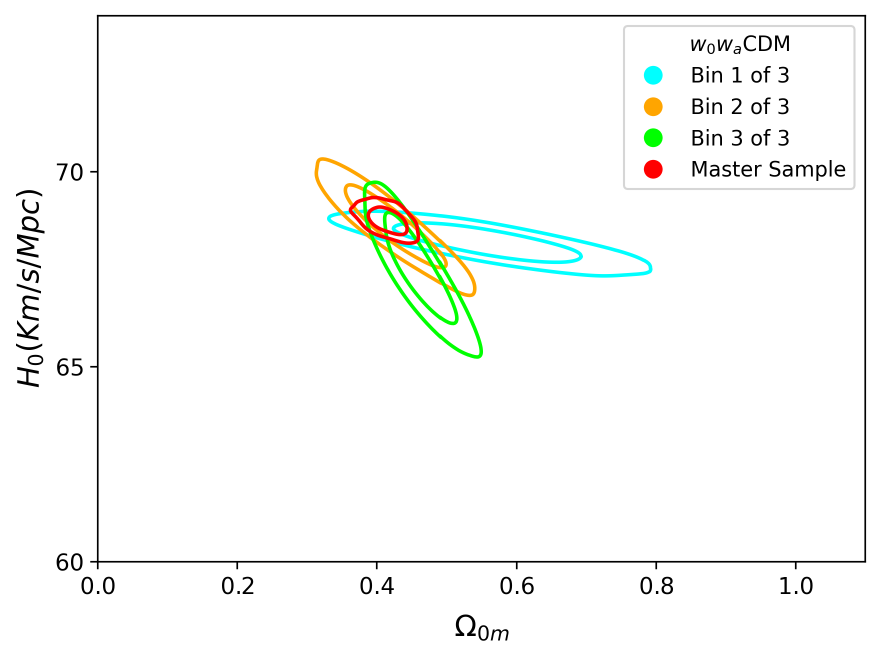}  
    \includegraphics[width=0.45\linewidth]{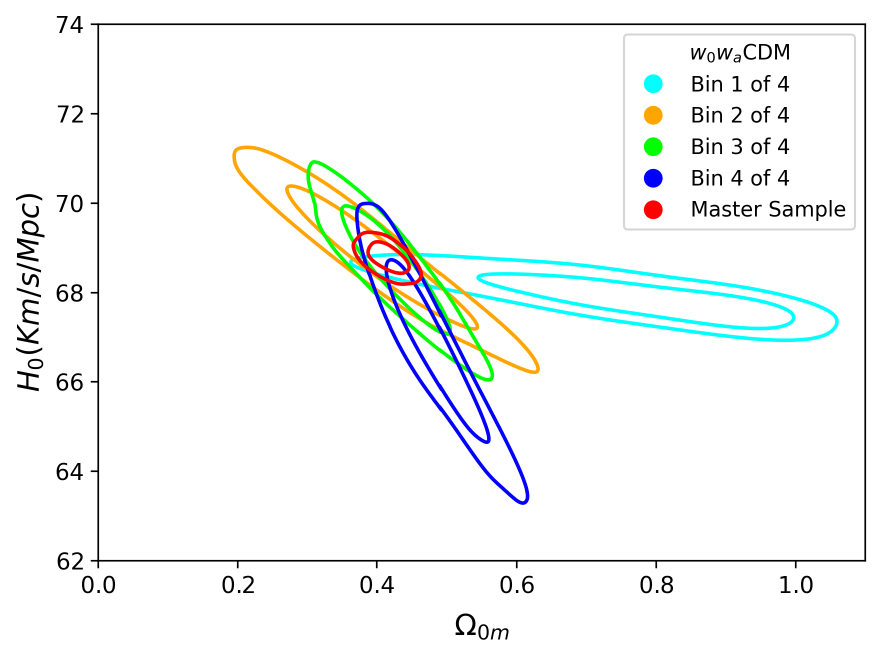}
    \caption{Contours for the 2D posterior analysis for the parameters $H_0,\Omega_M$ in the $\Lambda$CDM and $w_{0}w_{a}$CDM models with the Master Sample. The inner line of each contour represents the 1 $\sigma$ confidence interval ($68\%$) while the external one refers to the 2 $\sigma$ level ($95\%$). In red, the posterior distributions are shown on the whole Master Sample. \textbf{Top left}: 3 bins for the Master Sample, calibrated with $H_0=70$, in the $\Lambda$CDM. \textbf{Top right}: 4 bins for the Master Sample, calibrated with $H_0=70$, in the $\Lambda$CDM. \textbf{Bottom left}: 3 bins for the Master Sample, calibrated with $H_0=70$, in the $w_{0}w_{a}$CDM. \textbf{Bottom right}: 4 bins for the Master Sample, calibrated with $H_0=70$, in the $w_{0}w_{a}$CDM.}
    \label{fig:coloredcontours}
\end{figure}

The reported numbers reflect the results after removing duplicates from each catalog, ensuring a clean and reliable data set. The Master Sample provides an invaluable resource for future cosmological studies.
Such a large number of events without duplicates highlights the relevance of this sample in any cosmological analysis, whether a binning study or any other statistical approach is applied. We perform a preliminary analysis in this sample to show that varying $\Omega_M$ and $H_0$ together in large and flat prior intervals yields the same results for $H_0$ compared to when $\Omega_M$ is varied
within the range of 5 $\sigma$. Using the 3 and 4 bins of the Master Sample, we investigate the posterior distributions of $H_0$ and $\Omega_M$ in the $\Lambda$CDM and $w_{0}w_{a}$CDM models. The posteriors show how the 2 $\sigma$ confidence intervals are compatible among all the bins in both of our free parameters. We present this in figure \ref{fig:coloredcontours}, where the upper panels show the posterior distributions for $H_0=70$ calibration in $\Lambda$CDM and the lower panels report the $w_{0}w_{a}$CDM analysis.

\section{Methodology} \label{sec:methodology}

In this Section, all the binning techniques and likelihood constructions for SNe Ia analysis are detailed.  
\subsection{The binning techniques}
We explore three main binning techniques: 
\emph{equi-population}, \emph{MW}, and \emph{equispace in the log-scale of redshift} ($\log-z$). 
We anticipate that the reason for the multiple approaches is to show that the results remain the same regardless of the binning method and the number of bins in which the samples are divided. This is why we also use different bin numbers.

Furthermore, the $\log-z$ technique is chosen specifically to address the observational imbalance of these SNe Ia in several catalogs, where there is a missing population of SNe Ia at low fluxes and high-$z$. The SNe Ia, as any extragalactic object, are subject to the so-called Malmquist bias effect \citep{1920MeLuF..96....1M}. The binning approaches are described below.

\begin{itemize}
    \item {\textbf{Equi-populated bins:}}
The entire sample is divided into bins that have roughly the same number of SNe in each. The division needs to be such that each bin has sufficient SNe Ia to draw viable statistical conclusions. The equi-populated redshift-ordered bins are generated using the approach seen in \citet{Dainotti2021hubble}. However, this approach does not take into account that an equal number of SNe Ia favors regions of the larger population at low-$z$. Thus, we also use the following approaches.

\item {\textbf{MW:}}
Since we know that changing more parameters at the same time may not allow us to constrain a trend, if existing, we need to implement a new method, called the moving window, abbreviated as MW.
This approach consists of creating a bin that is allowed to move along the increasing redshift, remaining always equi-populated. For example, in a bin with $n$ SNe Ia, we add $j$ SNe Ia in the upper redshift boundary and the same $j$ SNe Ia in the lower adjacent boundary. The $j$ is called overlap, and is determined differently in each case according to the number of bins we produce and the number of data points present in each sample, ensuring at least a $j=25$ in the subsequent windows.

\item \textbf{Equi-spaced bin:} We also propose a binning based on the redshift range rather than on the population of every bin. The current approach predicts an equi-spaced binning considering the $\log_{10}$ scale of $z$. This allows an equivolume and it can be achieved by using the functionality of the \texttt{geomspace} command in the \texttt{numpy} package publicly available in Python. To create the bins, this method estimates the redshift values where each bin starts and ends. In general, it does not ensure that the bins are equally populated.
The full range of redshifts for a given SNe Ia sample is considered ($z_{min} - z_{max}$), and once we decide on the number of bins, we estimate the boundaries of the bins. The bins are equi-spaced in the $\log-z$ and not in linear scale: this allows us to compensate for the effect of reducing the number of SNe Ia with an increase in the redshift since the size of the bins in the linear scale will increase.
\end{itemize}

\subsubsection{Best fitting likelihoods}\label{newlike}
The likelihood of each of the bins is then tested, and the distribution and an inference of the MCMC parameter determines the $H_0$ value with its $1\sigma$ uncertainty in each bin, for a given median redshift. We here stress that in this analysis, large flat priors on the Hubble constant are adopted, namely $60 < H_0 < 80$, while the $\Omega_M$ and $w_a$ parameters are left free to vary in 1 $\sigma$ Gaussian priors, except for the case of Master Sample where the Gaussian priors are expanded up to 5 $\sigma$. Our analysis used the best-fit likelihoods likelihood verification for each bin. In the four SNe Ia catalogs investigated here, the $\chi^2$-likelihood minimization to constrain the cosmological parameters is the following:

\begin{equation}
    \chi^2_{\text{SNe}}=\Delta\mathbf{\mu}^{T}\mathbf{\mathcal{C}}^{-1}\Delta\mathbf{\mu},
    \label{eq:chi2default}
\end{equation}

\noindent where $\mathbf{\mathcal{C}}$ is the covariance matrix that includes both statistical and systematic uncertainties (represented by the matrices $D_{stat}$ and $C_{sys}$, respectively), $\mathbf{\Delta \mu=(\mu_{\text{th}}-\mu_{
\text{obs}})}$, and $\mathbf{\Delta \mu^{T}}$ its transpose. A similar form can be used to compute the residuals between $\mu_{\text{obs}}$ and $\mu_{th}$ SNe Ia and then test if these residuals follow a Gaussian distribution.
This approach follows the same method shown in \citet{DainottiBargiacchi2022c} and \citet{2025MNRAS.536..234L}.
Here we use the following formula to calculate the fit residuals for each bin \citep{2025MNRAS.536..234L}:
\begin{equation}
\label{res}
r=\mathbf{\mathcal{C}}^{-\frac{1}{2}}\Delta\mathbf{\mu}.
\end{equation}
Equation \ref{res} is used to test the Gaussianity assumption.
To investigate which distributions best fit the residuals, the \emph{FindDistribution} command in Wolfram Mathematica 13.1 is used: this is based on the Bayesian Information Criterion (BIC, \citealt{BIC}). For each bin, we select the best-fit distribution among the first 10 best-fits, guaranteeing that the contours of the cosmological parameters were properly constrained.

\subsubsection{\texorpdfstring{Marshall's likelihood and new high-$z$ Supernovae}{Marshall's likelihood and new high-z Supernovae}}\label{marshalllike}

An alternative approach to selecting different likelihoods in each bin according to the residuals is adopting the so-called \emph{Marshall's likelihood} \citep{2024ApJ...964...88M}, which suggests
a Poissonian contribution to the likelihood. This distribution assumes the following form:

\begin{equation}
    \mathcal{P}_{q}(m)=\frac{q^m}{m!}e^{-q},
    \label{eq:poisson}
\end{equation}

\noindent where $q$ is the average number of events per unit of time and $m$ is a natural number. The Poissonian distribution converges to a Gaussian distribution in the limit of large values for $q$. This allows approximating the Poisson distribution with a Gaussian one, as proven in \citet{2024ApJ...964...88M}. Under the same assumptions, SNe Ia in a given sample can be treated as discrete events labeled $j$. The Marshall's likelihood then assumes the following form:

\begin{equation}
    \log \mathcal{L}=-\frac{1}{2}\sum_{j}\biggr\{\log \Delta^{2}_{\mu,j}+\biggr(\frac{\mu_{th,j}-\mu_{obs,j}}{\Delta_{\mu,j}}\biggr)^{2} \biggr\},
    \label{eq:marshalllike}
\end{equation}

\noindent where $\Delta_{\mu,j}$ is the uncertainty on the value of the $j$-th observed $\mu_{obs,j}$. 
This method alters the log-likelihood function explicitly to account for the distance modulus residuals and their uncertainty.
Such a shape of the likelihood allows adding two further high-$z$ SNe Ia from recent discoveries, given that their $\mu_{\text{obs}}$ have been provided without covariance elements or systematic uncertainties. These SNe Ia are: SN 2023aeax at $z=2.15$ \citep{2024arXiv241111953P} and SN 2023adsy at $z=2.9$ \citep{2024arXiv241110427V}.
We here stress that Marshall's likelihood allows one to add a single SNe Ia without adding the covariance matrix. This is advantageous when the information is not present in the literature, but it leads to a less precise evaluation of the uncertainties.

\subsection{The fitting function and its extrapolation}\label{fittingfunction}

  The relationship between the values $H_0$ and $z$ in the bins is modeled by the following expression for $\mathcal{H}_0(z)$:

\begin{equation}
    \mathcal{H}_0(z) = \frac{\tilde{H}_0}{(1 + z)^\alpha},
    \label{eq:fittingfunction}
\end{equation}
where the parameters \( \tilde{H}_0 \) and \( \alpha \) are estimated by weighted least squares fitting. The $\tilde{H}_0$ is the value of the fitting function $\mathcal{H}_0(z)$ estimated at $z=0$, while $\alpha$ is the evolutionary coefficient. It is important to note that the $\tilde{H}_0$ value does not correspond exactly to the calibration value of $H_0$ in the considered sample, given that in the lowest redshift ranges we have SNe that are at low-$z$ but they are not exactly at $z=0$. Thus, the fitting function value at $z=0$ is different than the $H_0$ fitted for the whole sample. For this reason, the $\tilde{H}_0$ is a free parameter in the fitting procedure. The Levenberg-Marquardt method is applied to ensure a non-linear model fitting where the weights are given by the $1/\Delta_{H_0}^2$ values.
This approach prioritizes data points with higher measurement precision. In what follows, this fitting model is referred to as the power law.

We used median values of redshift bins, although the bin sizes do not enter the fitting as uncertainties on the independent variable $z$. 
The median $z$ is a choice that better represents the expected redshift in the bin. A preliminary test performed with the fitting of the $40$ bins of Pantheon in the $\Lambda$CDM (taken from \citealt{Dainotti2021hubble}) shows how the fitting procedure through the median and mean values of $z$ in the bins produces two $\alpha$ values compatible within $1 \sigma$. Horizontal bars on the redshift axis are only redshift ranges and not actual uncertainties; therefore, they are not used in the fitting procedure. The statistical significance of the uncertainties of the parameters and the confidence intervals is also calculated. The purpose of this project is to highlight any possible decreasing trend of the $H_0$ in a binned analysis of SNe Ia. To this end, the model in Equation \ref{eq:fittingfunction} has been assumed as the basic template without testing other plausible fitting models that could highlight the same evolutionary behavior for the Hubble constant. However, for comparison with another fitting model, the Akaike Information Criterion (AIC,  \citealt{Akaike1987}) has been computed for the power law model and the linear model, the latter being a fitting function of the form:

\begin{equation}
    \mathcal{H}_{0,l}(z)= A \cdot z + \tilde{H}_{0,l},
\label{eq:linearmodel}
\end{equation}

where $A$ is the slope of the fitting line and $\tilde{H}_{0,l}$ assumes the same meaning as $\tilde{H}_{0}$ in the power law case. 
To further understand if any of the two models are favored against the other, we introduce the log Bayes factor which is defined as:

\begin{equation}
\log BF_{P,L} = \frac{1}{2}(BIC_{L}-BIC_{P}) 
\label{eq:logbayesfactor}
\end{equation}

where $BIC_{L}$ and $BIC_{P}$ are the BIC values corresponding to the linear and power law models, respectively. The BIC is an asymptotic approximation of marginal likelihood, thus it can be used to replace the marginal likelihood in the Bayes factor formulation \citep{wagenmakers2007}. According to Jeffreys' scale (\citealt{Jeffreys1939-JEFTOP-5,Trotta01032008}) and through the adoption of the natural logarithm thresholds, evidence in favor of the power law over the linear model is inconclusive if $0< BF_{P,L}<1.2$, weak if $1.2< BF_{P,L}<2.3$, moderate if $2.3< BF_{P,L}<3.5$, and strong if $\log BF_{P,L}>3.5$. Conversely, in the case of a negative value of $\log BF_{P,L}$, evidence in favor of linear over the power law model is inconclusive if $-1.2< BF_{P,L}<0$, weak if $-2.3< BF_{P,L}<-1.2$, moderate if $-3.5<\log BF_{P,L}<-2.3$, and strong if $ BF_{P,L}<-3.5$.

After estimating the fitting function parameters, we extrapolate at $z = 1100$, which is the redshift value of the Last Scattering Surface (LSS). The purpose of extrapolating at this redshift is to compare the value of ${H_0}$ from the $\mathcal{H}_0(z)$ model, that is, $\mathcal{H}_0(z=1100)$, with the measurement of ${H_0}$ from the Planck 2018 analysis by \citet{Planck2018}: $H_{0, CMB}=67.4\pm0.5$. This extrapolation represents a simple test to verify that the fit function well represents the analyzed data. It provides an interesting comparison between the observed trend in the SNe Ia redshift $(z<3)$ and the cosmological value $(z=1100)$. The same test is performed by \citet{Dainotti2021hubble}. In this process, we do not consider the transition to the Dark Matter (DM) domination epoch at $z<1$. The only hypothesis considered in the estimation of $\mathcal{H}_0(z=1100)$ is the intrinsic nature of the observed decrease in ${H_0}$, with the ansatz that this trend may also apply to earlier epochs. Nevertheless, this assumption needs further tests with higher-$z$ cosmological probes \cite{2023MNRAS.521.3909B, 2023ApJ...951...63D} such as Gamma-ray Bursts (GRBs) and quasars (QSO) that naturally cover the $z>3$ part of the Hubble diagram.

\section{Results} \label{sec:results}
We have divided our results into \textit{Diamond} and \textit{Gold} with the nomenclature defined as follows:
\begin{itemize}
    \item \textbf{Diamond}: The Diamond cases consider all results where the trend is decreasing with at least $\alpha/\sigma_\alpha>1$ \textit{and} where $\mathcal{H}_0(z=1100)$ is compatible in 1 $\sigma$ with $H_{0, CMB}$.
    \item \textbf{Gold}: as in the Diamond cases, the Gold cases include all results where $\alpha/\sigma_\alpha>1$, \textit{but} $\mathcal{H}_0(z=1100)$ is not compatible in 1 $\sigma$ with $H_{0, CMB}$.
\end{itemize}

\begin{table*}[ht]
\centering
\begin{tabular}{|c|c|c|c|c|c|c|}
\hline
\multicolumn{7}{|c|}{{Fiducial Cosmological Parameters used in our analysis}}\tabularnewline
\hline
Sample & \textit{N} & Model & $\Omega_M$ & $w_0$ & $w_a$ & Source \\
\hline
Pantheon & 1048 & $\Lambda$CDM & $0.298 \pm 0.022$ & $\cdots$ & $\cdots$ & Tab.8 [1] \\ \hline
Pantheon & 1048 & $w_0w_a$CDM & $0.308 \pm 0.018$ & $-1.009 \pm 0.159$ & $-0.129 \pm  0.755$ & Tab.13$^{*}$ [1]\\ \hline
P+ & 1701 & $\Lambda$CDM & $0.334 \pm 0.018$ & $\cdots$ & $\cdots$ & Tab.3 [2] \\\hline
P+ & 1701 & $w_0w_a$CDM & $0.403 ^{+0.054}_{-0.098}$ & $-0.93 \pm 0.15$ & $-0.1^{+0.9}_{-2.0}$ & Tab.3 [2] \\\hline
JLA & 740 &  $\Lambda$CDM  & $0.295 \pm 0.034 $ & $\cdots$ & $\cdots$& Tab.10 [3]  \\ \hline
JLA & 740 & $w_0w_a$CDM  & $0.304 \pm 0.012 $ & $-0.957 \pm 0.124$ & $-0.336 \pm 0.552$ & Tab.10 [3]  \\ \hline
DES & 1829 & $\Lambda$CDM  & $0.352 \pm 0.017$ & $\cdots$ & $\cdots$ & Tab.2** [4] \\ \hline
DES & 1829 & $w_0w_a$CDM  & $0.495 \pm 0.038$ & $-0.36 \pm 0.33$ & $-8.8 \pm 4.1$ & Tab.2** [4] \\ \hline
Master & 3764 & $\Lambda$CDM  & $0.321 \pm 0.005$ & $\cdots$ & $\cdots$ & This paper\\ \hline
Master & 3764 & $w_0w_a$CDM  & $0.473 \pm 0.022$ & $-0.405 \pm 0.105$ & $-7.402 \pm 1.498$ & This paper\\ \hline
\end{tabular}
\caption{This table summarizes the fiducial values used by our analysis for several samples and the flat $\Lambda$CDM and $w_{0}w_{a}$CDM models. The first column represents the sample used, the second column is the total number of SNe Ia in a particular sample (\textit{N}), the third column titled `Model' indicates the assumed cosmological model, the fourth column shows the fiducial $\Omega_M$, and the successive columns indicate the values of $w_0$ and $w_a$ parameters from the CPL parametrization. The last column indicates the source from which we take the fiducial values. The symbol $^{*}$ denotes the SN+CMB. The symbol ** denotes the DES parameters alone without external probes. All the constraints quoted here consider values when both statistical and systematic uncertainties are included. Legend of Source column: [1]=\citep{Scolnic2018P}; [2]=\citep{Brout2022}; [3]=\citep{Betoule14};[4]=\citep{descollaboration2024darkenergysurveycosmology}}.
\label{tab:Omega}
\end{table*}

\begin{figure}[H]
    \centering
    \includegraphics[width=0.7\linewidth]{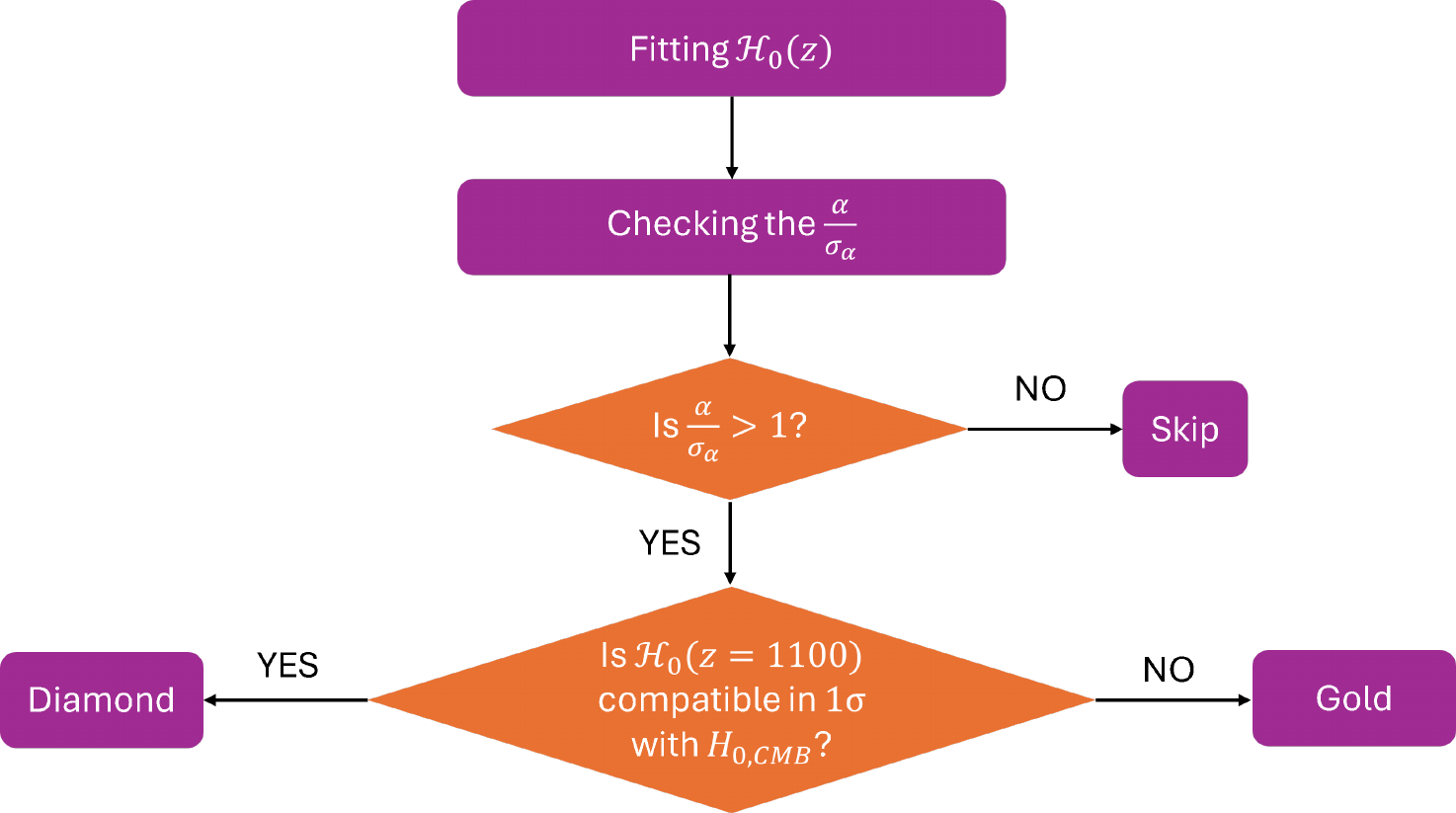}
    \caption{This flowchart summarizes the Diamond and Gold cases classification.}
    \label{fig:flowchart}
\end{figure}

A flowchart that summarizes the classification into Diamond and Gold cases is visible in figure \ref{fig:flowchart}. All analyses assume a flat isotropic universe described by the \(\Lambda\)CDM and the \(w_0w_a\)CDM models. The shown cases consider when $\Omega_M$ is fixed or when it is varied assuming 1 $\sigma$ Gaussian priors (according to the fiducial values tabulated in table \ref{tab:Omega}) within the $\Lambda$CDM framework. Along with these results, we show also the cases where $w_{0},w_{a},\Omega_M$ are fixed but also assuming Gaussian priors within 1 $\sigma$ for $\Omega_M$ and $w_a$ in the $w_{0}w_{a}$CDM model (again with the constraints in table \ref{tab:Omega}). When we consider the Master Sample, since we have more data available, we have expanded the parameters assuming 5 $\sigma$ priors.

\noindent Also for the Master sample, the fiducial values and 1 $\sigma$ intervals are all tabulated in table \ref{tab:Omega}.
We emphasize that the analysis presented here with the Pantheon sample differs from the one shown in \citet{Dainotti2021hubble}, where $H_0=73.5$ is used as a calibration value. Instead, we use $H_0=70$ to be consistent with the $M$ values of the calibration of several catalogs.
For the complete analysis that follows, we report the following cases:

\begin{itemize}
\item $\Lambda$CDM, varying only $H_0$ and fixing $\Omega_M$;
\item $\Lambda$CDM, varying $H_0$ and $\Omega_M$;
\item $w_{0}w_{a}$CDM, varying only $H_0$ and fixing $\Omega_M,w_{0},w_{a}$;
\item $w_{0}w_{a}$CDM, varying $H_0,\Omega_M,w_{a}$ and fixing $w_0$.
\end{itemize}

The results provide information on the evolution of \(\mathcal{H}_0(z)\) with redshift, and this analysis is in agreement with previous works \citep{Dainotti2021hubble,Dainotti2022hubble,DESIMONE2024}.
In all plots and analyses that follow, the red curve depicts the best-fit model, while the blue points with error bars show the values of \(H_0\) in each bin with their uncertainties of 1 $\sigma$. The blue color is also used for the horizontal $z$ intervals, and we stress that these are not uncertainties on the $z$ value but rather delimiters of the bin ranges. The same color coding will be adopted in all the following figures. In the plots, we refer to the $w_{0}w_{a}$CDM cases as $w_{a}$CDM given that in our analysis of this model the $w_{0}$ is fixed.
For tables \ref{tab:2Pantheon}, \ref{tab:tab3MW}, \ref{tab:4equispacelogz}, \ref{tab:5Marshall}, and \ref{tab:mastersample}, the fit parameters for $\mathcal{H}_0(z)$ in the Diamond and Gold cases are presented assuming a flat $\Lambda$CDM model and a flat $w_{0}w_{a}$CDM model. The first column indicates the sample, the second column shows the number of bins, and the third column denotes the assumed cosmology. The fourth and fifth columns denote the fit parameters, $\tilde{H}_0$ and $\alpha$, according to Equation \ref{eq:fittingfunction}.
The sixth column denotes the consistency of the evolutionary parameter $\alpha$ with zero in terms of $1\sigma$, represented by the ratio $\alpha/\sigma_\alpha$, and the seventh column denotes the extrapolated $\mathcal{H}_0(z = 1100)$, representing the $H_0$ in the LSS. The eighth and ninth columns represent the weighted averages among the bins for the $\Omega_M$ and $w_a$ parameters when they are free to vary. The $w_0$ is always fixed according to table \ref{tab:Omega}. The columns tenth and eleventh show the AIC for the power law model and the linear model, respectively.

\subsection{Equi-population binning}
\begin{figure} 
    \centering
    \includegraphics[width=0.45\textwidth]{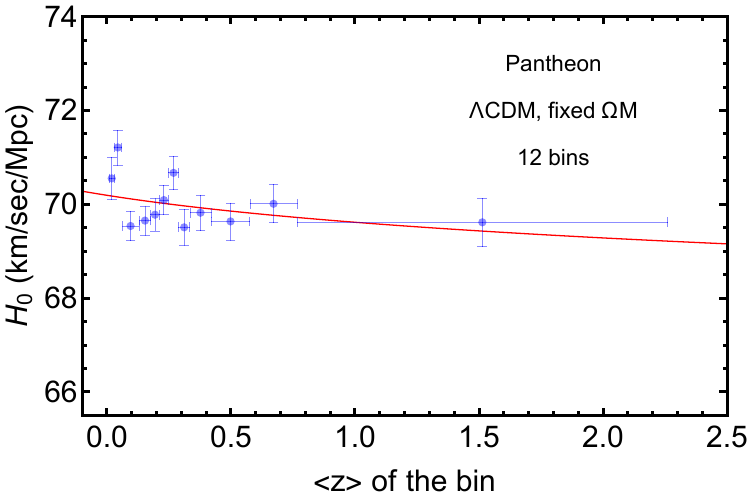}
    \includegraphics[width=0.45\textwidth]{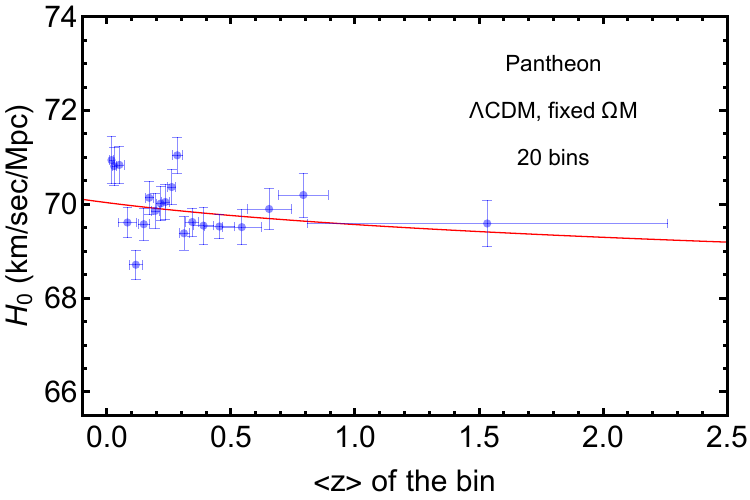}
    \includegraphics[width=0.45\textwidth]{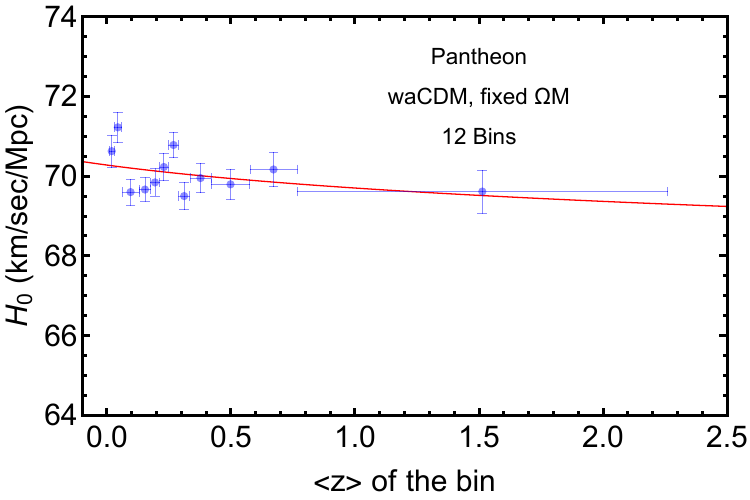}
    \includegraphics[width=0.45\textwidth]{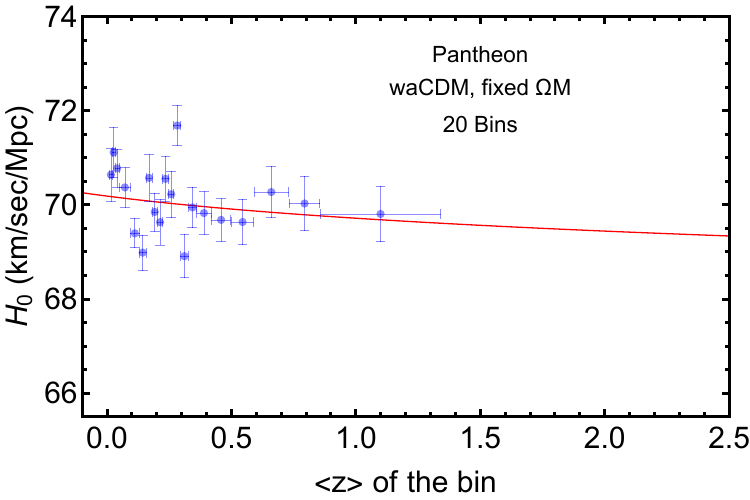}

    \caption{The fitting of $H_0$ values as a function of $z$ in the context of the equi-populated binning approach with fixed $\Omega_M$. 
    The Pantheon sample is calibrated with $H_0=70$ and $\Omega_M = 0.298$. \textbf{Top left and right:} show 12 and 20 bins within $\Lambda$CDM model, respectively. 
    {\bf Lower left and right:} show 12 and 20 bins within $w_0w_a$CDM model. The  results of these plots are summarized in table \ref{tab:2Pantheon} and the corresponding fiducial values are reported in table \ref{tab:Omega}.} 
    \label{fig:PantheonequipopulationfixedOm}
\end{figure}

\begin{figure}[H] 
    \centering
    \includegraphics[width=0.45\textwidth]{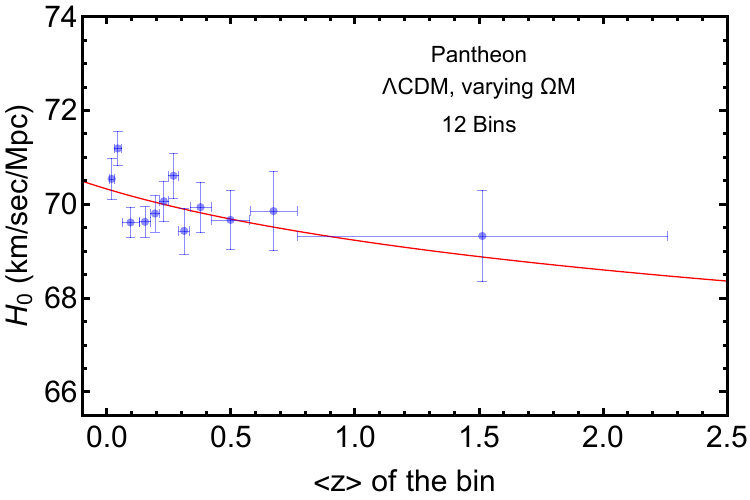}
    \includegraphics[width=0.45\textwidth]{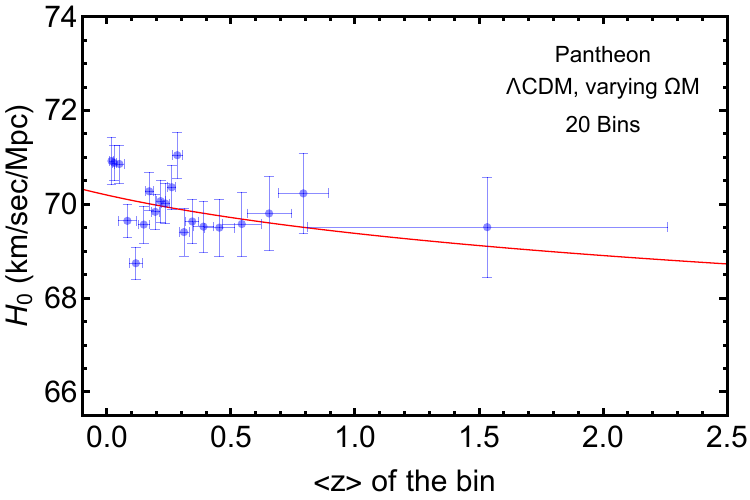}
    \includegraphics[width=0.45\textwidth]{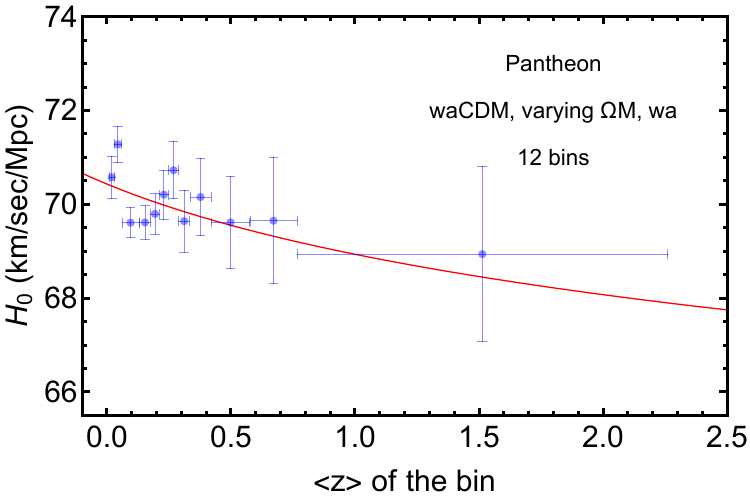}
    \includegraphics[width=0.45\textwidth]{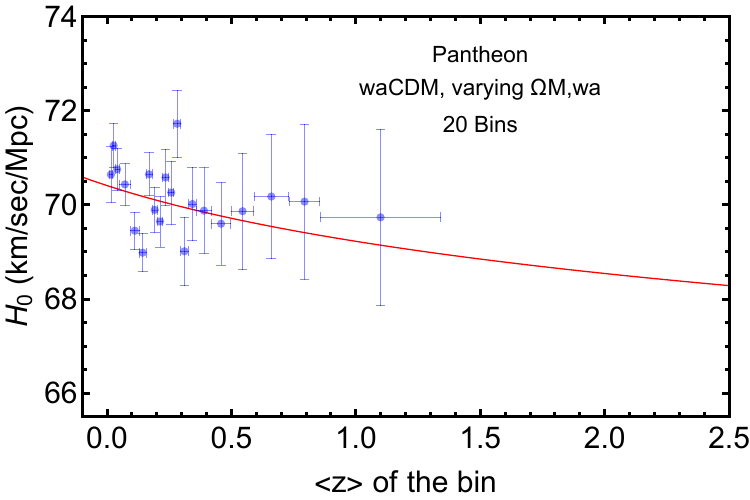}
    \caption{The fitting of $H_0$ values as a function of $z$ in the context of the equi-populated binning approach in the case of varying $\Omega_M$. 
    The Pantheon sample is calibrated with $H_0=70$. \textbf{Top left and right:} show 12 and 20 bins within $\Lambda$CDM model, respectively. \textbf{Bottom left and right:} show 12 and 20 bins within $w_{0}w_{a}$CDM model, respectively. The  results of these plots are summarized in magenta the middle and lower parts of table \ref{tab:2Pantheon} and the corresponding fiducial values are reported in table \ref{tab:Omega}.} \label{fig:PantheonequipopulationvaryingOm}
\end{figure}

This Section discusses the results of the equi-population binning.
Table \ref{tab:2Pantheon} presents the fit parameters \(\tilde{H}_0\) and \(\alpha\) for the Pantheon sample divided into equi-populated bins for the Diamond and Gold cases. The fitting functions for 12 and 20 bins are reported in figure \ref{fig:PantheonequipopulationfixedOm} for the cases of fixed $\Omega_M$ cases for both the \(\Lambda\)CDM and the \(w_0w_a\)CDM models, while the varying $\Omega_M$ cases are plotted in figure \ref{fig:PantheonequipopulationvaryingOm}.

For Diamond cases with fixed $\Omega_M$ for the \(\Lambda\)CDM and the \(w_0w_a\)CDM models in $12$ and $20$ bins, the fitted \(\tilde{H}_0\) ranges from \(70.04 \pm 0.14\) to \(70.28 \pm 0.17\). The parameter \(\alpha\) ranges from \(0.010 \pm 0.006\) to \(0.012 \pm 0.008\). The consistency of \(\alpha\) with zero, represented by \(\alpha/\sigma_\alpha\), ranges from $1.1$ to $1.5$, indicating mild evidence of evolution. The extrapolated \(\mathcal{H}_0(z = 1100)\), representing the $H_0$ in the LSS, ranges from \(64.60 \pm 3.54\) to \(65.58 \pm 4.00\). These values are significantly lower than the local \(H_0\) value and compatible in $1\sigma$ with the CMB value. There is only one Diamond case corresponding to varying $\Omega_M$, which is 20 bin $\Lambda$CDM. The $H_0$ value for it is given by $70.20 \pm 0.17$, while the $\alpha$ value is $0.017 \pm 0.010$. \(\alpha/\sigma_\alpha\) value for the case is $1.6$, while the extrapolated \(\mathcal{H}_0(z = 1100)\) is given  by $62.37\pm4.62$. 

The Gold cases, for the \(\Lambda\)CDM and the \(w_0w_a\)CDM models are all with varying $\Omega_m$. For these cases, \(\tilde{H}_0\) ranges from $70.32\pm0.20$ to $70.43\pm0.23$ while the $\alpha$ parameter ranges from $0.022\pm0.012$ to $0.031\pm0.018$. \(\alpha/\sigma_\alpha\) ranges from $1.4$ to $1.8$ hinting towards a mild variation. The extrapolated \(\mathcal{H}_0(z = 1100)\) has a range from $56.68\pm7.23$ to $60.03\pm5.18$.  

Considering the values reported in Table \ref{tab:2Pantheon}, the $\log BF_{P,L}$ values indicate that there is no statistically significant preference for either the linear model or the power law model. Both the \(\Lambda\)CDM and the \(w_0w_a\)CDM models show similar trends, demonstrating that this evolutionary effect exists regardless of the cosmological models and the variations in the DE EoS.
These findings confirm a slow decreasing trend of \(\mathcal{H}_0(z)\) for the Diamond and Gold samples. These cases suggest that the tension between late and early universe measurements of \(H_0\) remains. 

We performed an Orthogonal Distance Regression (ODR, \citealt{ODR}) fit for 20 bins of Pantheon, $\Lambda$CDM, varying $\Omega_M$ and compared with the non-weighted fitting of the power law model. The ODR results give $\tilde{H}_0=70.22 \pm 0.13$, $\alpha=0.018 \pm 0.008$ while the Levenberg-Marquardt fitting parameters are $\tilde{H}_0=70.20 \pm 0.17$, $\alpha=0.013 \pm 0.008$. For both the parameters, if we compare the $z-score$ we have: $z-score(\alpha)=0.4$, $z-score(\tilde{H}_0)=0.09$. This clearly show that there is no difference between the ODR and the weighted Levenberg-Marquardt fitting procedures.

The consistency of the results across the $\Lambda$CDM and $w_{0}w_{a}$CDM models reinforces the validity of these observations. 

\begin{table*}[ht!]
\renewcommand{\arraystretch}{1.1} 
\centering
\begin{adjustbox}{width=1.1\textwidth,center}
\begin{tabular}{|c|c|c|c|c|c|c|c|c|c|c|c|c|}
\hline
\multicolumn{13}{|c|}{{Equi-population, Diamond cases, Pantheon, best likelihood}}\tabularnewline
\hline
Bins & Model & $\tilde{H}_0$ & $\alpha$ & $\alpha / \sigma_{\alpha}$ & $\mathcal{H}_0(z=1100)$ & $\Omega_M$ & $w_a$ &  $AIC_{PL}$ & $AIC_{linear}$ & $BIC_{PL}$ & $BIC_{linear}$ & $\log BF_{P,L}$ \\
\hline

\multicolumn{13}{|c|}{{Fixed $\Omega_M$, $w_0$, $w_a$}}\tabularnewline
\hline
12 & $\Lambda$CDM & $70.19 \pm 0.17$ & $0.012 \pm 0.008$ & $1.5$ & $64.60 \pm 3.54$ &  & - & 25.1 & 25.6 & 26.1 & 26.6 & 0.25 \tabularnewline
\hline
20 & $\Lambda$CDM & $70.04 \pm 0.14$ & $0.010 \pm 0.006$ & $1.5$ & $65.46 \pm 3.00$ & - & - & 51.8 & 52.4 & 53.8 & 54.4 & 0.30 \tabularnewline
\hline
12 & $w_0w_a$CDM & $70.28 \pm 0.17$ & $0.012 \pm 0.008$ & $1.5$ & $64.67 \pm 3.63$ & - & - & 26.8 & 27.2 & 27.8 & 28.2 & 0.20 \tabularnewline
\hline
20 & $w_0w_a$CDM & $70.19 \pm 0.17$ & $0.010 \pm 0.009$ & $1.1$ & $65.58 \pm 4.00$ & - & - & 59.2 & 59.5 & 61.2 & 61.5 & 0.15 \tabularnewline
\hline
\multicolumn{13}{|c|}{{Varying $\Omega_M$, $w_a$}}\tabularnewline
\hline
20 & $\Lambda$CDM & $70.20 \pm 0.17$ & $0.017 \pm 0.010$ & $1.6$ & $62.37 \pm 4.62$ & $0.298 \pm 0.005$ & - & $52.1$ & $52.7$& 54.1 & 54.7 & 0.30 \tabularnewline
\hline
\multicolumn{13}{|c|}{{Equi-population, Gold cases, Pantheon, best likelihood}}\tabularnewline
\hline
\multicolumn{13}{|c|}{{Varying $\Omega_M$, $w_a$}} \tabularnewline
\hline
12 & $\Lambda$CDM & $70.32 \pm 0.20$ & $0.022 \pm 0.012$ & $1.8$ & $60.03 \pm 5.18$ & $0.299 \pm 0.006$ & - & 26.5 & 27.2 & 27.5 & 28.2 & 0.35 \tabularnewline
\hline
12 & $w_0w_a$CDM & $70.43 \pm 0.23$ & $0.031 \pm 0.018$ & $1.7$ & $56.68 \pm 7.23$ & $0.309 \pm 0.005$ & $-0.052 \pm 0.410$ & $31.6$ & $32.1$ & 32.6 & 33.1 & 0.25 \tabularnewline
\hline
20 & $w_0w_a$CDM & $70.41 \pm 0.23$ & $0.023 \pm 0.017$ & $1.4$ & $59.93 \pm 7.14$ & $0.308 \pm 0.004$ & $-0.153 \pm 0.165$ & 57.4 & 57.8 & 59.4 & 59.8 & 0.20 \tabularnewline
\hline
\end{tabular}
\end{adjustbox}
\caption{Fit parameters for $\mathcal{H}_0(z)$ in the Pantheon sample for Diamond and Gold cases of the equi-population binning method, assuming a flat $\Lambda$CDM model and a flat $w_{0}w_{a}$CDM model. The first column indicates the number of bins, the second column denotes the assumed cosmology, and the third and fourth columns denote the fit parameters, $\tilde{H}_0$ and $\alpha$, according to Equation \ref{eq:fittingfunction}.
The fifth column denotes the consistency of the evolutionary parameter $\alpha$ with zero in terms of $1\sigma$, represented by the ratio $\alpha/\sigma_{\alpha}$ and the sixth column denotes the extrapolated $\mathcal{H}_0(z = 1100)$, representing the $H_0$ in the LSS. The seventh and eight columns denote the values of $\Omega_M$, and $w_a$, respectively. The ninth and tenth columns denote the BIC for the power law and the linear model, respectively. The last column denotes the $\log BF_{P,L}$ value defined in Equation \ref{eq:logbayesfactor}. The upper panel reports the fixed $\Omega_M$ cases, while the middle and lower panels of the table show the variable $\Omega_M$ ones, considering that in the latter, the Gold cases only are reported. All the uncertainties are given in $1\sigma$. The fiducial values are the same as in table \ref{tab:Omega}.}
\label{tab:2Pantheon}
\end{table*}

\subsubsection{The MW binning approach}
\begin{figure}
   \centering
   \includegraphics[width=0.412\textwidth]{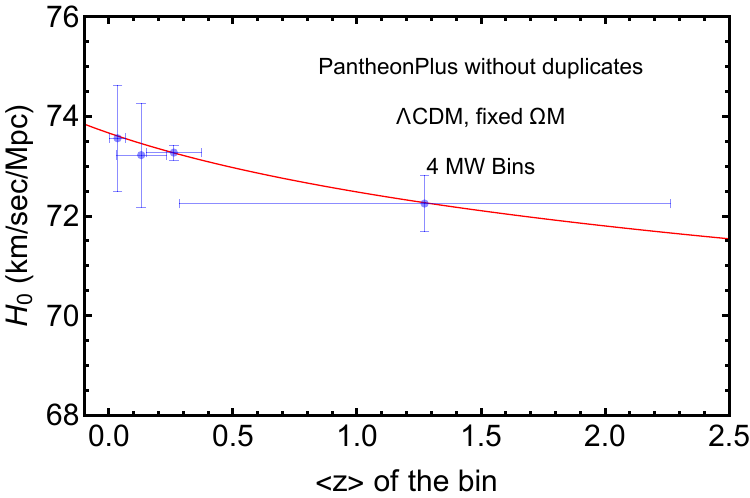}
    \includegraphics[width=0.412\textwidth]{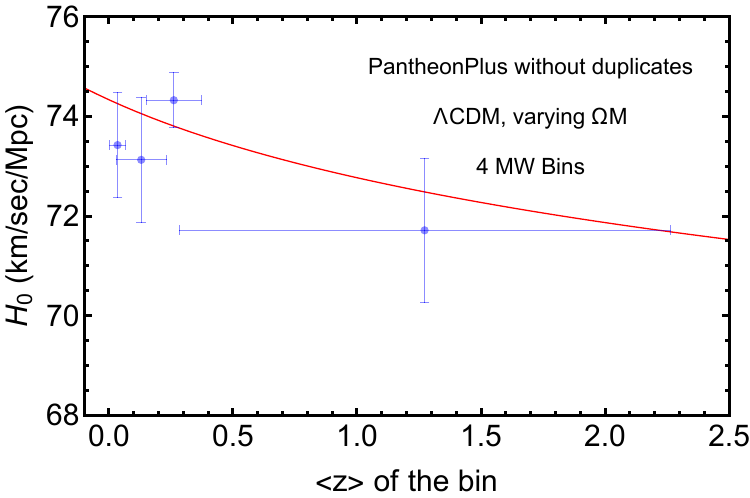}
    \includegraphics[width=0.412\textwidth]{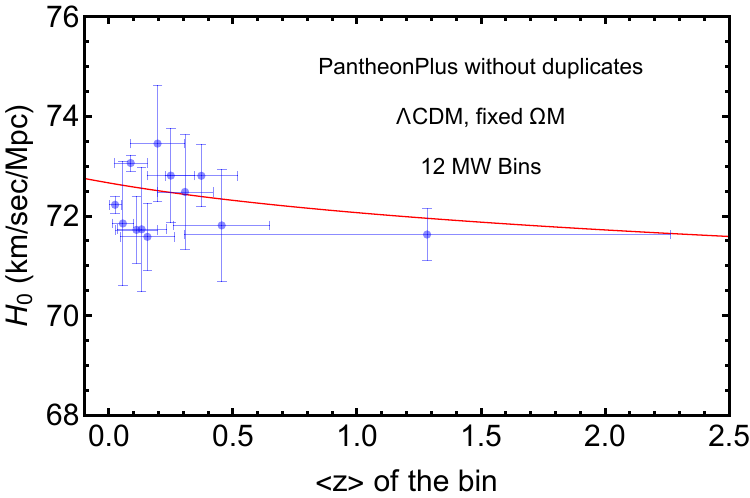}
    \includegraphics[width=0.412\textwidth]{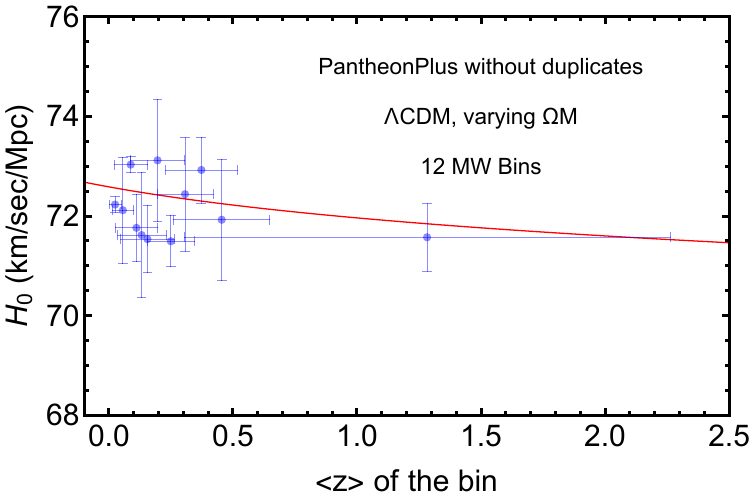}
    \includegraphics[width=0.412\textwidth]{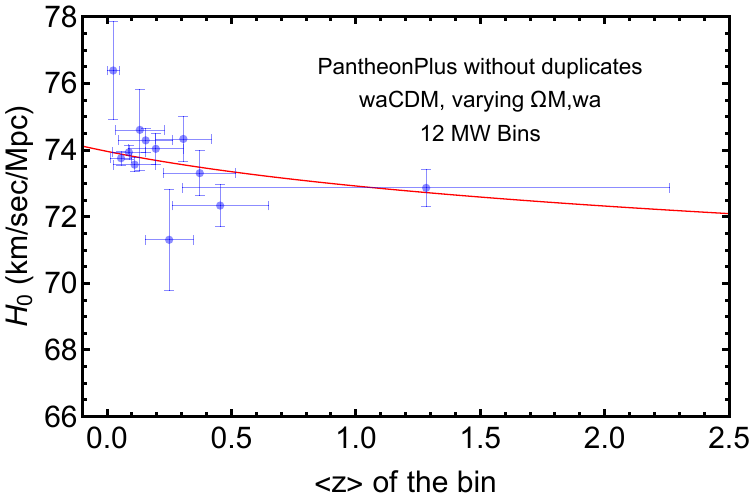}
    \caption{The fitting of the $H_0$ values in the Diamond cases for the equi-populated MW approach, adopting the best likelihoods, reported in the upper part of table \ref{tab:tab3MW}. {\bf Upper left and right}: P+ without duplicates with 4 bins for fixed and varying $\Omega_M$ in the $\Lambda$CDM, respectively. {\bf Middle}: the same as the Upper but for 12 bins.    
    {\bf Bottom}: P+ without duplicates with 12 bins for varying $\Omega_M$ in the $w_0w_a$CDM.}
    \label{fig:6}
\end{figure}

\begin{figure}
   \centering 
    \includegraphics[width=0.35\textwidth]{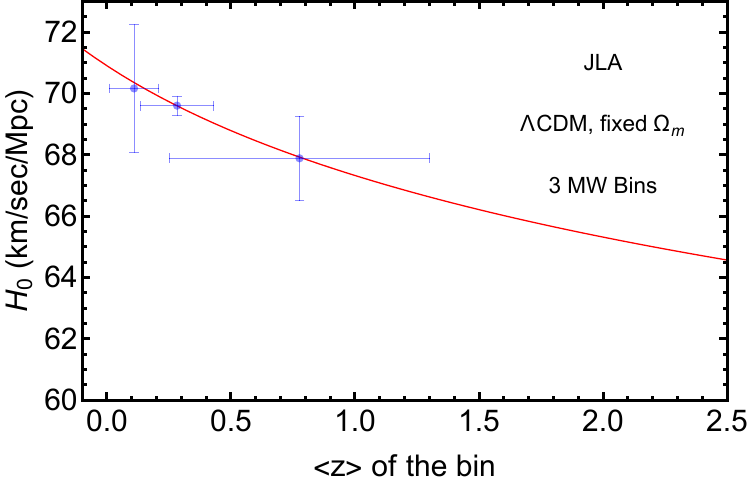}
    \includegraphics[width=0.35\textwidth]{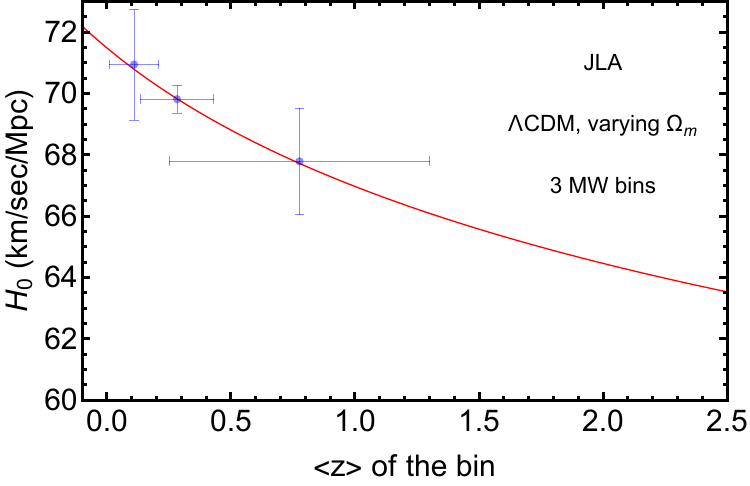}
    \includegraphics[width=0.35\textwidth]{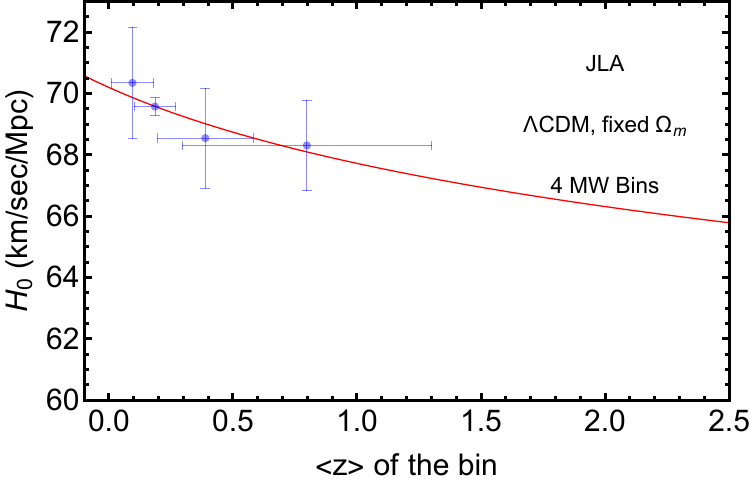}
    \includegraphics[width=0.35\textwidth]{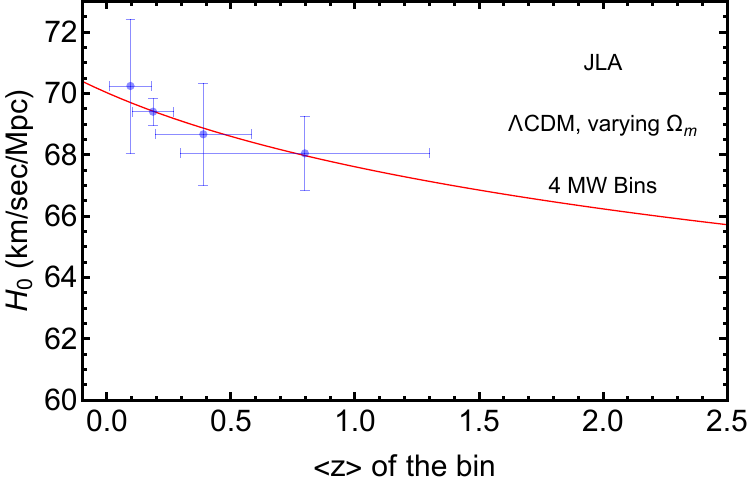}
    \includegraphics[width=0.35\textwidth]{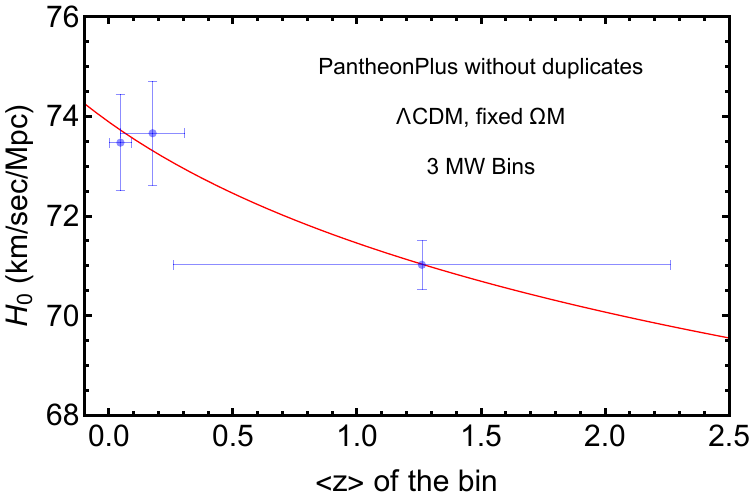}
    \includegraphics[width=0.35\textwidth]{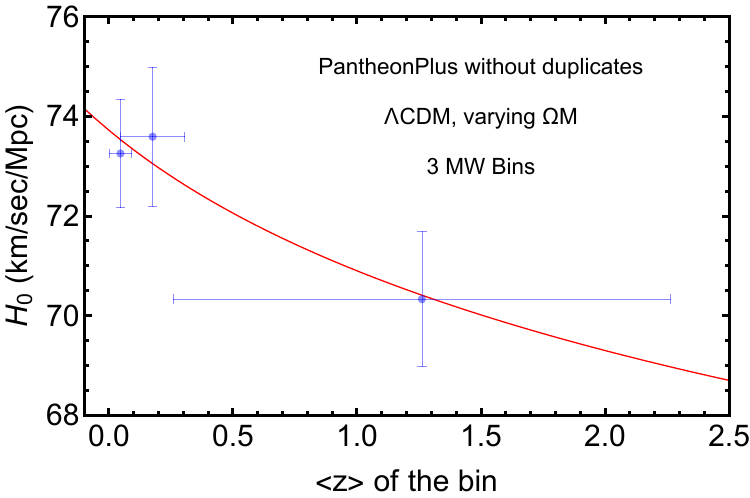}
    \includegraphics[width=0.35\textwidth]{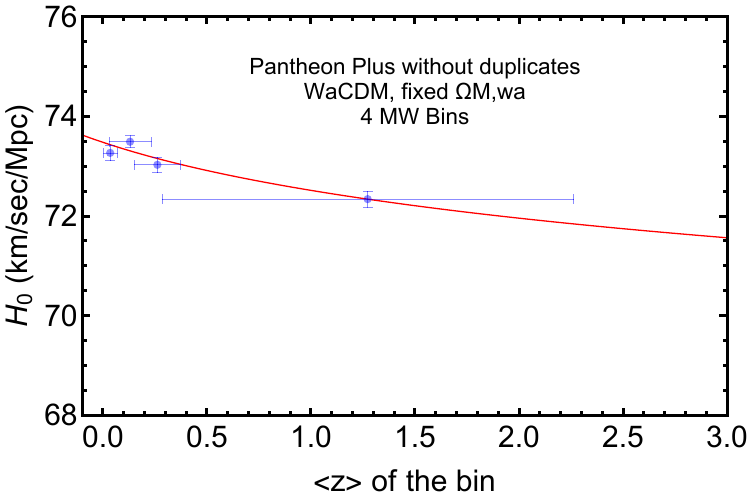}
    \includegraphics[width=0.35\textwidth]{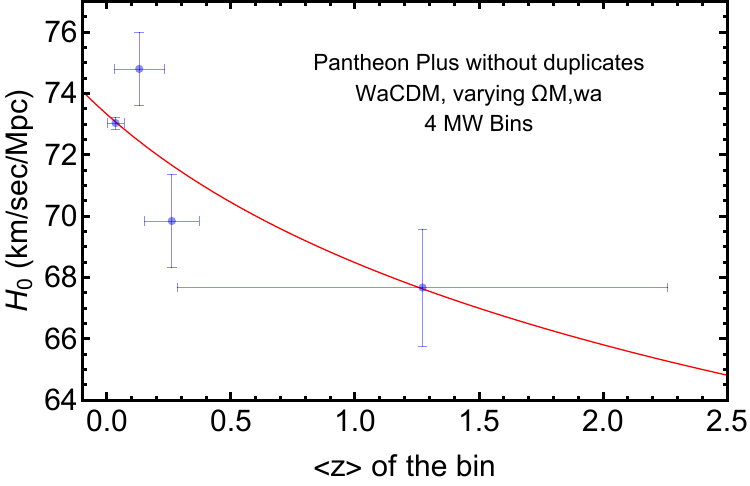}
     \includegraphics[width=0.35\textwidth]{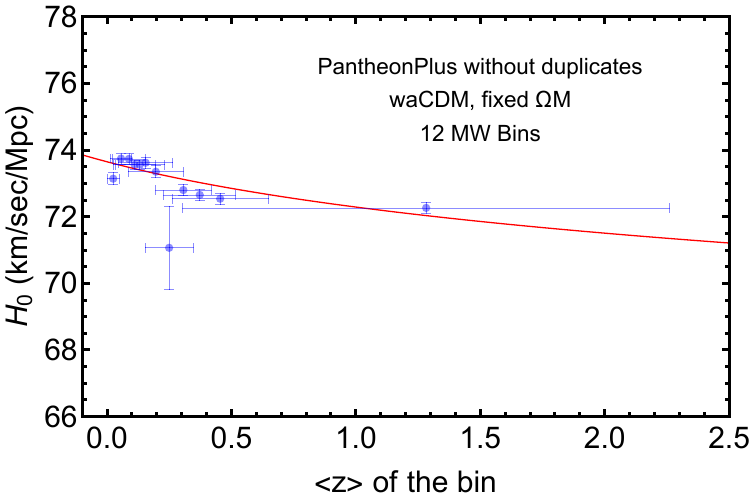}    
    \includegraphics[width=0.35\textwidth]{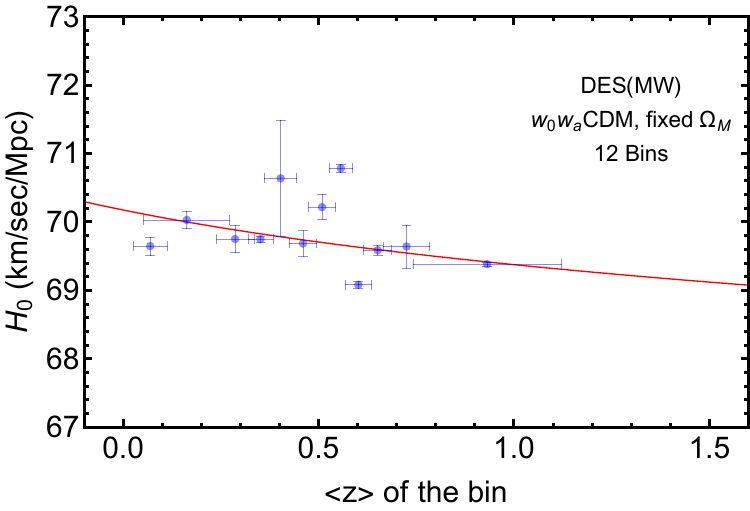}
   \caption{The fitting of $H_0$ values in the Gold cases for the equi-populated MW approach, adopting the best likelihoods, see the lower panel of table \ref{tab:tab3MW}. {\bf First row:} JLA in 3 bins with fixed (left) and varying $\Omega_M$ (right) within $\Lambda$CDM. {\bf Second row:} the same as the first row but in 4 bins. {\bf Third row:} P+ without duplicates in 3 bins with fixed (left) and varying $\Omega_M$ (right) within $\Lambda$CDM.  {\bf Fourth row}: P+ without duplicates in 4 bins with fixed (left) and varying $\Omega_M$ (right) within $w_0w_a$CDM. {\bf Bottom}: P+ without duplicates in 12 bins with fixed $\Omega_M$ within $w_{0}w_{a}$CDM(left) and DES in 12 bins with fixed $\Omega_M$ within $w_{0}w_{a}$CDM (right).}
    \label{fig:7}
\end{figure}

Since we know that changing more parameters simultaneously will reduce the constraining power on the cosmological parameters, we need to implement MW.
The MW procedure, chosen here, allows us to add a fixed number of SNe Ia in each bin on the lower and higher redshift boundaries.
This way, each bin will have a larger number of SNe Ia compared to the equi-population method. With this approach, we obtain 5 Diamond and 10 Gold cases. The results of diamond cases are in Figure \ref{fig:6} and the results of gold cases are in Figure \ref{fig:7} respectively.

Although this method is not needed for constraining the parameters in the equi-population scenario, we here test its reliability so that we can safely use it in the other cases, see table \ref{tab:tab3MW}.
In table \ref{tab:tab3MW}, with the MW, we have analyzed the P+ in 3 bins in the $\Lambda$CDM framework, in 4 and 12 bins in both $\Lambda$CDM and $w_{0}w_{a}$CDM scenarios, the JLA in 3 and 4 bins within the $\Lambda$CDM, and with DES data 12 bins within the $w_0w_a$CDM scenario.
With the MW, we note some steepening of the trend of $\alpha$: indeed, $\alpha$ ranges from $0.012 \pm 0.011$ to $0.094 \pm 0.073$.
We note that the trend of $\alpha/\sigma_\alpha$ varies widely from 1 to even 9.1. The reason for such a high value is that when non-Gaussian distributions are the best fit, sometimes there is no constraining power to obtain the values of $H_0$, $\Omega_M$, and $w_{a}$. Thus, successive fit distributions are chosen that guarantee the constraining power on our parameters. When we compare our power law model with the linear model, we have mixed preferences according to the AIC, shown in the last two columns of table \ref{tab:tab3MW}.
For most of the cases, there is no statistically significant evidence of preference for either linear or power law model. However, $\log BF_{P,L}$ value indicate that for P+ and DES samples in 12 bins $w_{0}w_{a}$CDM model (with fixed parameters), power law model is preferred over the linear one.

\subsection{Equi-spacing on the log-z}
\label{subsec5.2}
\begin{figure}
   \centering
   \includegraphics[width=0.412\textwidth] {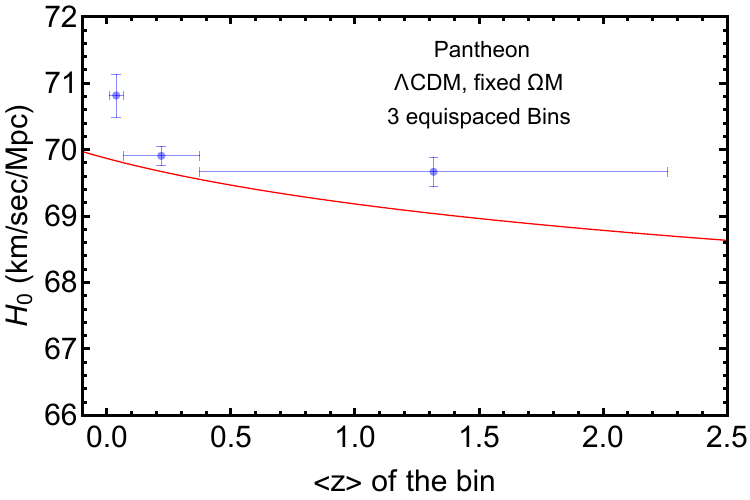}
     \includegraphics[width=0.412\textwidth]{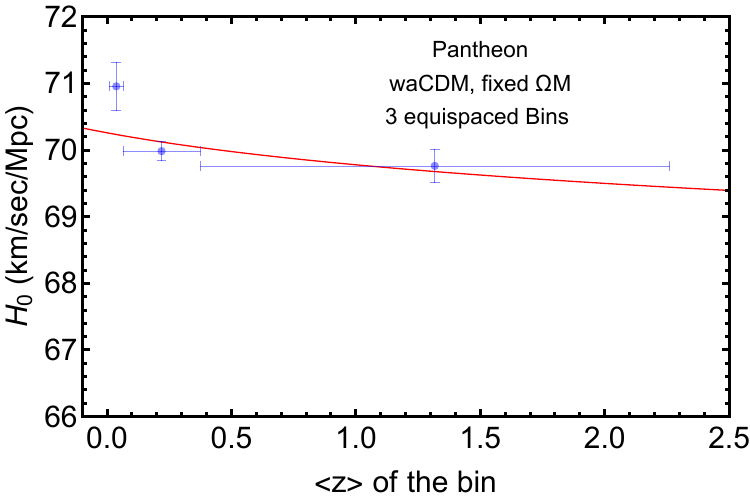}
     \includegraphics[width=0.412\textwidth]{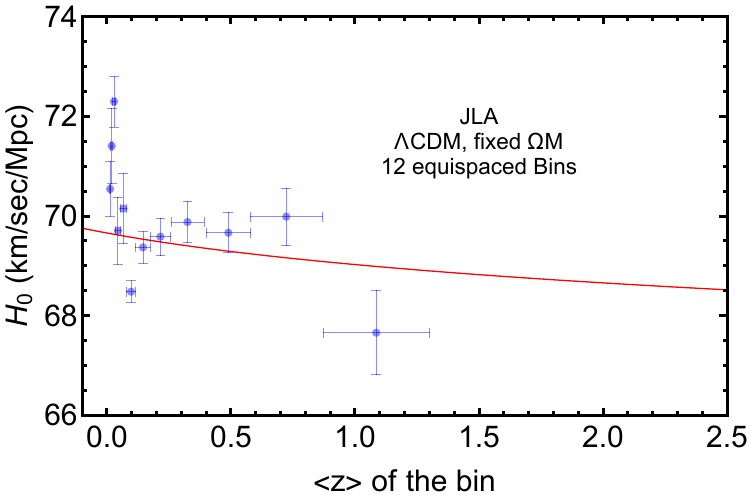}
     \includegraphics[width=0.412\textwidth]{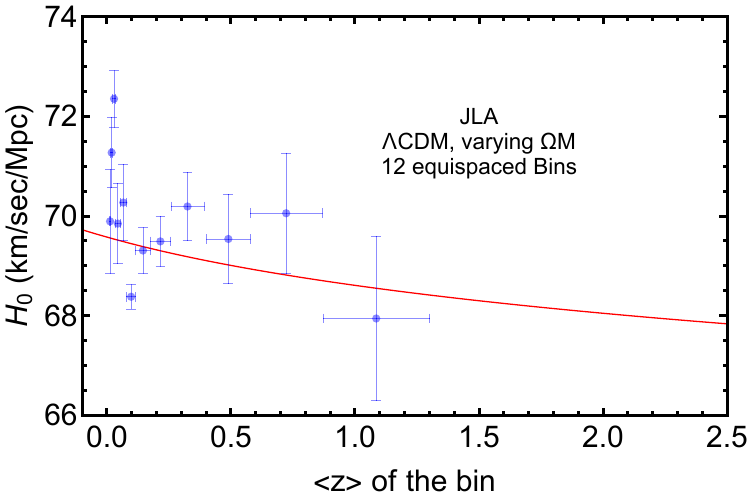}
     \includegraphics[width=0.412\textwidth]{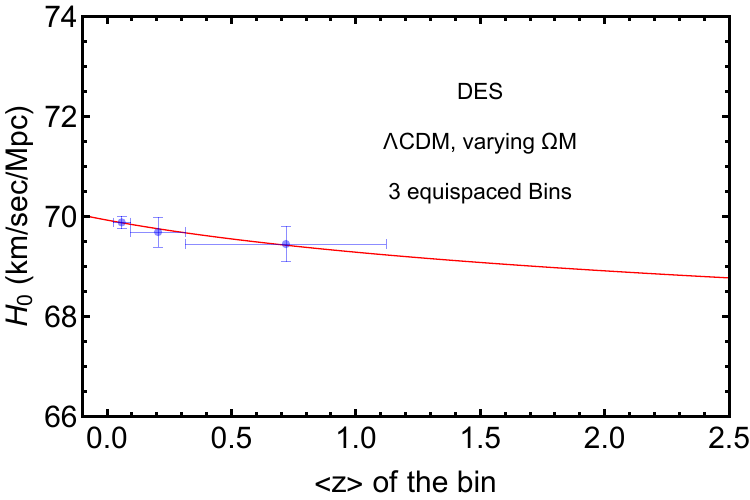}
    \includegraphics[width=0.412\textwidth]
    {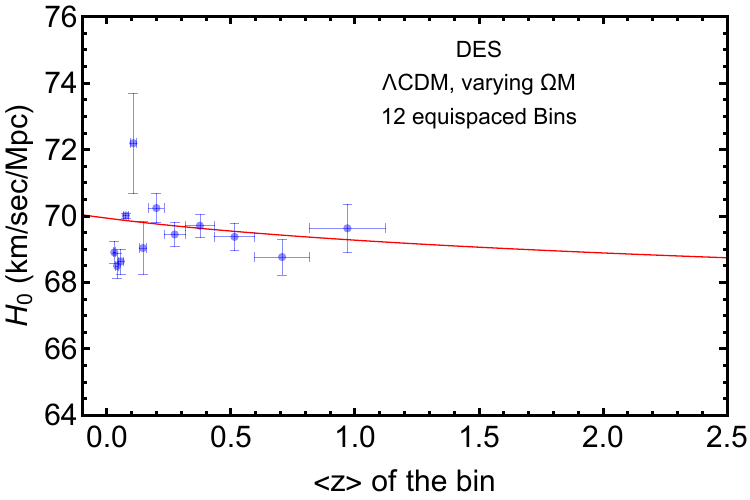}

   \caption{The fitting of $H_0$ values as a function of $z$ in the context of the equi-spacing binning approach on the $\log-z$. The Pantheon, JLA, and DES samples are calibrated with $H_0 = 70$. \textbf{Top left:} shows 3 bins within $\Lambda$CDM model for Pantheon, fixing $\Omega_M$. \textbf{Top right:} shows 3 bins within $w_{0}w_{a}$CDM model for Pantheon, fixing $\Omega_M$. \textbf{Middle left:} shows 12 bins within $\Lambda$CDM model for JLA, fixing $\Omega_M$. \textbf{Middle right:} shows 12 bins within $\Lambda$CDM model for JLA, varying $\Omega_M$. \textbf{Bottom left:} shows 3 bins within $\Lambda$CDM model for DES, varying $\Omega_M$. \textbf{Bottom right:} the same as the Bottom left panel but for 12 bins. The results are summarized in the upper part of table \ref{tab:4equispacelogz} and the corresponding fiducial values are reported in table \ref{tab:Omega}.}
    \label{fig:8}
\end{figure}

\begin{figure}
   \centering
       \includegraphics[width=0.24\textwidth]
    {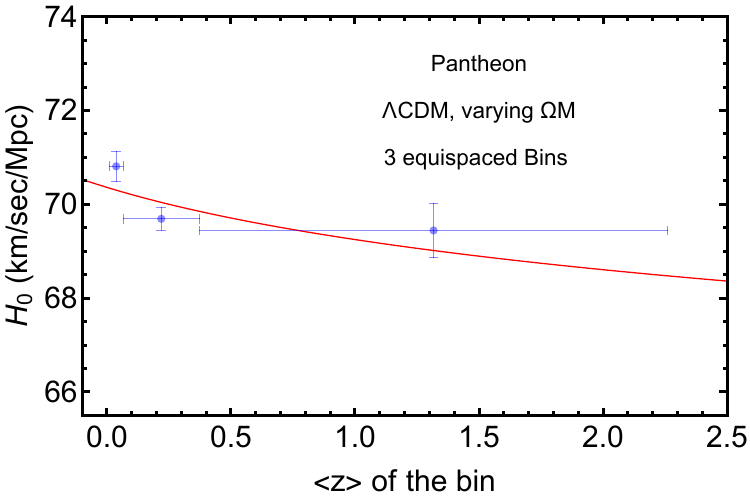}
    \includegraphics[width=0.24\textwidth]{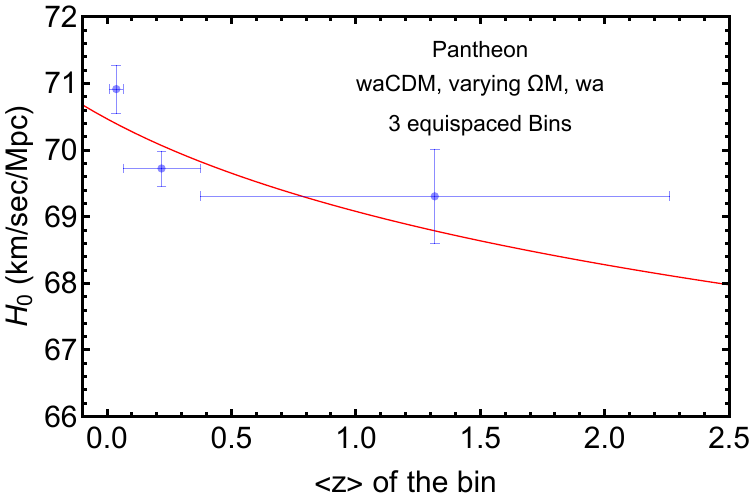}
    \includegraphics[width=0.24\textwidth]
    {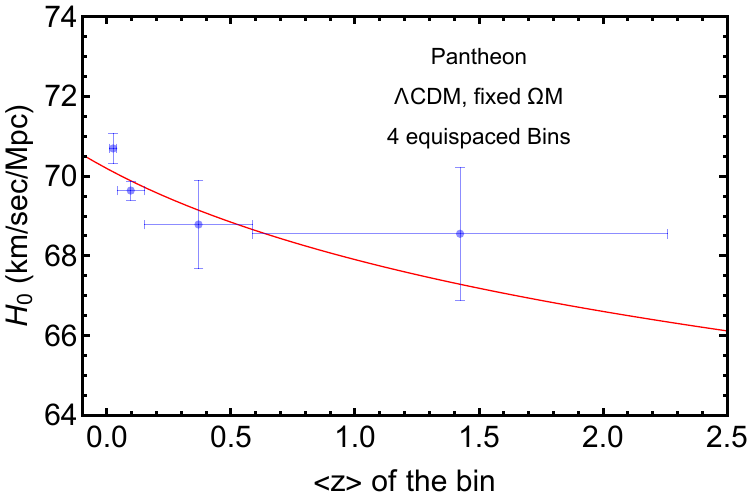}
    \includegraphics[width=0.24\textwidth]
    {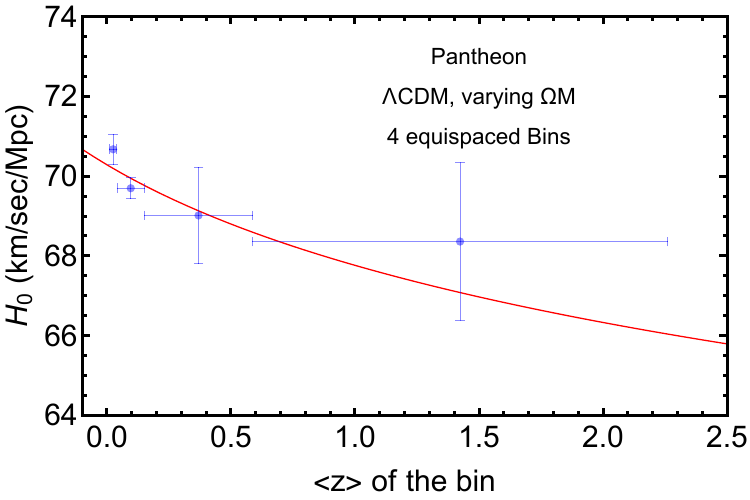}
    \includegraphics[width=0.24\textwidth]
    {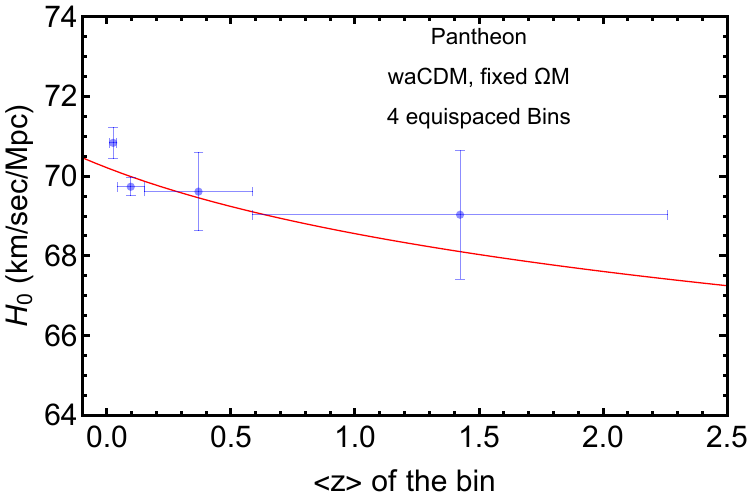}
    \includegraphics[width=0.24\textwidth]
    {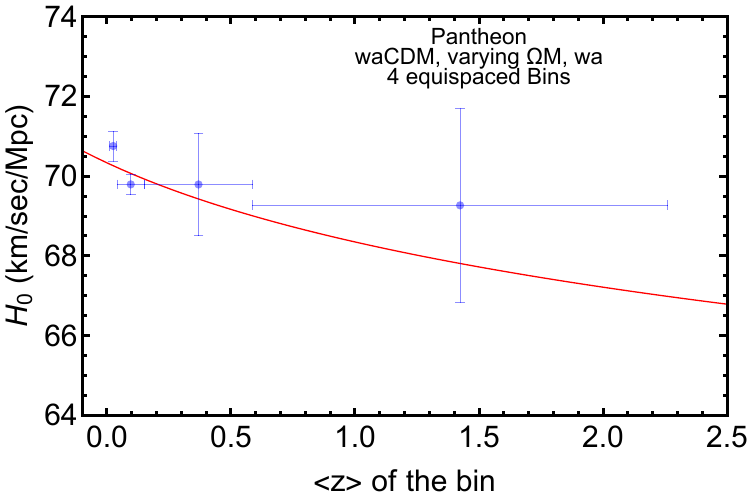}
    \includegraphics[width=0.24\textwidth]{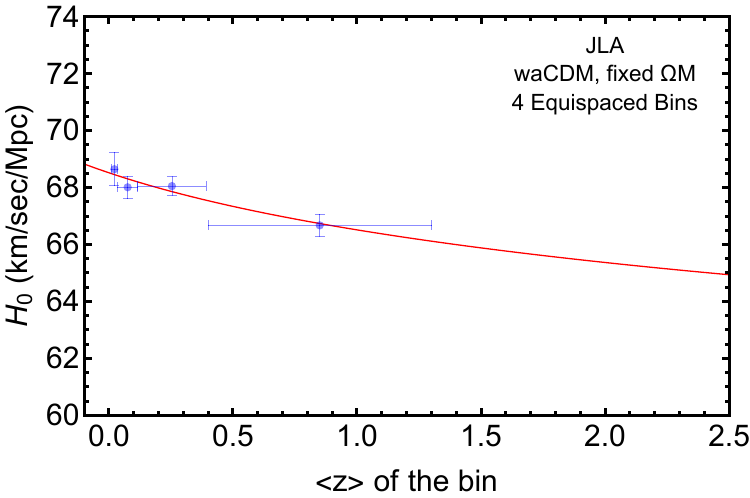}
    \includegraphics[width=0.24\textwidth]
    {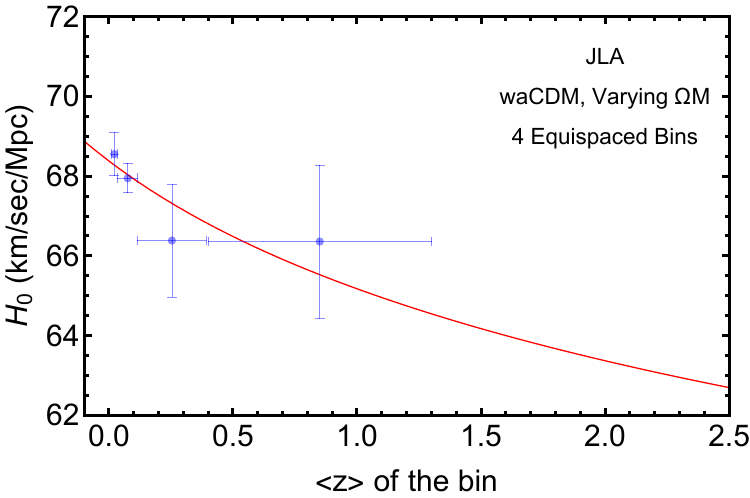}
    \includegraphics[width=0.24\textwidth]
    {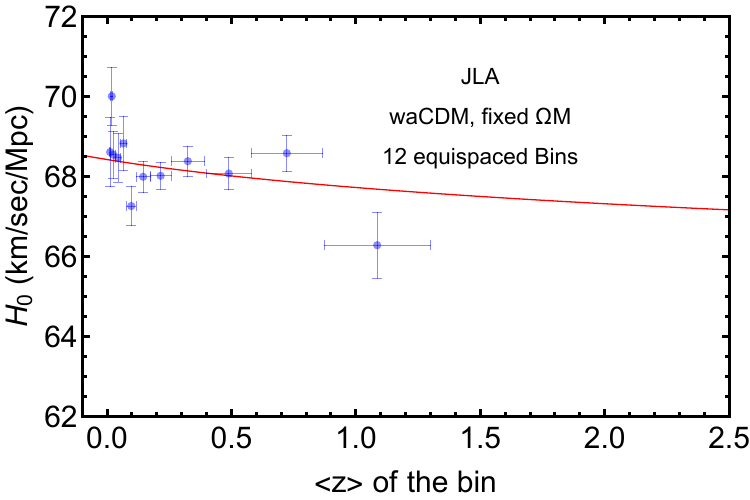}
    \includegraphics[width=0.24\textwidth]{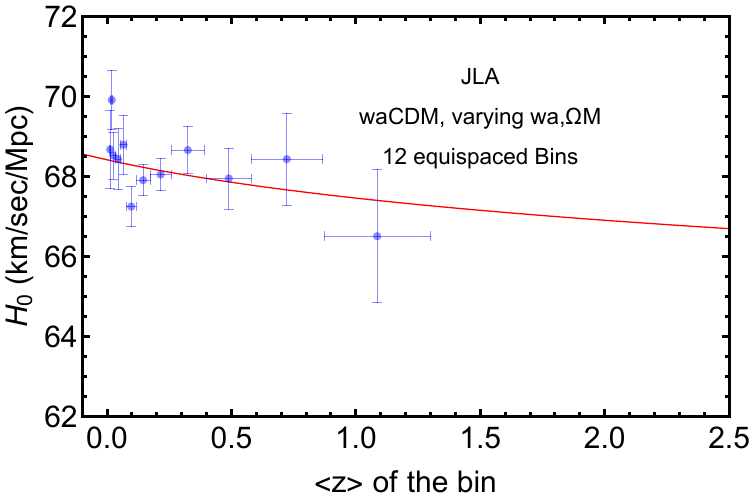}
    \includegraphics[width=0.24\textwidth]{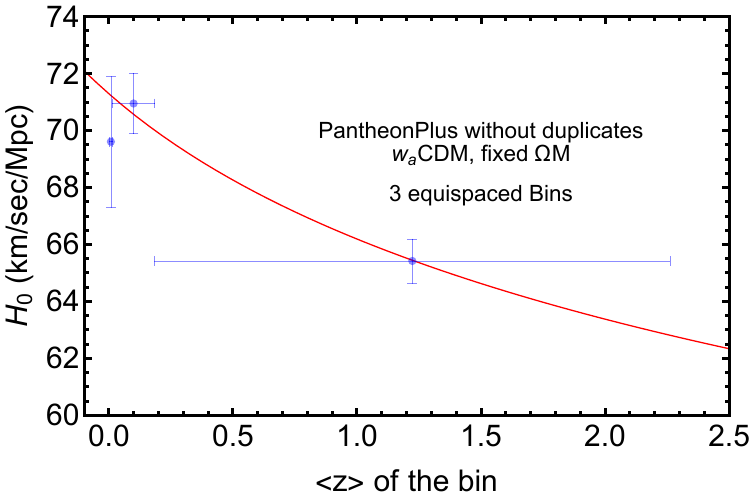}
    \includegraphics[width=0.24\textwidth]{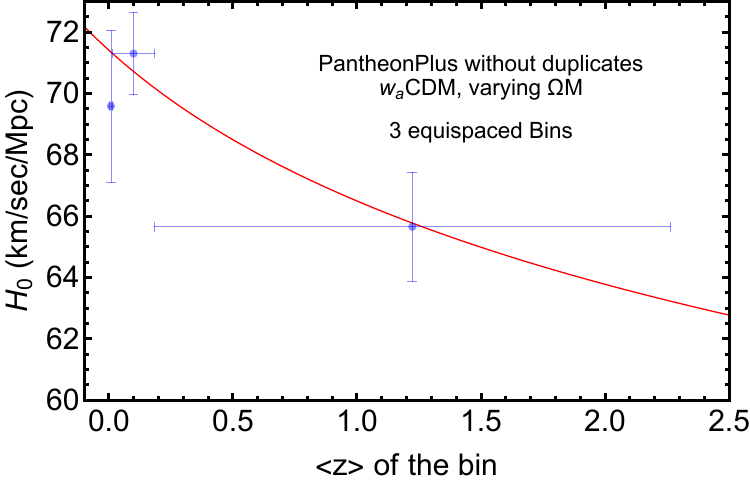}
    \includegraphics[width=0.24\textwidth]{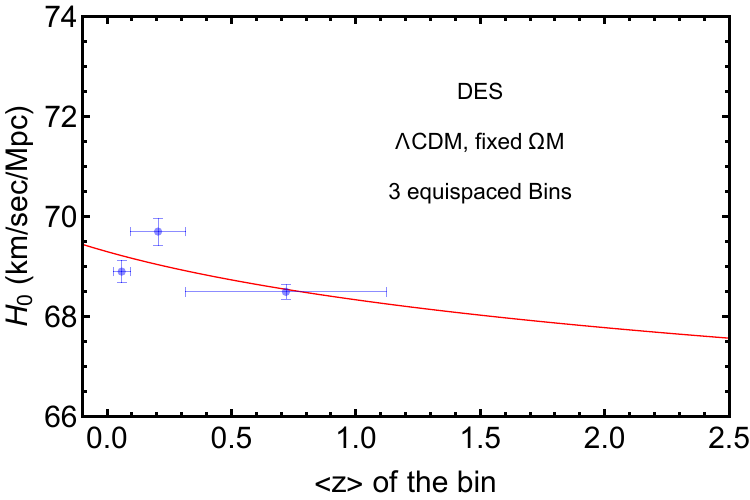}
    \includegraphics[width=0.24\textwidth]{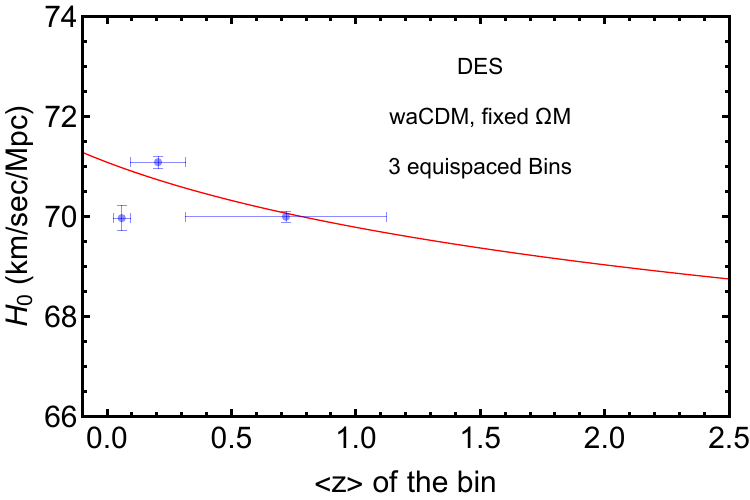}
    \includegraphics[width=0.24\textwidth]{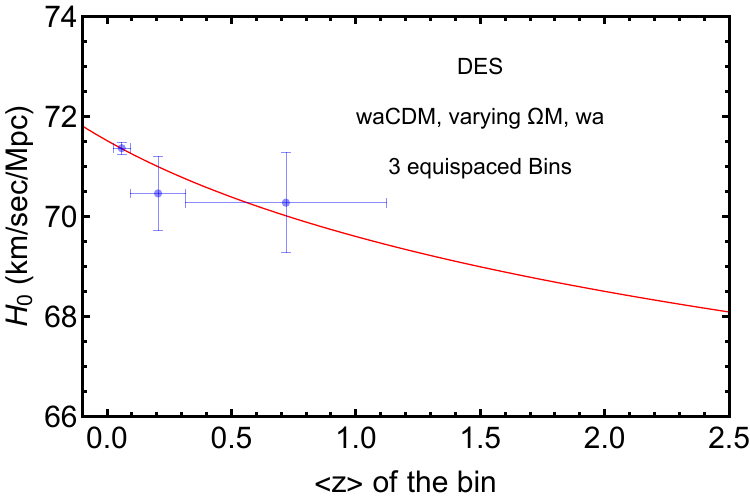}
    \includegraphics[width=0.24\textwidth]{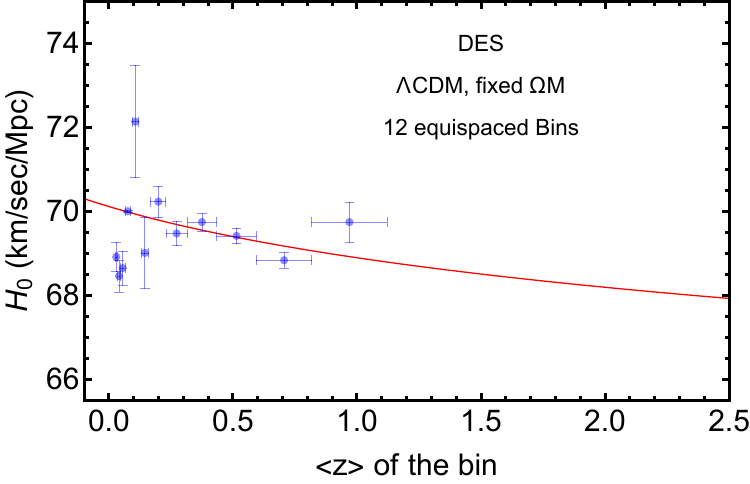}
    \caption{The fitting of $H_0$ values as a function of $z$ in the context of the equi-spacing binning approach on the $\log-z$. The Pantheon, JLA, and DES samples are calibrated with $H_0 = 70$, while P+ is calibrated with $H_0=73.04$. \textbf{First row:} shows the Pantheon sample from the left to the right, with 3 bins within $\Lambda$CDM model varying $\Omega_M$, 3 bins in the $w_{0}w_{a}$CDM models with varying parameters, respectively, 4 bins within $\Lambda$CDM with fixed and varying $\Omega_M$, respectively. \textbf{Second row:} shows the 4 bins of the Pantheon (first and second panels) and JLA (third and fourth panels) within the $w_{0}w_{a}$CDM model, with fixed and varying $\Omega_M$, respectively.   
        \textbf{Third row:} shows the JLA in 12 bins (first and second panels) and P+ samples in 3 bins (third and fourth panels) within the $w_0w_a$CDM model with fixed and varying $\Omega_M$, respectively.
        \textbf{Fourth row:} includes the DES sample in 3 bins (first panel) and 12 bins (fourth panel) within $\Lambda$CDM with fixed $\Omega_M$. The second and third panels show DES 3 bins with $w_{0}w_{a}$CDM with fixed and varying parameters, respectively.
        The results are summarized in table \ref{tab:4equispacelogz} and the corresponding fiducial values are reported in table \ref{tab:Omega}.}
    \label{fig:9}
\end{figure}

Here, we discuss the equi-spacing in $\log-z$. The Diamond cases are shown in the upper panel of table \ref{tab:4equispacelogz} and figure \ref{fig:9}. 
We report the following cases for the Diamond: the Pantheon sample in 3 bins, JLA in 12 bins, DES in 3 and 12 bins within $\Lambda$CDM, where Pantheon 3 bins are also present in the $w_{0}w_{a}$CDM model. Furthermore, for the Gold cases, we have the additional cases of the Pantheon in 4 bins through the $\Lambda$CDM and the $w_{0}w_{a}$CDM models, 4 bins for JLA and 3 bins for P+ without duplicates, both of them in the $w_{0}w_{a}$CDM model, together with DES 3 bins in $w_{0}w_{a}$CDM.
Regarding the Diamond cases, see the upper part of table \ref{tab:4equispacelogz} and figure \ref{fig:8}, the fit parameter $\tilde{H}_0$ ranges from \(69.58 \pm 0.24\) to \(70.26 \pm 0.18\), while \(\alpha\) ranges from \(0.010 \pm 0.006\) to \(0.020 \pm 0.020\). The ratio \(\alpha/\sigma_\alpha\) varies from 1.0 to 2.0, and the extrapolated \(\mathcal{H}_0(z=1100)\) ranges from \(60.40 \pm 8.51\) to \(65.56 \pm 2.73\).
Inspecting the $\log BF_{P,L}$ values, it can be said that there is no statistically significant preference for either linear or power law models.

Concerning the Gold results, see the lower part of table \ref{tab:4equispacelogz} and figure \ref{fig:9}, the smallest $\tilde{H}_0$ is \(68.38 \pm 0.38\) and the largest is \(72.52 \pm 0.18\). 
In the case of P+ without duplicates, we have also added the analysis in 5 $\sigma$ as an example to show that the results remain compatible in 1 $\sigma$ with previous cases, but only the error bars increase. The analysis extends to P+ without duplicates within the $w_{0}w_{a}$CDM model.
Regarding the trend, $\alpha$ ranges from $\alpha=0.015 \pm 0.012$ to $\alpha=0.107 \pm 0.025$. The latter value has been obtained with 5 $\sigma$, but similar results are obtained within 1 $\sigma$. The $\alpha/\sigma_\alpha$ ranges from 1.0 to 6.1, the latter found with the DES sample. 
Here we show that the Gold cases do not reach the CMB values because the $\alpha$ value is steeper than the other Diamond cases in several samples by $\sim 2$ to $5$ times.
Similar scenario as the Diamond emerges from the inspection of the $\log BF_{P,L}$. None of the models are preferred for most of the cases. For DES 3 bins $w_0w_a$CDM model (fixed parameters), however, the linear model is favored weakly according to $\log BF_{P,L}$.

\begin{table*}[ht!]
\arrayrulecolor{black}
\footnotesize
\renewcommand{\arraystretch}{1.1} 
\centering
\begin{adjustbox}{width=1.2\textwidth,center}
\begin{tabular}{|c|c|c|c|c|c|c|c|c|c|c|c|c|c|}
\hline
\multicolumn{14}{|c|}{Equi-population binning using MW, Diamond  best likelihood} \\
\hline
Sample & Bins & Model & $\tilde{H}_0$ & $\alpha$ & $\alpha / \sigma_{\alpha}$ & $\mathcal{H}_0(z=1100)$ & $\Omega_M$ &  $w_a$ & $AIC_{PL}$ & $AIC_{linear}$ & $BIC_{PL}$ & $BIC_{linear}$ & $\log BF_{P,L}$\\
& & & & & & & & & & & & & \\
\hline
P+ no dupl. & 4 & $\Lambda$CDM & $73.66 \pm 0.29$ & $0.023 \pm 0.013$ & $1.8$ & $62.56 \pm 5.74$ & - & - & $8.6$ & $8.6$ & 7.4 & 7.4 & 0.0  \\
 & & fixed $\Omega_M$ &  &  &  &  & & & & & & & \\
\hline
P+ no dupl. & 4 & $\Lambda$CDM & $74.33 \pm 0.69$ & $0.031 \pm 0.030$ & $1.0$ & $59.95 \pm 12.67$ & $0.299 \pm 0.028$ & $ - $ & $15.8$ & $15.4$ & 14.6 & 14.2 & $-0.20$  \\
(*) & & varying $\Omega_M$ &  &  &  &  & & & & & & & \\
\hline
P+ no dupl. & 12 & $\Lambda$CDM & $72.66 \pm 0.13$ & $0.012 \pm 0.009$ & $1.3$ & $66.86 \pm 4.26$ & - & - &$ 39.1$ & $38.5$ & 40.1 & 39.5 & $-0.30$  \\
 & & fixed $\Omega_M$  &  &  &  &  & & & & & & & \\
\hline
P+ no dupl. & 12 & $\Lambda$CDM & $72.58 \pm 0.13$ & $0.012 \pm 0.011$ & $1.1$ & $66.53 \pm 5.14$ & $0.334 \pm 0.01$ & - & $40.2$ & $39.9$ & 41.2 & 40.9  & $-0.15$  \\
 & & varying $\Omega_M$  &  &  &  &  & & & & & & & \\
 \hline
P+ no dupl. & 12 & $w_0w_a$CDM & $73.96 \pm 0.14$ & $0.020 \pm 0.010$ & $2.1$ & $64.10 \pm 4.30$ & $0.366\pm 0.13$ & $-0.109 \pm 0.041 $ & $28.2$ & $28.4$ & 29.2 & 29.4 & 0.10  \\
 & & varying $\Omega_M, w_a$  &  &  &  &  & & & & & &  &  \\
\hline
\multicolumn{14}{|c|}{Equi-population binning using MW, Gold best likelihood} \\
\hline
Sample & Bins & Model & $\tilde{H}_0$ & $\alpha$ & $\alpha / \sigma_{\alpha}$ & $\mathcal{H}_0(z=1100)$ & $\Omega_M$ &  $w_a$ & $AIC_{PL}$ & $AIC_{linear}$ & $BIC_{PL}$ & $BIC_{linear}$ & $\log BF_{P,L}$\\
& & & & & & & & & & & & & \\
\hline
JLA & 3 & $\Lambda$CDM & $70.90 \pm 1.16$ & $0.075 \pm 0.060$ & $1.2$ & $42.01 \pm 17.78$ & - & - & $11.3$ & $11.3$  & 9.5 & 9.5 & 0.0  \\ 
& & fixed $\Omega_M$  &  &  &  &  & & & & & & & \\
\hline
JLA & 3 & $\Lambda$CDM & $71.48 \pm 1.41$ & $0.094 \pm 0.073$ & $1.3$ & $36.96 \pm 18.82$ & $0.290 \pm 0.017$ &- & $12.2$ & $12.3$ & 10.4 & 10.5 & $0.05$  \\
& & varying $\Omega_M$  &  &  &  &  & & & & & & & \\
\hline
JLA & 4 & $\Lambda$CDM & $70.20 \pm 0.71$ & $0.052 \pm 0.050$ & $1.0$ & $48.83 \pm 16.96$ & - & - & $14.0$ & $14.1$  & 12.8 & 12.9 & 0.05  \\
& & fixed $\Omega_M$  &  &  &  &  & & & & & & & \\
\hline
JLA & 4 & $\Lambda$CDM & $70.02 \pm 0.78$ & $0.051 \pm 0.043$ & $1.2$ & $49.12 \pm 14.89$ & $0.300 \pm 0.016$ & - & $14.7$ & $14.7$ & 13.5 & 13.5 & 0.0  \\
& & varying $\Omega_M$  &  &  &  &  & & & & & & & \\
\hline
P+ no dupl.& 3 & $\Lambda$CDM & $73.89 \pm 0.81$ & $0.048 \pm 0.016$ & $2.9$ & $52.70 \pm 6.10$ & - & - & $10.3$ & $10.2$  & 8.5 & 8.4 & $-0.05$  \\
 & & fixed $\Omega_M$  &  &  &  &  & & & & & & & \\
\hline
P+ no dupl. & 3 & $\Lambda$CDM & $73.72 \pm 0.98$ & $0.056 \pm 0.031$ & $1.8$ & $49.70 \pm 10.68$ & $0.349 \pm 0.041$ & - & $13.2$ & $13.1$  & 11.4 & 11.3 & $-0.05$  \\
(*) & & varying $\Omega_M$  &  &  &  &  & & & & & & & \\
\hline
P+ no dupl. & 4 & $w_0w_a$CDM & $73.48 \pm 0.10$ & $0.019 \pm 0.003$ & $5.5$ & $64.30 \pm 1.56$ & - & - & $2.0$ & $2.1$  & 0.8 & 0.9 & 0.05  \\
 & & fixed $\Omega_M, w_0, w_a$  &  &  &  &  & & & & & &  & \\
 \hline
P+ no dupl. & 4 & $w_0w_a$CDM & $73.32 \pm 0.22$ & $0.098 \pm 0.034$ & $2.9$ & $36.80 \pm 8.72$ & $0.436 \pm 0.036$ & $-0.104 \pm 0.660 $ & $18.0$ & $18.1$  & 16.8 & 16.9 & 0.05  \\
 & & varying $\Omega_M, w_a$ &  &  &  &  & & & & & & & \\
\hline
P+ no dupl. & 12 & $w_0w_a$CDM & $73.65 \pm 0.07$ & $0.027 \pm 0.003$ & $9.1$ & $61.02 \pm 1.27$ & - & - & $22.6$ & $30.0$  & 23.6 & 31.0 & 3.70  \\
 & & fixed $\Omega_M, w_0, w_a$  &  &  &  &  & & & & & &  & \\
 \hline
DES & 12 & $w_0w_a$CDM  & $ 69.90 \pm 0.06$ & $0.012 \pm 0.002$  & $7.1$ & $64.02 \pm 0.80$ & - & - & $104.2$ & $107.9$  & 105.2 & 108.9 & 1.85  \\
 & & fixed $\Omega_M, w_0, w_a$  &  &  &  &  & & & & & &  & \\

\hline
\end{tabular}
\end{adjustbox}
\caption{Fit parameters for $\mathcal{H}_0(z)$ in the Diamond (upper panel) and Gold cases (lower panel) considering the MW binning assuming a flat $\Lambda$CDM model and a flat $w_{0}w_{a}$CDM model. The first column indicates the sample, the second column indicates the number of bins, and the third column denotes the assumed cosmology. The fourth and fifth columns denote the fit parameters, $\tilde{H}_0$ and $\alpha$,  according to Equation \ref{eq:fittingfunction}.
The sixth column denotes the consistency of the evolutionary parameter $\alpha$ with zero in terms of $1\sigma$, represented by the ratio $\alpha/\sigma_\alpha$, and the seventh column denotes the extrapolated $\mathcal{H}_0(z = 1100)$, representing the $H_0$ in the LSS. 
The seventh and eighth columns denote the values of $\Omega_M$ and $w_a$, respectively. The ninth and tenth columns denote the AIC for the power law and the linear model, respectively.
The eleventh and twelfth columns denote the BIC for the power law and the linear model, respectively. The last column denotes the $\log BF_{P,L}$ value defined in Equation \ref{eq:logbayesfactor}. The asterisk (*) denotes the cases where the Gaussian priors have been expanded up to 5 $\sigma$. All the uncertainties are given in $1\sigma$. The fiducial values are the same as in table \ref{tab:Omega}.}
\label{tab:tab3MW}
\end{table*}

\begin{table*}[ht!]
\arrayrulecolor{black}
\footnotesize
\renewcommand{\arraystretch}{1.1} 
\centering
\begin{adjustbox}{width=1.2\textwidth,center}
\begin{tabular}{|c|c|c|c|c|c|c|c|c|c|c|c|c|c|}
\hline
\multicolumn{14}{|c|}{Equi-spacing in $\text{log}\,z$, Diamond best likelihood} \\
\hline
Sample & Bins & Model & $\tilde{H}_0$ & $\alpha$ & $\alpha/\sigma_{\alpha}$ & $\mathcal{H}_0(z=1100)$ & $\Omega_M$ & $w_a$ & $AIC_{PL}$ & $AIC_{linear}$ & $BIC_{PL}$ & $BIC_{linear}$ & $\log BF_{P,L}$  \\
\hline
Pantheon & 3 & $\Lambda$CDM & $70.22 \pm 0.18$ & $0.011 \pm 0.006$ & $2.0$ & $65.13 \pm 2.52$  & - & - & $7.5$  & $8.0$ & 5.7 & 6.2 & $0.25$  \\
 & & fixed $\Omega_M$ &  &  &  &  & & & & & & & \\
\hline
Pantheon & 3 & $w_0w_a$CDM & $70.26 \pm 0.18$ & $0.010 \pm 0.006$ & $1.7$ & $65.56 \pm 2.73$ & - & - & $7.8$ & $8.3$ & 6.0 & 6.5 & $0.25$  \\
 & & fixed $\Omega_M, w_0, w_a$ &  &  & & & &  & & & & & \\
\hline
JLA & 12 & $\Lambda$CDM & $69.66 \pm 0.18$ & $0.013 \pm 0.011$ & $1.2$ & $63.52 \pm 4.98$ & - & - & $79.3$ & $79.3$ & 80.3 & 80.3 & 0.0  \\
 & & fixed $\Omega_M$ &  &  &  &  & & & & & & & \\
\hline
JLA & 12 & $\Lambda$CDM & $69.58 \pm 0.24$ & $0.020 \pm 0.020$ & $1.0$ & $60.40 \pm 8.51$ & $0.295 \pm 0.010$ & - & $74.4$ & $74.7$ & 75.4 & 75.7 & $0.15$  \\ 
 & & varying $\Omega_M$ &  &  &  &  & & & & & & & \\
\hline
DES & 3 & $\Lambda$CDM & $69.93 \pm 0.14$ & $0.013 \pm 0.011$  & $1.2$ & $63.71 \pm 4.78$ & $0.348 \pm 0.009$ & - & $2.8$ & $2.8$ & 1.0 & 1.0 & 0.0 \\
 & & varying $\Omega_M$ &  &  &  & & & & & & & & \\
\hline
DES & 12 & $\Lambda$CDM & $69.94 \pm 0.09$ & $0.014 \pm 0.009$ & $1.6$ & $63.53 \pm 3.90$ & $0.353 \pm 0.005$ & - & $53.9$ & $53.6$ & 54.9 & 54.6 & $-0.15$  \\
 & & varying $\Omega_M$ &  &  &  & & & & & & & & \\
\hline
\multicolumn{14}{|c|}{Equi-spacing in $\text{log}\,z$, Gold best likelihood} \\
\hline
Sample & Bins & Model & $\tilde{H}_0$ & $\alpha$ & $\alpha/\sigma_{\alpha}$ & $\mathcal{H}_0(z=1100)$ & $\Omega_M$ & $w_a$ & $AIC_{PL}$ & $AIC_{linear}$ & $BIC_{PL}$ & $BIC_{linear}$ & $\log BF_{P,L}$  \\
\hline
Pantheon & 3 & $\Lambda$CDM & $70.36 \pm 0.25$ & $0.023 \pm 0.012$ & $1.9$ & $59.90 \pm 5.02$ & $0.309 \pm 0.011$ & - & $10.5$ & $11.4$ & 8.7 & 9.6 & $0.45$  \\
& & varying $\Omega_M$ &  &  & & & &  & & & & & \\
\hline
Pantheon & 3 & $w_0w_a$CDM & $70.47 \pm 0.29$ & $0.029 \pm 0.014$ & $2.0$ & $57.63 \pm 5.80$ & $0.311 \pm 0.010$ & $0.200 \pm 0.313$ & $10.4$ & $11.4$ & 8.6 & 9.6 & $0.50$  \\
& & varying $\Omega_M, w_a$ &  & & &  &  &  & & & & & \\
\hline
Pantheon & 4 & $\Lambda$CDM & $70.19 \pm 0.26$ & $0.048 \pm 0.027$ & $1.8$ & $50.25 \pm 9.49$ & - & - & $13.9$ & $14.8$ & 12.7 & 13.6 & $0.45$  \\
& & fixed $\Omega_M$ &  &  & & & &  & & & & & \\
\hline
Pantheon & 4 & $\Lambda$CDM & $70.29 \pm 0.28$ & $0.053 \pm 0.031$ & $1.7$ & $48.58 \pm 10.57$ & $0.295 \pm 0.011$ & - & $13.4$ & $14.2$ & 12.2 & 13.0 & $0.40$  \\
& & varying $\Omega_M$ &  &  & & &  &  & & & & & \\
\hline
Pantheon & 4 & $w_0w_a$CDM & $70.21 \pm 0.25$ & $0.034 \pm 0.026$ & $1.3$ & $55.19 \pm 9.93$ & - & - & $14.2$ & $14.7$ & 13.0 & 13.5 & $0.25$  \\ 
& & fixed $\Omega_M, w_0, w_a$ & & & &  &  &  & & & & & \\
\hline
Pantheon & 4 & $w_0w_a$CDM & $70.33 \pm 0.30$ & $0.041 \pm 0.036$ & $1.2$ & $52.67 \pm 13.25$ & $0.306 \pm 0.009 $ & $-0.282 \pm 0.371$  & $14.3$ & $14.7$ & 13.1 & 13.5 & $0.20$ \\
& & varying $\Omega_M, w_a$ &  & & & &  &  & & & & & \\
\hline
JLA & 4 & $w_0w_a$CDM & $68.52 \pm 0.31$ & $0.043 \pm 0.013$ & $3.3$ & $50.76 \pm 4.62$ & - & - & $7.2$ & $7.0$ & 6.0 & 5.8 & $-0.10$  \\
& & fixed $\Omega_M, w_0, w_a$ & & & &  &  &  & & & & & \\
\hline
JLA & 4 & $w_0w_a$CDM & $68.38 \pm 0.38$ & $0.069 \pm 0.048$ & $1.5$ & $42.09 \pm 14.10$ & $ 0.349 \pm 0.041 $ & $ -0.386 \pm 0.271 $ & $13.0$ & $13.3$ & 11.8 & 12.1 & $0.15$  \\
& & varying $\Omega_M, w_a$ &  & & & &  &  & & & & & \\
\hline
JLA & 12 & $w_0w_a$CDM & $68.43 \pm 0.23$ & $0.015 \pm 0.012$ & $1.3$ & $61.68 \pm 4.99$ & - & - & $30.0$ & $30.0$ & 31.0 & 31.0 & 0.0  \\
& & fixed $\Omega_M, w_0, w_a$ &  & & & &  &  & & & & & \\
\hline
JLA & 12 & $w_0w_a$CDM & $68.42 \pm 0.27$ & $0.020 \pm 0.020$ & $1.0$ & $59.34 \pm 8.21$ & $ 0.303 \pm 0.003 $ & $-0.338 \pm 0.154 $ & $31.4$ & $31.5$ & 32.4 & 32.5 & $0.05$  \\
& & varying $\Omega_M, w_a$ &  &  & & & &  & & & & & \\
\hline
P+ no dupl. & 3 & $w_{0}w_{a}$CDM & $71.29 \pm 1.07$ & $0.107 \pm 0.025$ & $4.3$ & $33.67 \pm 5.88$ & - & - & $13.4$ & $13.2$ & 11.6 & 11.4 & $-0.10$  \\
 & & fixed $\Omega_M, w_a$ &  &  & & & &  & & & & & \\
\hline
P+ no dupl. & 3 & $w_{0}w_{a}$CDM & $71.41 \pm 1.33$ & $0.103 \pm 0.044$ & $2.4$ & $34.74 \pm 10.60$ & $0.384 \pm 0.107$ & $-0.052 \pm 0.410$ & $15.8$ & $15.6$ & 14.0 & 13.8 & $-0.10$  \\
(*) & & varying $\Omega_M$ &  & & & &  &  & & & & & \\
\hline
DES & 3 & $\Lambda$CDM & $69.30 \pm 0.22$ & $0.020 \pm 0.007$  & $2.9$ & $60.16 \pm 3.13$ & - & - & $10.1$ & $9.6$ & 8.3 & 7.8 & $-0.25$  \\ 
 & & fixed $\Omega_M$ &  &  &  & & &  & & & & & \\
\hline
DES & 3 & $w_0w_a$CDM & $71.09 \pm 0.16$ & $0.027 \pm 0.005$  & $5.4$ & $58.98 \pm 2.34$ & - & - & $26.1$ & $23.3$ & 24.3 & 21.5 & $-1.40$  \\
 & & fixed $\Omega_M, w_0, w_a$ &  &  & & &  &  & & & & & \\
\hline
DES & 3 & $w_0w_a$CDM & $72.52 \pm 0.18$ & $0.039 \pm 0.028$  & $1.4$ & $54.35 \pm 10.68$ & $0.490 \pm 0.013$ & $-9.425 \pm 1.497$ & $7.2$ & $7.4$ & 5.4 & 5.6 & $0.10$  \\
 & & varying $\Omega_M, w_a$ &  &  & & &  &  & & & & & \\
\hline
DES & 12 & $\Lambda$CDM & $70.11 \pm 0.04$ & $0.025 \pm 0.004$  & $6.1$ & $58.72 \pm 1.70$ & - & - & $52.9$ & $52.5$ & 53.9 & 53.5 & $-0.20$  \\
 & & fixed $\Omega_M$ &  &  &  & & &  & & & & & \\
\hline

\end{tabular}
\end{adjustbox}
\caption{Fit parameters for $\mathcal{H}_0(z)$ in the Diamond (upper part) and Gold cases (lower part) considering the equi-spacing binning on the $\log-z$ method, assuming a flat $\Lambda$CDM model and a flat $w_{0}w_{a}$CDM model. The column headers are the same as in table \ref{tab:tab3MW}. The asterisk (*) denotes the case where the Gaussian priors have been expanded up to 5 $\sigma$. All the uncertainties are given in $1\sigma$. The fiducial values are the same as in table \ref{tab:Omega}.}
\label{tab:4equispacelogz}

\end{table*}

\subsection{Equispacing with Marshall's Likelihood}\label{subsec5.3}

\begin{figure}
\centering
    \includegraphics[width=0.33\textwidth]{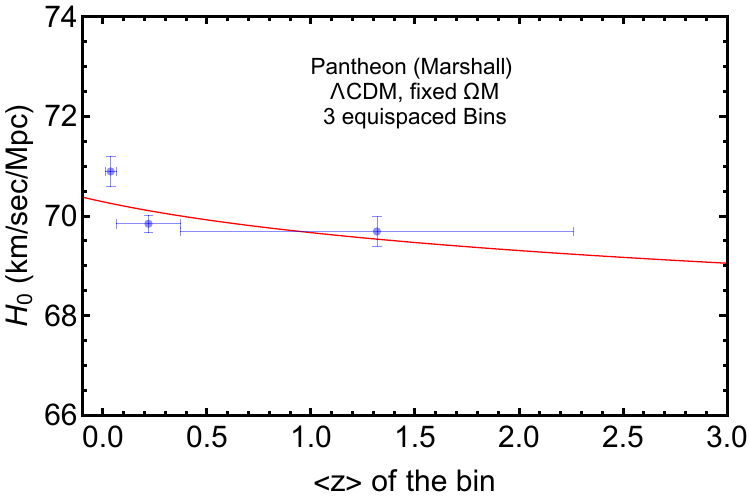}
    \includegraphics[width=0.33\textwidth]{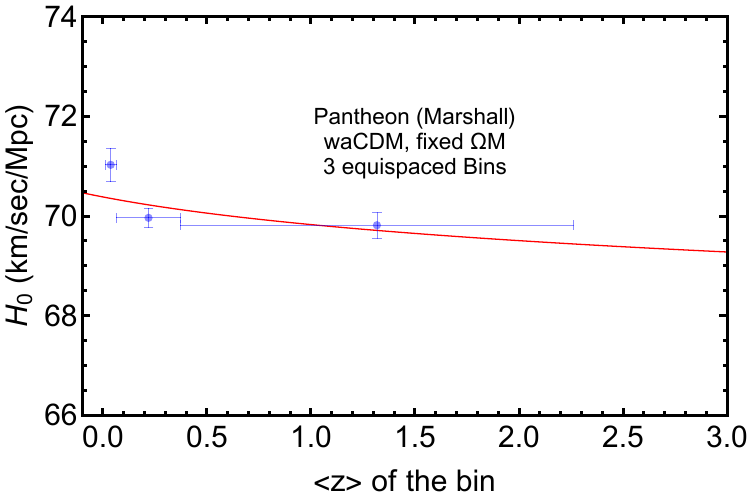}
    \includegraphics[width=0.33\textwidth]{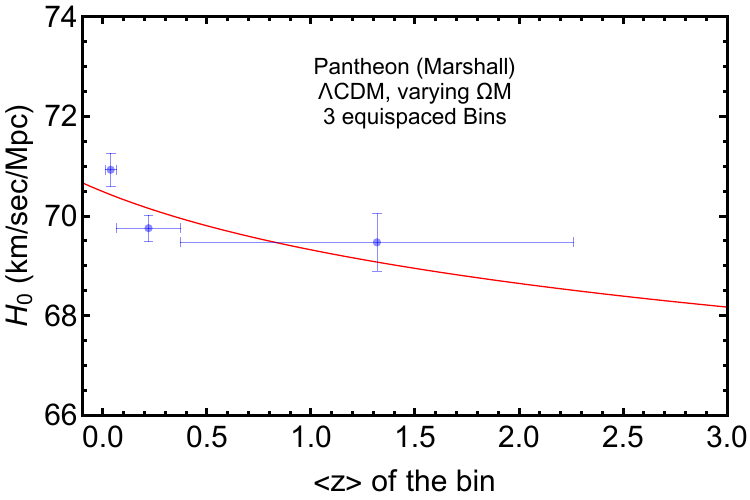}
    \includegraphics[width=0.33\textwidth]{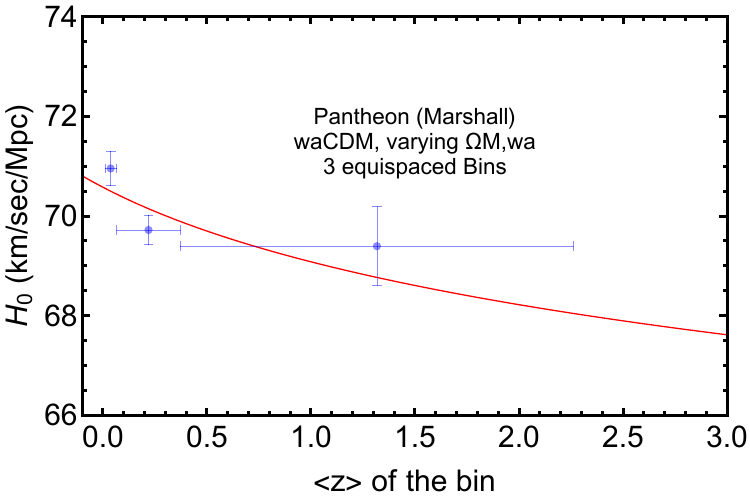}
    \includegraphics[width=0.33\textwidth]{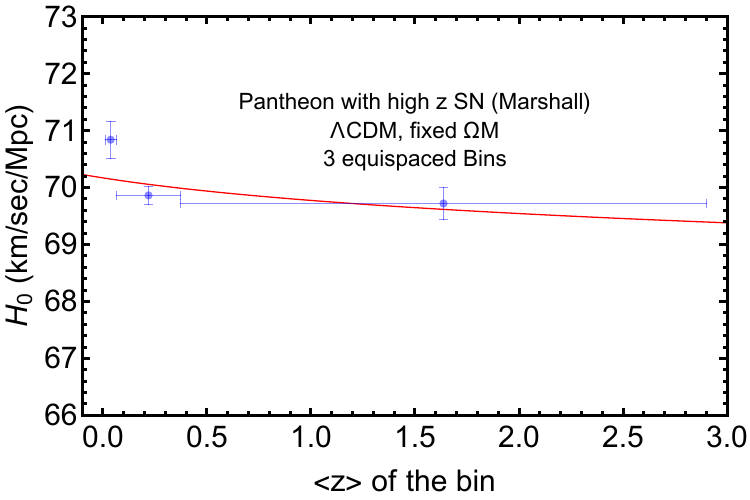}
    \includegraphics[width=0.33\textwidth]{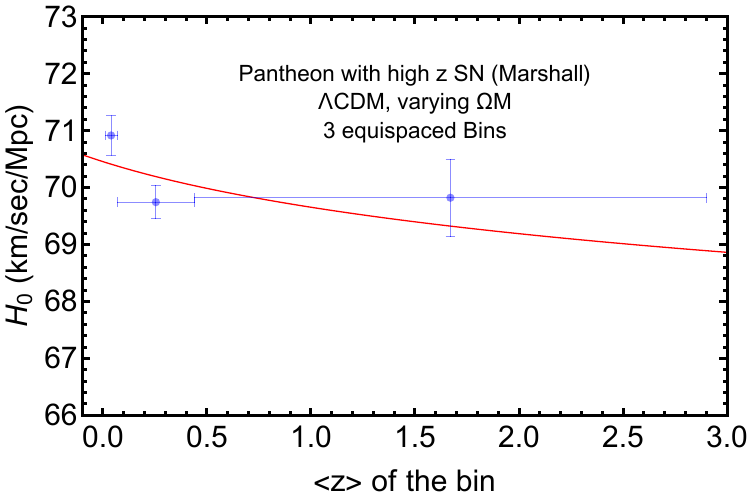}
    \includegraphics[width=0.33\textwidth]{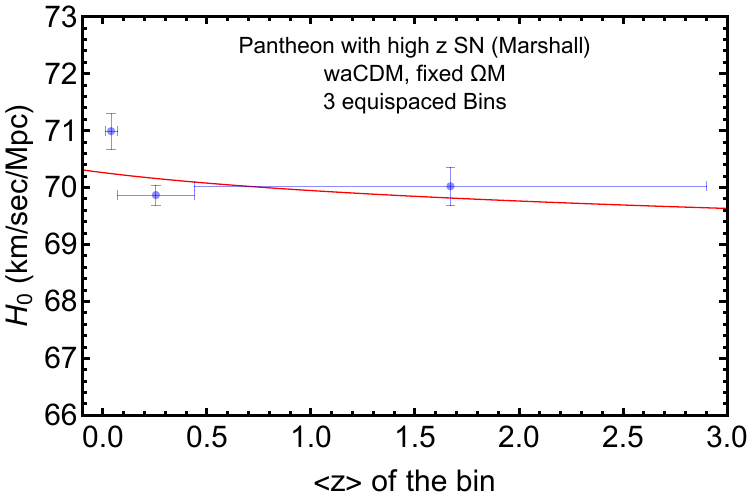}
    \includegraphics[width=0.33\textwidth]{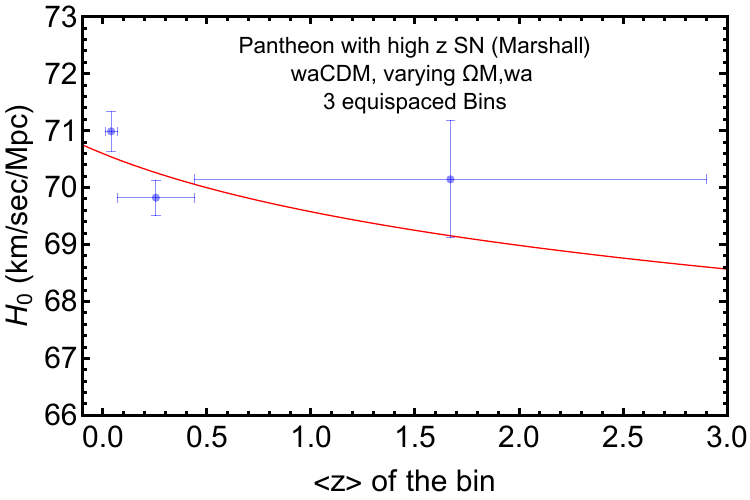}
    \includegraphics[width=0.33\textwidth]{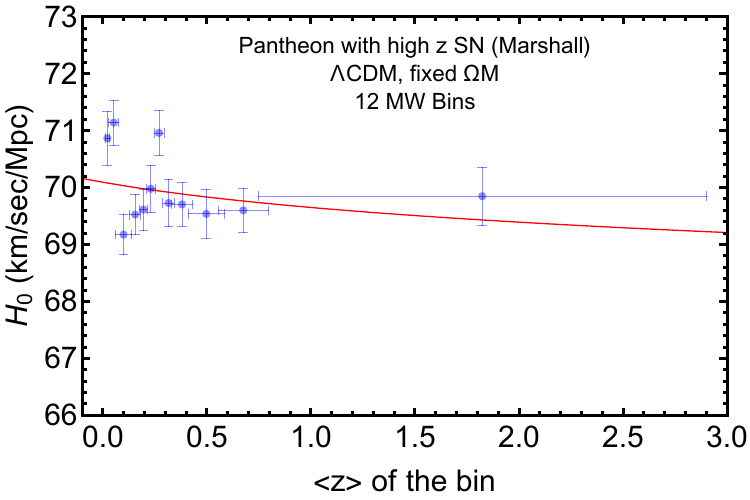}
    \includegraphics[width=0.33\textwidth]{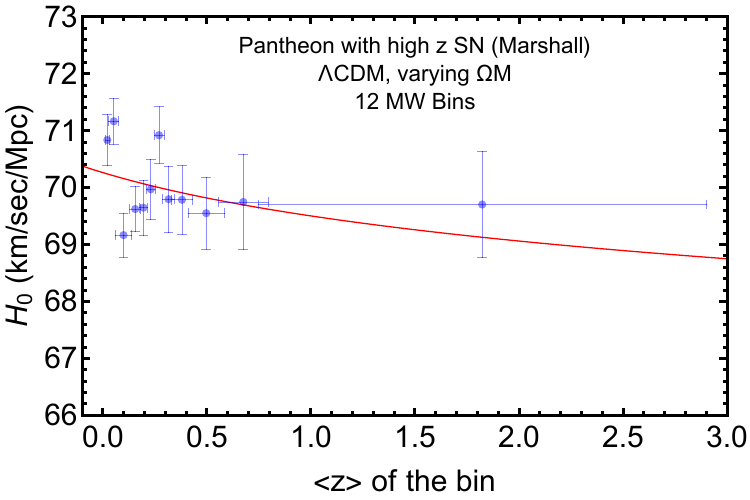}

    \caption{The fitting of $H_0$ values as a function of $z$ in the context of the equi-spacing binning on the $\log-z$ treated with Marshall's likelihood for the Pantheon sample with and without high-$z$ SN. This sample is calibrated with $H_0 = 70$. The first 3 rows of the Pantheon are divided into 3 bins, while the last row is divided into 12 bins.
    \textbf{First row:} shows fixed $\Omega_M$ corresponding to $\Lambda$CDM (left), $w_0w_a$CDM (middle) and  varying $\Omega_M$ corresponding to $\Lambda$CDM (right) cosmology. \textbf{Second row:} shows varying $\Omega_M$ corresponding to $w_0w_a$CDM (left), high-$z$ SNe Ia corresponding to $\Lambda$CDM model for both fixed $\Omega_M$ (middle) and varying $\Omega_M$ (right) cosmology. \textbf{Third row:} reports high-$z$ SNe Ia corresponding to the $w_0w_a$CDM model for both fixed $\Omega_M$ (left) and varying $\Omega_M$ (right) \textbf{Fourth row:} includes high-$z$ in 12 bins equi-populated with MW, corresponding to the $\Lambda$CDM model for both fixed $\Omega_M$ (left) and varying $\Omega_M$ (right). The results of these plots are summarized in table \ref{tab:5Marshall}, and the corresponding fiducial values are reported in table \ref{tab:Omega}.}
    \label{fig:logz_marshall_Diamond}
\end{figure}

We here adopt Marshall's likelihood and equi-spaced binning in the $\log-z$ on the Pantheon sample, without and with the high-$z$ SNe Ia contribution. The results for Diamond cases without the high-$z$ SNe Ia are reported in the upper part of the table \ref{tab:5Marshall}, and the corresponding plots are in the first row of figure \ref{fig:logz_marshall_Diamond}. The results of Gold cases without high-$z$ SNe Ia are added in the upper middle of the table \ref{tab:5Marshall} and plotted in the second row of figure \ref{fig:logz_marshall_Diamond}.
Diamond cases with high-$z$ SNe Ia are reported in the lower middle part of table \ref{tab:5Marshall}, and the plots are shown in the third and fourth rows of figure \ref{fig:logz_marshall_Diamond}. The bottom part of the table \ref{tab:5Marshall} includes the results of Diamond cases with high-$z$ SNe Ia, which have been analysed with equi-populated binning using the MW: the plots for these cases are shown in the fifth row of figure \ref{fig:logz_marshall_Diamond}. For equi-spaced binning, the Pantheon sample with and without high-$z$ SNe Ia is analyzed in $3$ bins for both $\Lambda$CDM and $w_0w_a$CDM models. The equi-populated binning with MW approach is used for Pantheon samples with high-$z$ in 12 bins for the $\Lambda$CDM model.

In the Diamond cases without high-$z$ SNe Ia, $\tilde{H}_0$ and $\alpha$ values for the $\Lambda$CDM and $w_0w_a$CDM models are compatible in 1 $\sigma$. The ratios \(\alpha/\sigma_\alpha\) take the corresponding values $1.9$ and $1.8$, respectively.

In the Gold cases without high-$z$ SNe Ia, $\tilde{H}_0$ values are compatible within 1 $\sigma$ with the previous Diamond case. $\alpha$ values are $0.024 \pm 0.012$ to $0.031 \pm 0.015$. Although the models differ between the two cases, no direct comparison can be drawn.
The values of the ratio \(\alpha/\sigma_\alpha\) are $2.0$ for both cases.

Concerning the Diamond cases that include the high-$z$ SNe Ia, $H_0$ ranges from $70.17 \pm 0.18$ to $70.60 \pm 0.31$, while $\alpha$ varies from $0.006\pm 0.006$ to $0.021\pm 0.016$. 
Indeed, the third case shows very large error bars and the $\alpha/\sigma_{\alpha}$=1, showing that this case may be due more to fluctuations rather than an intrinsic decreasing trend.
This case is similar to the JLA 12 bins in the $\Lambda$CDM and $w_ow_a$CDM with varying $\Omega_M$ in table \ref{tab:4equispacelogz}.

Also for the Diamond cases including SNe Ia at high-z, the $\tilde{H}_0$ are compatible among themselves in 1 $\sigma$, while \(\alpha/\sigma_\alpha\) has a range from $1.0$ to $1.5$. In Diamond cases that include high-$z$ SNe Ia, conducted with MW, within the $\Lambda$CDM model, $\tilde{H}_0$ takes the values of $70.09\pm0.18$ for $\Lambda$CDM with fixed $\Omega_M$ and $70.26\pm0.22$ for $\Lambda$CDM with variable $\Omega_M$. The corresponding $\alpha$ values are $0.009\pm0.007$ and $0.016\pm0.012$ for the fixed and varying $\Omega_M$ cases, respectively.
Thus, the values of $\tilde{H}_0$ and $\alpha$ are compatible within 1 $\sigma$. The trend of $\alpha$ is compatible within 1 $\sigma$, also with the non-MW case.

Looking at the $\log BF_{P,L}$ values, a preference for power law is not statistically significant for any of the cases reported in table \ref{tab:5Marshall}. Interestingly, all the high-$z$ SNe Ia cases in table \ref{tab:5Marshall} are Diamond when treated with Marshall likelihood. In general, these findings about $\alpha$ show that trend estimates are insensitive to binning strategies and cosmological models, reflecting the persistence of this decreasing trend for $H_0$.

\begin{table*}[ht!]
\arrayrulecolor{black}
\footnotesize
\renewcommand{\arraystretch}{1.1} 
\centering
\begin{adjustbox}{width=1.2\textwidth,center}
\begin{tabular}{|c|c|c|c|c|c|c|c|c|c|c|c|c|}
\hline
\multicolumn{13}{|c|}{Equi-spacing in $\text{log}\,z$, Marshall likelihood, Diamond cases for the Pantheon sample} \\
\hline
Bins & Model & $\tilde{H}_0$ & $\alpha$ & $\alpha / \sigma_{\alpha}$ & $\mathcal{H}_0(z=1100)$ & $\Omega_M$ & $w_a$ &  $AIC_{PL}$  & $AIC_{linear}$ & $BIC_{PL}$ & $BIC_{linear}$ & $\log BF_{P,L}$\\
\hline
3 & $\Lambda$CDM & $ 70.29 \pm 0.20$ & $0.013 \pm 0.007$ & $1.9$ & $64.27 \pm 3.10$ &-&-& $10.1$ & $10.8$ & $8.3$ & $9.0$ & $0.35$ \\
 & fixed $\Omega_M$ & & & & & & & & & & & \\
\hline

3 & $w_0w_a$CDM & $70.38 \pm 0.21$ & $0.011 \pm 0.006$ & $1.8$ & $65.16 \pm 2.74$ & - & - & $9.4$ & $9.9$ & $7.6$ & $8.1$ & $0.25$ \\
 & fixed $\Omega_M, w_0, w_a$ & & & & & & & & & & & \\
\hline

\multicolumn{13}{|c|}{Equi-spacing in $\text{log}\,z$, Marshall likelihood, Gold cases for the Pantheon sample} \\
\hline
Bins & Model & $\tilde{H}_0$ & $\alpha$ & $\alpha / \sigma_{\alpha}$ & $\mathcal{H}_0(z=1100)$ & $\Omega_M$ & $w_a$ &  $AIC_{PL}$  & $AIC_{linear}$ & $BIC_{PL}$ & $BIC_{linear}$ & $\log BF_{P,L}$ \\
\hline
3 & $\Lambda$CDM & $ 70.49 \pm 0.26$ & $0.024 \pm 0.012$ & $2.0$ & $59.54 \pm 4.93$ & $ 0.305\pm 0.011$ & - &  $10.6$  & $11.6$ & $8.8$ & $9.8$ & $0.50$\\
 & varying $\Omega_M$ & & & & & & & & & & & \\
\hline
3 & $w_0w_a$CDM & $70.58 \pm 0.29$ & $0.031 \pm 0.015$ & $2.0$ & $56.84 \pm 6.16$ & $0.312 \pm 0.010$ & $0.096 \pm 0.330$ & $10.9$ & $11.9$ & $9.1$ & $10.1$ & $0.50$\\
 & varying $\Omega_M, w_a$ & & & & & & & & & & & \\
\hline

\multicolumn{13}{|c|}{Equi-spacing in $\text{log}\,z$ with high-$z$, Marshall likelihood, Diamond cases for the Pantheon sample} \\
\hline
Bins & Model & $\tilde{H}_0$ & $\alpha$ & $\alpha / \sigma_{\alpha}$ & $\mathcal{H}_0(z=1100)$ & $\Omega_M$ & $w_a$ &  $AIC_{PL}$  & $AIC_{linear}$ & $BIC_{PL}$ & $BIC_{linear}$ & $\log BF_{P,L}$ \\
\hline
3 & $\Lambda$CDM & {$70.17 \pm 0.18$} & $0.008 \pm 0.005$ & $1.5$ & $66.27 \pm 2.54$ & - & - & $9.1$& $9.6$ & $7.3$ & $7.8$ & $0.25$\\
 & fixed $\Omega_M$ & & & & & & & & & & & \\
\hline
3 & $\Lambda$CDM & $70.46 \pm 0.28$ & $0.016 \pm 0.012$ & $1.4$ & $62.77 \pm 5.13$ & $0.300 \pm 0.012$ & - & $11.2$ & $11.9$ & $9.4$   & $10.1$ & $0.35$\\
 & varying $\Omega_M$ & & & & & & & & & & & \\
\hline
3 & $w_0w_a$CDM & $70.26 \pm 0.20$ & $0.006 \pm 0.006$ & $1.0$  & $67.14 \pm 3.00$ & - & - & $12.2$ & $12.6$ & $10.4$ & $10.8$ & $0.20$\\
 & fixed $\Omega_M, w_0, w_a$ & & & & & & & & & & & \\
\hline
3 & $w_0w_a$CDM & $70.60\pm 0.31$ & $0.021 \pm 0.016$ & $1.3$ & $60.93 \pm 6.88$ & $0.309 \pm 0.010$ & $-0.132 \pm 0.350$ & $11.7$ &$12.5$ & $9.9$  & $10.7$ & $0.40$ \\
 & varying $\Omega_M, w_a$ & & & & & & & & & & & \\
\hline
\multicolumn{13}{|c|}{Equi-population binning with MW, with high-$z$, Marshall likelihood, Diamond cases for Pantheon sample} \\
\hline
Bins & Model & $\tilde{H}_0$ & $\alpha$ & $\alpha / \sigma_{\alpha}$ & $\mathcal{H}_0(z=1100)$ & $\Omega_M$ & $w_a$ &  $AIC_{PL}$  & $AIC_{linear}$ & $BIC_{PL}$ & $BIC_{linear}$ & $\log BF_{P,L}$ \\
\hline
12 & $\Lambda$CDM & $70.09 \pm 0.18$ & $0.009 \pm 0.007$ & $1.2$ & $65.74 \pm 3.39$ & -  & - & $33.7$ & $34.3$ & $34.7$ & $35.3$ & $0.30$\\
 & fixed $\Omega_M$ & & & & & & & & & & & \\
\hline
12 & $\Lambda$CDM & $70.26 \pm 0.22$ & $0.016 \pm 0.012$ & $1.3$ & $62.94 \pm 5.17$ & $0.298 \pm 0.006$  & - & $33.8$ & $34.4$ & $34.8$ & $35.4$ & $0.30$ \\
 & varying $\Omega_M$ & & & & & & & & & & & \\
\hline
\end{tabular}
\end{adjustbox}
\caption{Fit parameters for $H_0(z)$ in the Diamond and Gold cases equi-spacing binning on the $\log z$ method treated with Marshall's likelihood, assuming a flat $\Lambda$CDM model and a flat $w_0w_a$CDM model, without and with high-$z$ SNe Ia with the Pantheon sample. The column headers are the same as in table \ref{tab:tab3MW}. All uncertainties are $1\sigma$; fiducial values match table \ref{tab:Omega}.}
\label{tab:5Marshall}
\end{table*}

\clearpage

\subsection{The Master Sample analysis}\label{subsec5.4}
\begin{figure}
   \centering
    \includegraphics[width=0.3\textwidth]{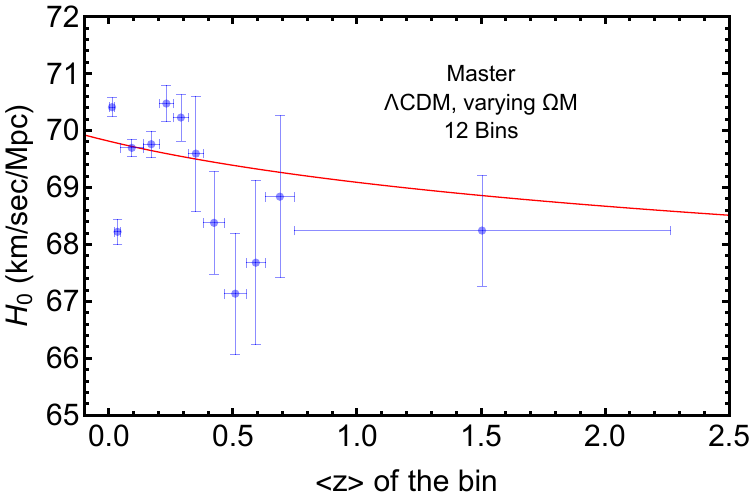}
    \includegraphics[width=0.3\textwidth]{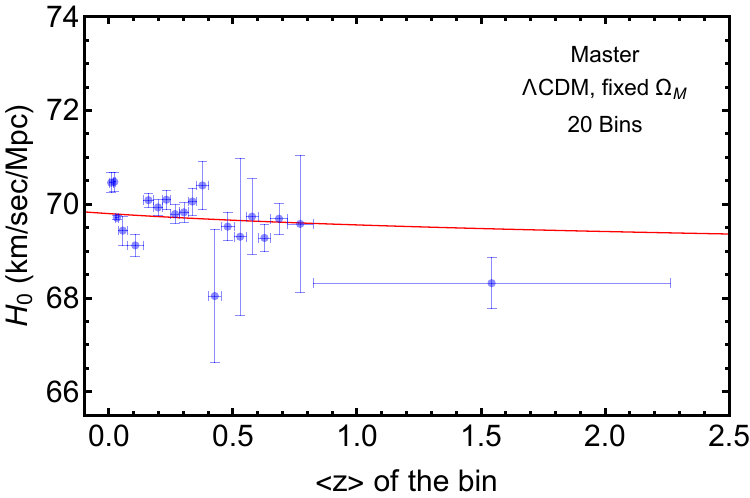}
    \includegraphics[width=0.3\textwidth]{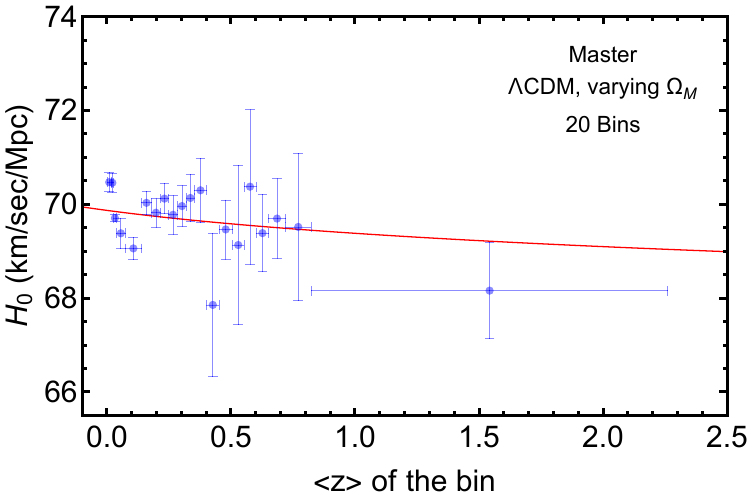}
    \includegraphics[width=0.48\textwidth]{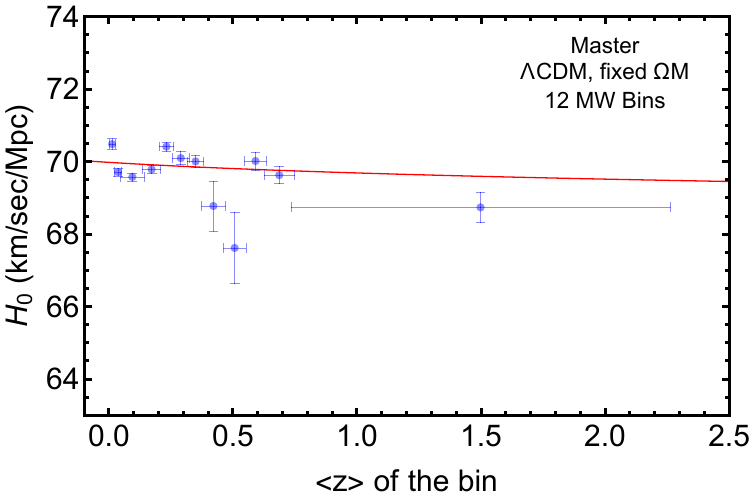}
    \includegraphics[width=0.48\textwidth]{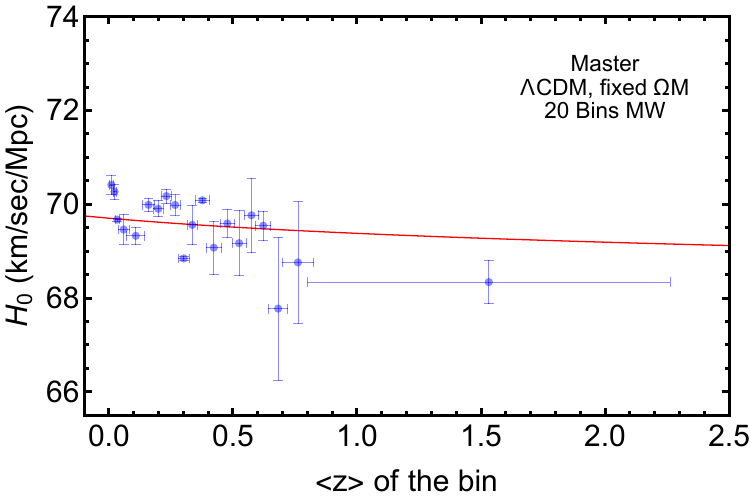}
    \includegraphics[width=0.48\textwidth]{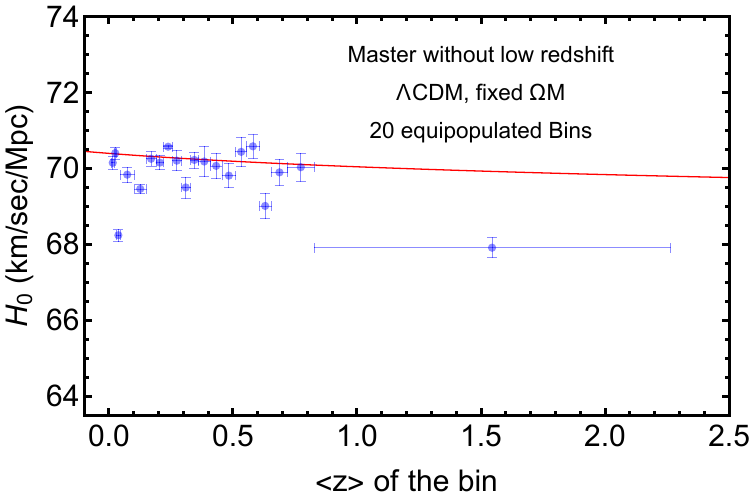}
    \caption{The fitting of $H_0$ values as a function of $z$ in the context of equi-population and MW binning of the Master Sample, with $H_0=70$ for the Diamond sample within the $\Lambda$CDM. \textbf{First row:} the left panel shows 12 bins with varying $\Omega_M$, the middle and right panels show 20 bins with fixed and varying $\Omega_M$, respectively. \textbf{Second row:} shows 12 and 20 bins with fixed $\Omega_M$, respectively. 
    \textbf{Third row:}
    This panel shows the 20 bins with fixed $\Omega_M$ without the low-$z$ SNe Ia.
    The results of these plots are summarized in table \ref{tab:mastersample} and the corresponding fiducial values are reported in table \ref{tab:Omega}.}
   \label{fig:masterdiamondreg}
\end{figure}

\begin{figure}
   \centering
    \includegraphics[width=0.3\textwidth]{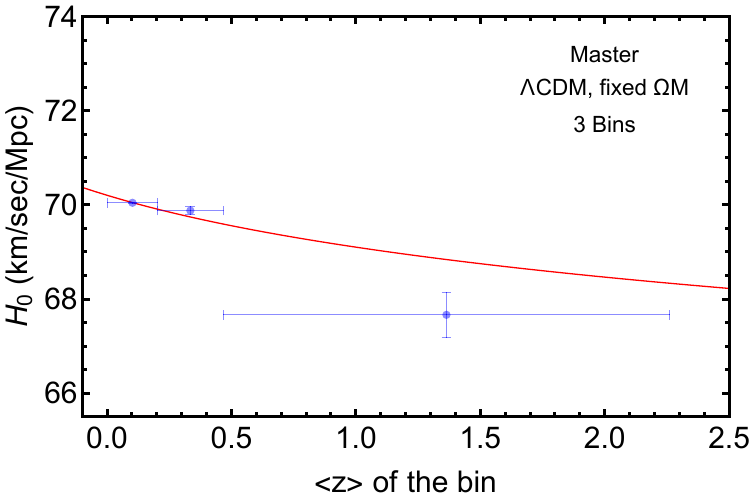}
    \includegraphics[width=0.3\textwidth]{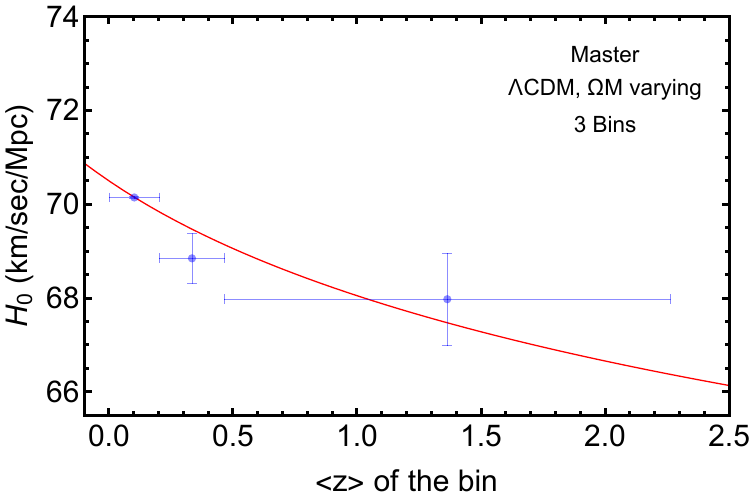}
    \includegraphics[width=0.3\textwidth]{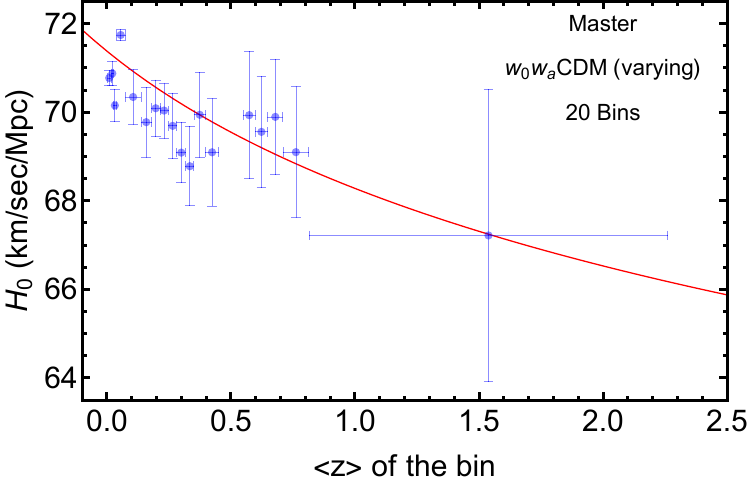}
    \includegraphics[width=0.3\textwidth]{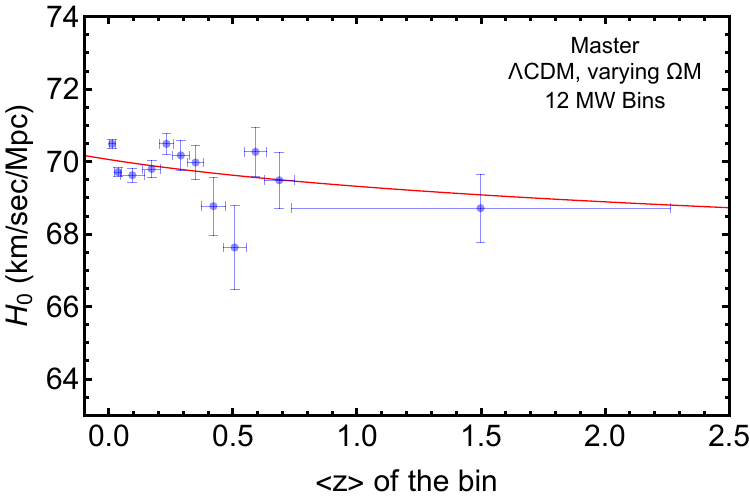}
    \includegraphics[width=0.3\textwidth]{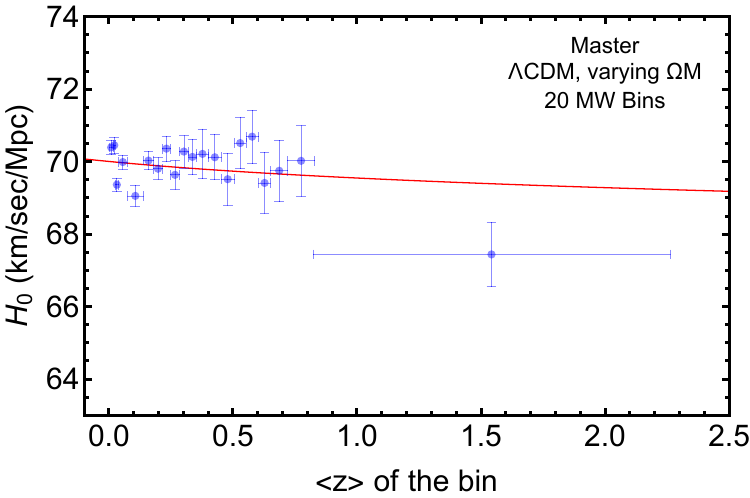}
    \includegraphics[width=0.3\textwidth]{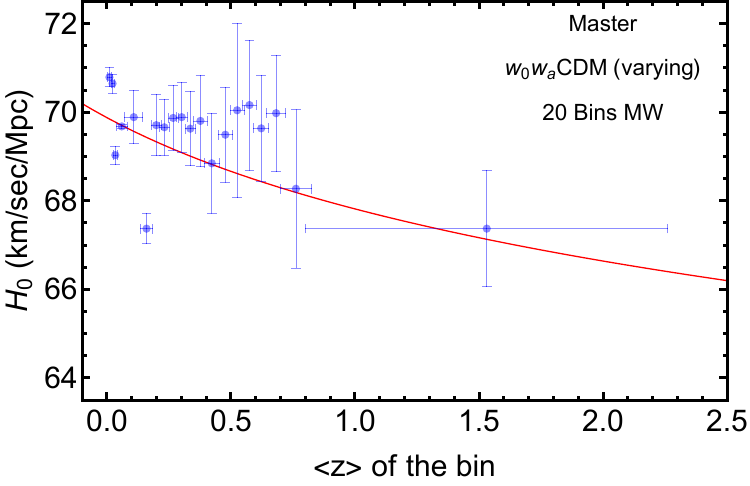}
    \includegraphics[width=0.3\textwidth]{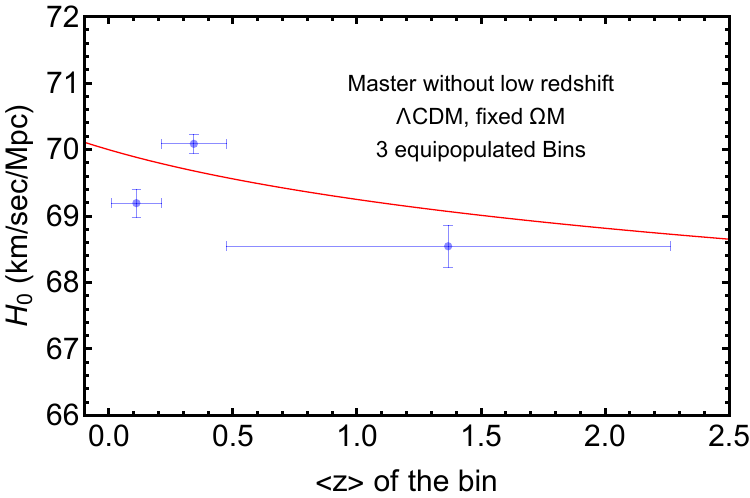}
    \includegraphics[width=0.3\textwidth]{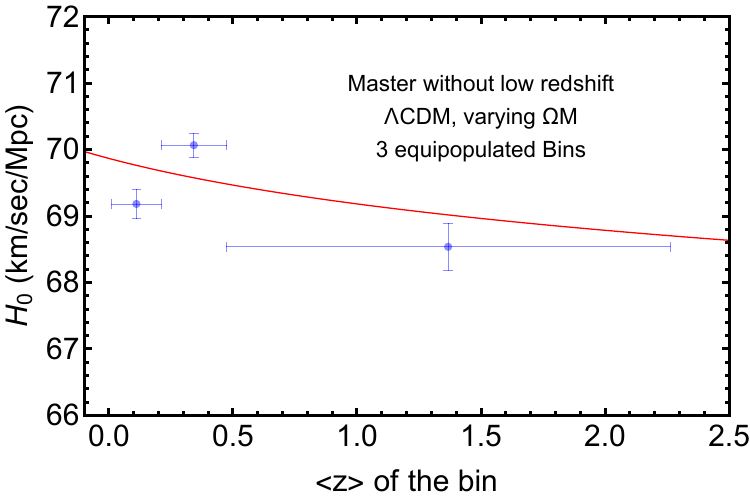}
    \includegraphics[width=0.3\textwidth]{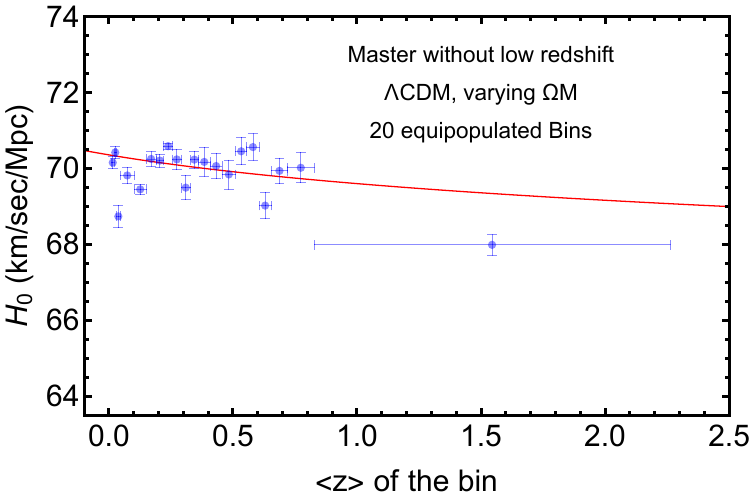}
    
   \caption{The fitting of $H_0$ values as a function of $z$ in the context of equi-population and MW binning of the Master Sample, with $H_0=70$ for the Gold sample with the $\Lambda$CDM and $w_0w_a$CDM. 
   \textbf{First row:} the left and middle panels shows 3 bins in the $\Lambda$CDM with fixed and varying $\Omega_M$, respectively. The right panel shows 20 bins with $w_0w_a$CDM and varying $\Omega_M$. 
   \textbf{Second row:} shows MW binning. The left and middle panels show 12 and 20 bins, respectively, both in the $\Lambda$CDM with varying $\Omega_M$. The right panel shows 20 bins with $w_0w_a$CDM with the varying $\Omega_M$. 
   \textbf{Third row:} shows equi-population binning without low-z. The left and middle panels show 3 bins within $\Lambda$CDM with fixed $\Omega_M$ and varying $\Omega_M$, respectively. The right panel shows 20 bins with $\Lambda$CDM with the varying $\Omega_M$. 
   The results of these plots are summarized in table \ref{tab:mastersample} and the corresponding fiducial values are reported in table \ref{tab:Omega}.}
   \label{fig:mastersamplenewLCDM}
\end{figure}

In this Section, we detail the findings of the analysis performed on the Master Sample, a collection of DES, P+, Pantheon, and JLA data with duplicates removed. The results with $H_0=70$ calibration are summarized in table \ref{tab:mastersample} and are plotted in figure \ref{fig:masterdiamondreg} and figure \ref{fig:mastersamplenewLCDM}.
Since we have shown that trends are similar independently of binning techniques, we use the equi-population and the MW binning here. We used 3, 12, and 20 redshift-ordered bins. The 3 and 12 bins are investigated within the $\Lambda$CDM model, and the 20 bins are also investigated in the $w_{0}w_{a}$CDM model. In addition, in the case of the Master Sample, the check for the distributions of best-fitting residuals has been performed. 

The Diamond cases, see the upper part of figure \ref{fig:masterdiamondreg}, have a $\tilde{H}_0$ around $69.80$ and are all compatible within 1 $\sigma$. The $\alpha$ values range from $0.005 \pm 0.004$ to $0.015 \pm 0.011$, thus also being compatible within 1 $\sigma$. The $\log BF_{P,L}$ values show no statistical evidence for preferring one model versus the other. Continuing on the Diamond sample for the MW case, see the middle panel of figure \ref{fig:masterdiamondreg}, we obtain compatible $\alpha$ values within 1 $\sigma$ in the $\Lambda$CDM for 12 and 20 bins with fixed $\Omega_M$ to the case with the equi-population binning. The $\tilde{H}_0$ for this case are compatible within 3 $\sigma$ with each other and with the previous case of Diamond. We note that the $\alpha$ values are all compatible within 1 $\sigma$ with the equi-population binning. There is no preference for any model according to $\log BF_{P,L}$ values.

Discussing the Diamond sample in the MW case, but excluding SNe Ia at $z \leq 0.01$, see the lower panel of figure \ref{fig:masterdiamondreg}, we note that only one case belongs to the Diamond: the $\Lambda$CDM with fixed $\Omega_M$. The $\tilde{H}_0$ value for this case is the highest among the Diamond cases, carrying $70.40 \pm 0.06$. Since this configuration is present in all three Master Diamond cases, it shows that the $\Lambda$CDM with fixed $\Omega_M$ favors a good extrapolation to the redshift of the LSS. Unlike the previous Diamond cases, the $\log BF_{P,L}$ suggest that linear model is favored compared to the power law.

The equi-population for the Gold cases, see the upper panel of figure \ref{fig:mastersamplenewLCDM}, has a $\tilde{H}_0$ range from $70.21 \pm 0.04$ to $71.38 \pm 0.11$ and the $\alpha$ values ranging from $0.023 \pm 0.005$ to $0.064 \pm 0.014$.
The $\log BF_{P,L}$ values do not show any distinguishable preference between the two models. The values of $\alpha/\sigma_\alpha$ range from 3.0 to 4.6. This ratio is between 2.5 and 3.8 times higher than the Diamond cases, and the $\alpha$ parameter is from 1.5 to 12.8 times steeper than the Diamond cases, although there is no direct correspondence of the cases; thus, a direct comparison cannot be drawn.
In this case, the $\log BF_{P,L}$ values do not point out a clearly preferred model.

Moving forward with the Gold sample for the MW case, see the middle panel of figures \ref{fig:mastersamplenewLCDM}, we have 12 and 20 bins for $\Lambda$CDM varying $\Omega_M$ and 20 bins for $w_0w_a$CDM with varying $\Omega_M$. The $\tilde{H}_0$ range from $69.80 \pm 0.05$ to $70.06 \pm 0.09$ and the $\alpha$ values range from $0.015 \pm 0.009$ to $0.043 \pm 0.012$. The $\alpha$ values are compatible within 1 $\sigma$ within the $\Lambda$CDM model, while the $\alpha$ of the $w_0w_a$CDM model is compatible only in 2 $\sigma$. In the cases of 12 bins and 20 bins within $\Lambda$CDM, the $\log BF_{P,L}$ values show no evidence for preference between the two models. However, there is a slight preference for power law model considering 20 bins in $w_0w_a$CDM as indicated through the $\log BF_{P,L}$. The $\alpha/\sigma_\alpha$ values are smaller for the MW than for the equi-population bins.

For the Gold cases with removal of low-$z$ SNe Ia, see the bottom panel of figures \ref{fig:mastersamplenewLCDM}, the $\tilde{H}_0$ range from $69.87 \pm 0.21$ to $70.36 \pm 0.07$ and the $\alpha$ values range from $0.014 \pm 0.008$ to $0.016 \pm 0.004$. The $\alpha$ values are compatible within 1 $\sigma$ and the linear model is favored over the power law for the 20 bins $\Lambda$CDM, varying $\Omega_M$ as suggested by the $\log BF_{P,L}$ value. For the 3 bins $\Lambda$CDM cases with fixed and varying $\Omega_m$, there is no clear preference for any model as indicated by the $\log BF_{P,L}$ value.
The peculiar velocities of SNe Ia host galaxies can contribute up to the 10\% of the total apparent recession velocity effect observed at redshift $z \sim 0.01$, as reported for the P+ in \citealt{Peterson2022}: for this reason, the test our results with the removal of low-$z$ SNe Ia from the Master Sample has been conducted through the $\Lambda$CDM cosmology to weigh the effect of the peculiar velocities. The results show that a decreasing trend is still visible, and there is no strong evidence of statistical difference between the $\alpha$ values within the error bars and the $\alpha/\sigma_\alpha$ among the overall sample of Gold and Diamond.
Overall, the results of the Master Sample indicate that, in general, the combination of the four SNe Ia catalogs still shows a slow decreasing trend for the $H_0$.

\begin{table*}
\arrayrulecolor{black}
\footnotesize
\renewcommand{\arraystretch}{1.1} 
\centering
\begin{adjustbox}{width=1.2\textwidth,center}
\begin{tabular}{|c|c|c|c|c|c|c|c|c|c|c|c|c|}
\hline
\multicolumn{13}{|c|}{Equi-population binning, Master Sample $H_0=70$, Diamond cases} \\
\hline
Bins & Model & $\tilde{H}_0$ & $\alpha$ & $\alpha/\sigma_{\alpha}$ & $\mathcal{H}_0(z=1100)$ & $\Omega_M$ & $w_a$ & $AIC_{PL}$ & $AIC_{linear}$ & $BIC_{PL}$ & $BIC_{linear}$ & $\log BF_{P,L}$  \\
\hline
12 & $\Lambda$CDM & $69.81 \pm 0.11$ & $0.015 \pm 0.011$ & $1.4$ & $62.85 \pm 4.78$ & $0.320 \pm 0.007$ & - & $94.0$ & $93.4$ & $95.0$ & $94.4$ & $-0.30$\\
 & varying $\Omega_M$ &  &  &  &  &  &  & & & & & \\
\hline
20 & $\Lambda$CDM & $69.80 \pm 0.04$ & $0.005 \pm 0.004$ & $1.2$ & $67.40 \pm 1.92$ & - & - & $52.9$ & $51.7$ & $54.9$ & $53.7$ & $-0.60$\\
 & fixed $\Omega_M$ &  &  &  &  &  & & & & & & \\
\hline

20 & $\Lambda$CDM & $69.87 \pm 0.07$ & $0.010 \pm 0.008$ & $1.3$ & $65.09 \pm 3.63$ & $0.324 \pm 0.005$ & - & 56.3 & 55.9 & 58.3 & 57.9 & $-0.20$\\
 & varying $\Omega_M$ &  &  &  &  &  &  &  & & & & \\
\hline
\multicolumn{13}{|c|}{Equi-population binning, Master Sample $H_0=70$, Gold cases} \\
\hline
Bins & Model & $\tilde{H}_0$ & $\alpha$ & $\alpha/\sigma_{\alpha}$ & $\mathcal{H}_0(z=1100)$ & $\Omega_M$ & $w_a$ & $AIC_{PL}$ & $AIC_{linear}$ & $BIC_{PL}$ & $BIC_{linear}$ & $\log BF_{P,L}$\\
\hline
3 & $\Lambda$CDM & $70.21 \pm 0.04$ & $0.023 \pm 0.005$ & $4.4$ & $59.83 \pm 2.16$ & - & - &54.4  &  55.1 & 52.6 & 53.3 & 0.35\\
 & fixed $\Omega_M$ &  &  &  &  &  &  &  &  & & & \\
\hline
3 & $\Lambda$CDM & $70.50 \pm 0.05$ & $0.051 \pm 0.017$ & $3.0$ & $49.18 \pm 5.93$ & $0.299 \pm 0.005$ & - & $4.2$ & $5.2$ & $2.4$ & $3.4$ & 0.50 \\
 & varying $\Omega_M$ &  &  &  &  &  & & & &  &  & \\
\hline

20 & $w_0w_a$CDM & $71.38 \pm 0.11$ & $0.064 \pm 0.014$ & $4.6$ & $45.58 \pm 4.45$ & $0.473 \pm 0.005$ & $-7.361 \pm 1.338$ & 84.0 & 83.8 & 86.0 & 85.8 & $-0.10$\\
 & varying $\Omega_M,w_a$ &  &  &  &  &  &  &  & & & & \\
\hline

\multicolumn{13}{|c|}{Equi-population binning using MW, Master Sample, $H_0=70$, Diamond cases} \\
\hline
Bins & Model & $\tilde{H}_0$ & $\alpha$ & $\alpha/\sigma_{\alpha}$ & $\mathcal{H}_0(z=1100)$ & $\Omega_M$ & $w_a$  & $AIC_{PL}$ & $AIC_{linear}$ & $BIC_{PL}$ & $BIC_{linear}$ & $\log BF_{P,L}$ \\
\hline

12 & $\Lambda$CDM & $69.98 \pm 0.07$ & $0.006 \pm 0.004$ & $1.4$ & $67.08 \pm 2.02$ & - & - & $54.3$ & $53.0$ &$55.3$ & $54.0$ & $-0.65$ \\
 & fixed $\Omega_M$ &  &  &  &  &  &  &  & & & & \\
\hline

20 & $\Lambda$CDM & $69.70 \pm 0.04$ & $0.007 \pm 0.002$ & $2.8$ & $66.51 \pm 1.13$ & - & - & 363.6 & 363.6 & 365.6 & 365.6 & 0.0\\
 & fixed $\Omega_M$ &  &  & & & &  &  &  &  & & \\
\hline

\multicolumn{13}{|c|}{Equi-population binning using MW, Master Sample, $H_0=70$, Gold cases} \\
\hline
Bins & Model & $\tilde{H}_0$ & $\alpha$ & $\alpha/\sigma_{\alpha}$ & $\mathcal{H}_0(z=1100)$ & $\Omega_M$ & $w_a$ & $AIC_{PL}$ & $AIC_{linear}$ & $BIC_{PL}$ & $BIC_{linear}$ & $\log BF_{P,L}$ \\
\hline
12 & $\Lambda$CDM & $70.06 \pm 0.09$ & $0.015 \pm 0.009$  & $1.8$ & $62.94 \pm 3.85$ & $0.324 \pm 0.005$ & - & $38.2$ & $38.2$  &$39.2$ &$39.2$ & 0.0\\
 & varying $\Omega_M$ &  &  &  &  & & & &  &  &  & \\
\hline
20 & $\Lambda$CDM & $69.80 \pm 0.05$ & $0.016 \pm 0.008$ & $2.2$ & $62.18 \pm 3.27$ & $0.324 \pm 0.006$ & - & 62.7 & 62.5 & 64.7 & 64.5 & $-0.10$\\
 & varying $\Omega_M$ &  &  &  &  & & & &  &  & & \\
\hline
20 & $w_0w_a$CDM & $69.88 \pm 0.06$ & $0.043 \pm 0.012$ & $3.7$ & $51.64 \pm 4.22$ & $0.473 \pm 0.005$ & $-7.269 \pm 0.318$ & 120.1 & 122.7 & 122.1 & 124.7 & $1.30$ \\
 & varying $\Omega_M,w_a$ &  &  &  &  &  & & &  &  & & \\
\hline

\multicolumn{13}{|c|}{Equi-population binning, Master Sample without low-$z$ SN, $H_0=70$, Diamond cases} \\
\hline
Bins & Model & $\tilde{H}_0$ & $\alpha$ & $\alpha/\sigma_{\alpha}$ & $\mathcal{H}_0(z=1100)$ & $\Omega_M$ & $w_a$ & $AIC_{PL}$ & $AIC_{linear}$ & $BIC_{PL}$ & $BIC_{linear}$ & $\log BF_{P,L}$ \\
\hline

20 & $\Lambda$CDM & $70.40 \pm 0.06$ & $0.007 \pm 0.003$ & $2.1$ & $66.91 \pm 1.62$ & - & - & $369.3$ & $354.3$ & $371.3$ & $356.3$ & $-7.50$ \\
 & fixed $\Omega_M$ &  &  &  &  &  &  & & & & & \\
\hline

\multicolumn{13}{|c|}{Equi-population binning, Master Sample without low-$z$ SN, $H_0=70$, Gold cases} \\
\hline
Bins & Model & $\tilde{H}_0$ & $\alpha$ & $\alpha/\sigma_{\alpha}$ & $\mathcal{H}_0(z=1100)$ & $\Omega_M$ & $w_a$ & $AIC_{PL}$ & $AIC_{linear}$ & $BIC_{PL}$ & $BIC_{linear}$ & $\log BF_{P,L}$ \\
\hline
3 & $\Lambda$CDM & $70.00 \pm 0.19$  & $0.015 \pm 0.007$ & $2.2$ & $62.80 \pm 3.16$ & - & - & $24.4$ & $21.9$ & $22.6$ & $20.1$ & $-1.25$ \\
 & fixed $\Omega_M$ &  &  & & &  &  &  &  &  & & \\
\hline
3 & $\Lambda$CDM & $69.87 \pm 0.21$ & $0.014 \pm 0.008$ & $1.8$ & $63.25 \pm 3.52$ & $0.322 \pm 0.003$ & - & $19.2$ & $17.5$ & $17.4$ & $15.7$ & $-0.85$ \\
 & varying $\Omega_M$ &  &  &  &  &  &  & & & & & \\
\hline
20 & $\Lambda$CDM & $70.36 \pm 0.07$ & $0.016 \pm 0.004$ & $4.2$ & $63.08 \pm 1.65$ & $0.321 \pm 0.001$ & - & $141.1$ & $129.3$ & $143.1$ & $131.3$ & $-5.90$ \\
 & varying $\Omega_M$ &  &  &  &  &  &  & & & & & \\
\hline

\end{tabular}
\end{adjustbox}
\caption{Fit parameters for $\mathcal{H}_0(z)$ in the equi-population binning (upper part), MW binning (middle part), and without low-$z$ SNe Ia equi-populated binning (lower part) of the Master Sample, assuming flat $\Lambda$CDM and $w_{0}w_{a}$CDM models. The columns indicate the same quantities of tables \ref{tab:tab3MW} and \ref{tab:4equispacelogz}.
All the uncertainties are given in $1\sigma$. The fiducial values are the same as table \ref{tab:Omega}.}
\label{tab:mastersample}
\end{table*}

\section{Discussion}{\label{sec:discussion}}

Several cases in the updated analysis show high values of $\alpha/\sigma_\alpha$ ($\geq 3$) under different binning strategies and cosmological models, see Table \ref{tab:3sigma}.

Notably, in the equi-population binning using the MW method under the Gold best likelihoods (see table \ref{tab:tab3MW}) shows 3 cases. The P+ sample without duplicates shows a sharp increase in $\alpha/\sigma_\alpha$, which gives $5.5$ and $9.1$ for 4 and 12 bins within the $ w_0w_a$CDM model, fixing $\Omega_M$. Similarly, the DES sample for 12 bins also yields a significant $\alpha/\sigma_\alpha$ value of $7.1$ under the same model.

Through the $\log-z$ equi-spacing (table \ref{tab:4equispacelogz}), we show 4 cases of high $\alpha/\sigma_\alpha$ for JLA, P+, and DES samples in the configuration of fixed parameters. The DES sample shows consistently high values of $\alpha/\sigma_\alpha$ for 3 and 12 bins within the $w_{0}w_{a}$CDM and $\Lambda$CDM models, respectively, with the first reaching $5.4$ and the latter $6.1$. The JLA sample, although relatively moderate, still indicates a significant $\alpha/\sigma_\alpha = 3.3$ in $4$ bins under the $w_{0}w_{a}$CDM model. P+, instead, has $\alpha/\sigma_\alpha=4.3$ in 3 bins of $w_{0}w_{a}$CDM.

The Master Sample (table \ref{tab:mastersample}) reveals 5 cases of similar high $\alpha/\sigma_\alpha$. Under equi-population binning, all Gold cases have $\alpha/\sigma_\alpha$ as high as $4.6$ for 20 bins within the $w_{0}w_{a}$CDM varying parameters scenario, indicating an extreme deviation, followed by the cases of $\alpha/\sigma_\alpha= 4.4, 3.7, 3$ with 3 bins within the $\Lambda$CDM with fixed $\Omega_M$, with 20 bins within the $w_{0}w_{a}$CDM varying parameters scenario and  $\Lambda$CDM varying $\Omega_M$, respectively.

Similar effects are seen when we remove low-$z$ SNe Ia, with $\alpha/\sigma_\alpha=4.2$ in 20 bins of $\Lambda$CDM with varying $\Omega_M$. This suggests that the observed decreasing behaviour is unlikely to be associated with local SNe Ia.

One can argue that the inclusion of high-$z$ SNe Ia highlights this trend and future additions of SNe Ia from the Subaru Telescope \citep{1992ESOC...42...43K}, the Roman Space Telescope \citep{2021MNRAS.507.1746E}, and the James Webb Telescope \citep{2023PASP..135f8001G} might confirm the observed decreasing behavior for $H_0$. If we have a less unbalanced sample of SNe Ia at high-$z$, we can highlight this trend more consistently across samples. This may shed light on possible biases in the stretch parameters, as shown in \citealt{Nicolas2021}, other hidden biases, or deviations from the $\Lambda$CDM and $w_0w_a$CDM models. 

\begin{table}[ht!]
\centering
\scriptsize
\begin{tabular}{|>{\raggedright\arraybackslash}p{6cm}|>{\centering\arraybackslash}p{2cm}|>{\centering\arraybackslash}p{3cm}|>{\centering\arraybackslash}p{3cm}|}
\hline
\multicolumn{4}{|c|}{Summary table of the cases with $\alpha/\sigma_{\alpha} \geq 3$} \\
\hline
\multicolumn{4}{|c|}{Equi-population binning using MW, Gold best likelihood} \\
\hline
Sample & Bins & Model & $\alpha/\sigma_\alpha$ \\
\hline
P+ no dupl. & 4  & $w_{0}w_{a}$CDM              & 5.5  \\
            &    & \text{fixed} $\Omega_M, w_{0}, w_{a}$      &       \\
\hline
P+ no dupl. & 12 & $w_{0}w_{a}$CDM              & 9.1  \\
            &    & \text{fixed} $\Omega_M, w_{0}, w_{a}$      &      \\
\hline
DES         & 12 & $w_{0}w_{a}$CDM              & 7.1  \\
            &    & \text{fixed} $\Omega_M, w_{0}, w_{a}$      &       \\
\hline
\multicolumn{4}{|c|}{Equi-spacing in log $z$, Gold best likelihood} \\
\hline
Sample & Bins & Model & $\alpha/\sigma_\alpha$ \\
\hline
JLA         & 4  & $w_{0}w_{a}$CDM                       & 3.3  \\
            &    & \text{fixed} $\Omega_M, w_{0}, w_{a}$               &       \\
\hline
P+ no dupl. & 3  & $w_{0}w_{a}$CDM                       & 4.3  \\
            &    & \text{fixed} $\Omega_M, w_0,w_a$      &       \\
\hline
DES         & 3  & $w_{0}w_{a}$CDM                       & 5.4 \\
            &    & \text{fixed} $\Omega_M, w_{0}, w_{a}$               &      \\
\hline
DES         & 12 & $\Lambda$CDM                           & 6.1 \\
            &    & \text{fixed} $\Omega_M$               &      \\
\hline
\multicolumn{4}{|c|}{Equi-population binning, Master Sample $H_0 = 70$, Gold cases} \\
\hline
Sample & Bins & Model & $\alpha/\sigma_\alpha$ \\
\hline
Master      & 3   & $\Lambda$CDM                           & 4.4  \\
            &     & \text{fixed} $\Omega_M$               &       \\
\hline
Master      & 3  & $\Lambda$CDM                        & 3.0 \\
            &     & \text{varying} $\Omega_M$      &       \\
\hline
Master      & 20  & $w_{0}w_{a}$CDM                        & 4.6  \\
            &     & \text{varying} $\Omega_M, w_{a}$             &       \\
\hline
\multicolumn{4}{|c|}{Equi-population binning using MW, Master Sample, $H_0 = 70$, Gold cases} \\
\hline
Sample & Bins & Model & $\alpha/\sigma_\alpha$ \\
\hline
Master      & 20  & $w_{0}w_{a}$CDM                        & 3.7   \\
            &     & \text{varying} $\Omega_M, w_{a}$             &        \\
\hline
\multicolumn{4}{|c|}{Equi-population binning, Master Sample without low-$z$ SN, $H_0 = 70$, Gold cases} \\
\hline
Sample & Bins & Model & $\alpha/\sigma_\alpha$ \\
\hline
Master      & 20  & $\Lambda$CDM                     & 4.2  \\
            &     & \text{varying} $\Omega_M$          &        \\
\hline

\end{tabular}
\caption{Summary of the cases where $\alpha/\sigma_{\alpha} \geq 3$. The first column specifies the SNe Ia sample, the second reports the number of bins, the third contains the cosmological model, and the fourth shows the value of $\alpha/\sigma_{\alpha}$.}
\label{tab:3sigma}
\end{table}

\section{Summary and conclusions} \label{sec:conclusions}
We have investigated the Hubble tension in four main SNe Ia catalogues: Pantheon, P+, JLA, and DES, as well as their total duplicate-free combination, here called the Master Sample. We applied three different binning techniques: the first considering the same number of SNe Ia in each bin, the second using an equi-populated moving binning technique along with the increasing $z$, and the third choosing $\log-z$ equi-spacing for each bin within the flat $\Lambda$CDM and the $w_{0}w_{a}$CDM models.

We performed an MCMC analysis in all bins to estimate the best values of $H_0$ and its uncertainty in $1\sigma$. We allow both to fix and vary $\Omega_M$ together with $H_0$ in the $\Lambda$CDM and $\Omega_M,w_a$ with $H_0$ in the $w_{0}w_{a}$CDM. After the values of $H_0$ are calculated in all the bins, we fit them with a decreasing function of the redshift. The results highlight how, in the majority of the SNe Ia catalogues and regardless of the binning approach, we observe a slow redshift-evolution towards lower values of the $H_0$.

According to the compatibility of the $\alpha$ parameter with zero and the compatibility of $H_0$ in $1\sigma$ between the CMB value and the fitting extrapolation at $z=1100$, we have defined two main categories of results identified as the Diamond and Gold samples. 
The Diamond cases show that their ratio of $\alpha/\sigma_{\alpha}>1$ and the compatibilities in 1 $\sigma$ with the Planck CMB measurements are ensured. 
The Diamond cases represent a significant fraction of results in our analysis (28/70=40\%), which is with equi-population, MW, and equi-spacing in the $\log-z$ binning for the Pantheon, P+, JLA, DES, and Master Sample. 

In general, the $\alpha/\sigma_\alpha$ ranges from 1.0 $\sigma$ (5 cases) to the $9.1 \sigma$ of 12 bins in the P+ without duplicates, analyzing the $w_{0}w_{a}$CDM model (Gold case, MW, fixed parameters). In all cases, except two cases of the Master sample including low-z SNe Ia, $\alpha \sim 0.01$ is compatible within 1 $\sigma$ within its error, in agreement with \citet{Dainotti2021hubble,Dainotti2022hubble,DESIMONE2024}.
These two cases of the Master sample, including low-z, are analyzed in 20 bins in the $\Lambda$CDM fixed $\Omega_M$ scenario without and with the MW, and carry $\alpha=0.005 \pm 0.004$ and $\alpha=0.007 \pm 0.002$, respectively. It is interesting to further discuss these cases in a forthcoming analysis.
Moreover, we have shown that 11 cases (16\% of the total) present $\alpha > 3 \sigma$, regardless of the model and the samples.

In this work, Marshall's likelihood has been tested in the binning approach, and it was revealed to be crucial since it allows adding high-$z$ SNe Ia that are not combined in literature with other SNe Ia, namely, those that do not have covariance matrices associated with them. The $\alpha$ and $\alpha/\sigma_\alpha$ values of Marshall's likelihood cases are compatible with the other results of the current analysis.

It is worth stressing that the $\log BF_{P,L}$ analysis does not mark any preference for any of the two fitting models, except for the cases discussed here. Considering the power law model, it is preferred in: P+ no duplicates 12 MW bins $w_{0}w_{a}$CDM with fixed parameters, DES 12 MW bins $w_{0}w_{a}$CDM with fixed parameters, and Master 20 MW bins $w_{0}w_{a}$CDM with varying parameters. For what it concerns the linear model, instead, it is preferred in what follows: DES 3 equispaced bins $w_{0}w_{a}$CDM with fixed parameters and Master 20 equipopulated bins without low-$z$ in $\Lambda$CDM (fixed and varying parameters).
Furthermore, while this linear decaying behavior seems to be, in some cases, a good \emph{ansatz}, nonetheless, it is hardly able to reconcile the observed decaying trend with the
required one to approach the Planck detected values, i.e., to deal with a Diamond case. This is because the parameter $a$ is always negative and its absolute value is in the order of $10^{-2}-10^{-1}$: when $z$ approaches the recombination value $z\simeq 1100$, the term $A \cdot z$ implies very small, or even negative, values of the effective Hubble constant, see Equation \ref{eq:linearmodel}. This observation leads us to claim that the linear decreasing trend is appropriate for a low-$z$-value representation only, to be thought of as a Taylor expansion of the power law profile.

The newly obtained combination of the SNe Ia catalogs serves as a significant resource for the scientific community. It offers an extended sample, useful for redshift binning analysis and other research requiring diverse SNe Ia samples. The Master Sample will be made available upon request following the publication of this paper. The Master Sample is free of duplicates, and it proves itself reliable for cosmological analysis that may involve combining probes at any redshift range. 
 
The observed $H_0$ trend can highlight a possible hidden evolutionary effect for SNe Ia parameters \citep{Nicolas2021}. On the other hand, a caveat for the presence of many SNe Ia at low-$z$ must be placed since we face velocity uncertainties of the host galaxies for local SNe Ia. In particular, for the case of P+, the peculiar velocities can weigh up to 10\% to the total recession velocity for SNe Ia hosts \citep{Peterson2022}, this effect being more pronounced for local SNe Ia with $z\sim0.01$.

With an equi-populated binning analysis of the Master without low-$z$ SNe Ia, within the $\Lambda$CDM model with fixed $\Omega_M$, the trend persists, and this result opens the discussion on the possible hidden effects behind the decrease of $H_0$.

Indeed, the observed trend in the variety of SNe Ia catalogs suggests that certain selection biases or astrophysical evolutions for SNe Ia parameters remain unidentified, or new physics must be invoked. 
In this sense, the forthcoming contribution of surveys that observe SNe Ia at $z>1$ will shed more light and overcome the selection biases due to missing observations at high-$z$. 

If SNe Ia observations are confirmed to be free from biases or selection effects, the observed results may be explained through cosmological models alternative to the $\Lambda$CDM and the $w_{0}w_{a}$CDM ones. These include modified gravity theories, early- or late-time cosmological changes, and new perspectives on the nature of dark energy and dark matter, or their interaction. Many such ideas have been proposed within the scientific community over the years (for more details, see the Appendix \ref{sec:H0literature}).

However, since neither the $\Lambda$CDM nor
the $w_0w_a$CDM models appear to consistently reproduce a constant value of $H_0$ in a generic binning setup,
our analysis suggests that they have difficulties accounting for the
cosmological dynamics in the redshift interval of SNe Ia, as we have already witnessed with the DES and P+ samples, which have a steeper $\alpha \geq 6 \sigma_\alpha$. 
Given that the literature is rich in interesting proposals for tackling the Hubble tension, the current results represent a novel way for discriminating among different cosmological models. 

\section*{Acknowledgements}
We are very grateful to H. Marshall for providing invaluable suggestions for adding the use the Marshall's likelihood. B.D.S. acknowledges the support for the accommodation from the National Astronomical Observatory of Japan (NAOJ) and the financial support from: the University of Salerno, INFN - Gruppo Collegato di Salerno, and the FARB funding. K.K. is supported by KAKENHI Grant No. JP23KF0289, No. JP24H01825, and  No. JP24K07027. S.N. is supported by the ASPIRE project for top scientists, JST ’RIKEN-Berkeley Mathematical Quantum Science Initiative. We are grateful to K. Mandar, A. Bidlan, and R. Chakraborty for their help in proofreading the manuscript. We acknowledge E. O' Colgain and E. Di Valentino for their constructive questions and comments in the first arXiv submission. We have clarified some sentences in the text and about the analysis performed in this work by adding a flow chart. We also thank B. L. Frye for suggesting to add another data point [22, \citealt{2024arXiv240318902P}] in figure \ref{fig:H0probes}. We are grateful to T. Hamana for the discussion about the effect of low-redshift Supernovae Ia and E. Fazzari for the insights on the Bayes factor.   


\section{Appendix I: Overview on the Hubble Tension}\label{sec:H0literature}
Here is a review of some of the relevant probes and proposed solutions for the $H_0$ tension problem.
To shed more light on the $H_0$ tension, more precise measurements of $H_0$ are needed: to this end, standard candles are crucial, as well as current standard rulers and future standardizable probes. 

\subsubsection*{The Probes For Investigating The $H_0$ Tension}
\begin{itemize}
    \item {\bf SNe Ia.} Among the best standard candles, {\bf SNe Ia} plays a central role in the estimation of $H_0$. In past works, various techniques have been employed using diverse data sets, such as SDSS-II, SNLS, and Pan-STARRS1, incorporating improved standardization methods for SNe Ia \citep{2022EPJC...82..610M,2022A&A...665A.123M}. Enhanced analyses, such as the P+ data set with weak lensing corrections and progenitor model comparisons, have refined cosmological constraints \citep{2023MNRAS.520L..68S}. Including C-corrections to reduce systematic uncertainties in SNe Ia and {\bf Tip of the Red Giant Branch} ({\bf TRGB}) calibration, along with new distance ladder methods and Near Infrared (NIR) SNe Ia observations, provides insights into $H_{0}$ tension and Dark Matter (DM) distribution \citep{2023arXiv230702434C}. New analyses of anisotropies and DM constraints offer further understanding of the DM composition, excluding primordial black holes as the dominant contributors \citep{2023MNRAS.524.5762D}. NIR measurements of SNe Ia enhance $H_{0}$ precision, while the P+ reanalysis suggests that timescape cosmology may better align with observations than $\Lambda$CDM \citep{2023arXiv231101438L,2023arXiv230401831M}. After the release of P+, new updated SNe Ia studies include Hubble tension insights from further SNe Ia data \citep{2024ApJ...964L...4C,2024PhRvD.109f3519Z} together with the information provided by the CMB \citep{Planck2018,2023JCAP...10..045A,2023JCAP...05..032H,2023arXiv230713083L,2023arXiv230810738P,2023arXiv230109988Y}.

\item {\bf Cepheids.} The best standard candles observed in the local universe are the {\bf Cepheid stars} \citep{2022ApJ...939...89B,2023PhLB..84037886T,2024arXiv240302801A}, which are often used as local anchors to calibrate SNe Ia distances.

\item The {\bf Baryon Acoustic Oscillations} ({\bf BAOs}) are periodic density fluctuations in the distribution of baryonic matter caused by sound waves in the plasma of the early universe, which are frozen today in the Large Scale Structures \citep{Eisenstein2005}. Given their characteristic size (around $150\, Mpc$), they represent one of the main geometrical probes used nowadays to constrain the cosmological parameters \citep{Sharov_2018,2024arXiv240704322D,2024PhLB..85839027F,2024arXiv240602019J,VANPUTTEN2025194}.

\item {\bf High-$z$ probes.} Since SNe Ia and BAOs currently have a redshift span $z<3$, further probes at higher $z$ are needed to surpass this limit. In this respect, {\bf GRBs} are playing an essential role in the development of future cosmology \citep{Cardone2009,2017ApJ...848...88D,Cao2022, 2023MNRAS.518.2201D,2023arXiv231209440Z,2024Univ...10...75S}, as well as {\bf QSO} \citep{2023JCAP...07..041S,2024CQGra..41f5003A}. GRBs have been observed up to $z=9.4$ \citep{Cucchiara2011}, while QSO up to $z=10.1$ \citep{2024ApJ...960L...1N}.

\item {\bf Constraints from galaxies.} Concerning the cosmological analysis performed with galaxies, many approaches rely on observations of galaxy clusters \citep{alestas2022,2024MNRAS.531.1021P}, galaxies parallax \citep{Ferree2021}, early galaxies properties and {\bf Cosmic Chromometers} ({\bf CC}, \citealt{2023JApA...44...22G,2024Univ...10...48M,2024MNRAS.527.2152N,2024EPJP..139..711W}), the application of the {\bf Tully-Fisher relation} \citep{2023MNRAS.524.1885W,2024MNRAS.532.2234H}, and {\bf Active Galactic Nuclei} ({\bf AGNs}, \citealt{Lu2021}).

\item {\bf Gravitational probes.} An important contribution to the $H_0$ estimation is expected from {\bf Gravitational Waves} ({\bf GW}) and {\bf Dark Sirens} \citep{2021arXiv210402728G,Gray2021,Li2021,Mozzon2021,Palmese2021,2023arXiv230710940B,2023JCAP...03..024C,2023arXiv230209764D,2023MNRAS.524.3537G,2023JCAP...08..070J,2023arXiv231204632M,2023PhRvD.107d3027T,2023MNRAS.526.6224T,2023ApJ...943...13W,2024SCPMA..6710413B,Soni2024,2024ApJS..270...24Y}, and a likewise important contribution can be available within the gravitational lensing scenario
\citep{2023ApJ...955..140S,2023RAA....23c5001Z,2024arXiv240213476L}.

\item {\bf Other probes.} Other cosmological objects and observables are able to constrain the $H_0$ values, such as the {\bf Harrison-Zeldovich} effect measurements \citep{2024MNRAS.527L..54J}, the {\bf Velocity Acoustic Oscillations} ({\bf VAOs}, \citealt{2023PhRvD.107b3524S}), {\bf Fast Radio Bursts} ({\bf FRBs},  \citealt{2023ApJ...955..101W,2023SCPMA..6620412Z,2024arXiv240703532F}), {\bf Pulsar Timing Array} \citep{2022arXiv220509261R,2023arXiv231103391B,2024SCPMA..6710413B}, {\bf Supernovae Type II} ({\bf SNe II}, \citealt{2022MNRAS.514.4620D}), {\bf Planetary Nebula Luminosity Function} ({\bf PNLF}, \citealt{2021ApJ...916...21R}), Solar System proper motion \citep{Horstmann2021}, {\bf TRGB} \citep{2024arXiv240317048L}, Infrared Surface Brightness Fluctuations ({\bf IR SBF}, \citealt{2023ApJ...953...35G}), and the {\bf Old Astrophysical Objects} ({\bf OAOs}, \citealt{2023ApJ...953..149C,2023EPJC...83..875C}).

In the future, {\bf Large-Scale Structure} surveys will provide invaluable information for testing cosmological models. The Euclid Deep Survey (EDS) and Large Synoptic Survey Telescope (LSST) will have access to a statistically relevant number of {\bf Superluminous Supernovae} in the next decade \citep{Andreoni2021, fanizza2021precision}.

\item {\bf Combined probes.} To maximize the precision that can be reached to infer cosmological parameters, a common approach in the literature is to use many probes simultaneously in the same analysis. For instance, SNe Ia, CMB, BAOs, Weak Gravitational Lensing, Large-Scale Structure, Cosmic Chronometers (CC), GWs, GRBs, {\bf HII Galaxies}, {\bf Starburst Galaxies}, QSO, AGNs, Strong Lensing, galaxy clusters, and Multi-Messenger Probes \citep{Freedman2021,   2023EPJC...83..274B,2023JCAP...04..023B,2023A&A...672A.157B,2023JCAP...02..041C,2023MNRAS.521.4963D, 2023Symm...15..259G, 2023ApJ...959...35K, 2023EPJC...83..949L,2023PhRvD.108h3519S,  2023PhRvD.108j3525S,  2023arXiv231116377R, 2023MNRAS.519.4938Y, 2023ChPhC..47c5103Z,    2024arXiv240605049C, 2024MNRAS.528L.168C, 2024MNRAS.531L..52C,  2024JCAP...04..035D, 2024A&A...682A.148F, 2024ApJ...960..103L,2024PhLB..85438717L,2024PhRvD.109b3519P, 2024arXiv240600634R,  2024PDU....4501522S,     Taule2024}.

\end{itemize}

\subsubsection*{The Theoretical Proposals}
Let us now refer to some of the most recent and successful approaches in the field of theoretical models to tackle cosmological tensions. Since many groups independently found this $H_0$ tension and the tension persists regardless of the used probes and the sample sizes, a new intrinsic physics may come into play. 

\begin{itemize}

 \item {\bf Modified gravity theories.} In the literature, many researchers propose alternatives to the standard $\Lambda$CDM model through {\bf modified gravity theories} \citep{Gurzadyan2021, Khosravi2021, Farrugia2021, Mehdi2021, Palle2021, Petronikolou2021, Reyes2021,2022ApJ...935..156B,2022EPJC...82..418G, 2022InJPh..96.1289S, 2022EPJC...82.1039Y,2023PDU....3901153B, 2023JCAP...06..039D, 2023PTEP.2023k3E01K,   2023NuPhB.99316285M, 2023PDU....4201264P,   2023EPJC...83..330V, 2023MNRAS.523..453W,2024MNRAS.528.2711B,2024PhRvD.109l3528J, 2024arXiv240505843M, 2024MNRAS.527.7626R, 2024JCAP...10..052S, 2024PDU....4301407S, Bag2021, Nilsson2021, 2022PDU....3701112R, 2023PhRvD.108b4007H, 2023Univ....9..311E, 2023PhRvD.108j4037H, 2023PhLB..83637621J,2023EPJP..138..929S}. The variation of fundamental physics \citep{FranchinoVinas2021, Prat2021, Sola2021, 2022EPJC...82..480H,2022arXiv221113616S, 2023PhRvD.107d3505H,  2023PhRvL.130p1003L,2023PhRvD.107j4028Q, 2023PhRvD.107h3512S, 2023PDU....3901150T,  2024arXiv240701173T, 2022PDU....3801137H, 2022PhRvD.105l3512O, 2024SCPMA..6710413B, Zhang2021, 2021arXiv210509790K}, as well as the modification of the speed of light \citep{2020arXiv201010292N} or the gravitational constant \citep{Alestas2021b, Marra2021, sakr2021, alestas2022, 2024arXiv240211947E, Montani:2024buy, 2022PhRvD.106d3528P}, represent other possible explanations for the cosmological tensions.
 
 \item {\bf Teleparallel gravity.} A particular subgroup of non-Einsteinian gravity theories is composed of the so-called {\bf teleparallel gravity approaches}, where the torsion field is included in the geometrical setting \citep{Najera2021fitting,  2021arXiv210414065N, Ren2021, 2022CQGra..39s5021K}.  
 
 \item {\bf Further approaches.} In this bullet point, we mention some of the alternative approaches to construct a modified cosmological dynamics in a generalized framework: scalar field DE \citep{Pereira2021, Shrivastava2021}, Quasi Steady State Cosmology \citep{2023AnPhy.45469345S}, String swampland criteria \citep{2023Ap.....66..423K}, and Holographic models \citep{2023JCAP...02..045C}. Theoretical evidence of easing $H_{0}$ tension using conformal field theory algebras and other quantum field theory approaches had been considered \citep{Ambjorn2021, MorenoPulido2021}.

\item {\bf Dark Energy models.} Among the most common approaches proposed in the literature, modifications to the standard DE formulation are of particular relevance. The idea is to explore a scenario with a new exotic energy density that behaves like a cosmological constant at early times and then decays quickly at some critical redshift \(z_c\). Such kind of energy density like this is motivated by some string-axiverse-inspired scenarios for DE, \citep{2016PhRvD..94j3523K, Ghosh2017}, interactions between DE and other components, \citep{Divalentino2018,Aghaei2021,2023arXiv230805807H, 2023JCAP...01..042K,2023ARep...67..537A}, the Generalized Uncertainty Principle (GUP) and Extended Uncertainty Principle (EUP) modified Hubble parameters \citep{Aghababaei2021, Perivolaropoulos2021phantom, Allali2021, Artymowski2021, Bag2021, Banihashemi2021, Blinov2021, Cai2021, Cuesta2021, DiValentinodark, Gariazzo2021, Ghosh2021, Niedermann2021,2021arXiv210400596S, Perivolaropoulos2021Lagrangians,  2021arXiv210303815Y, Ye2021resolving, Zhou2021,2022JCAP...04..042A,2022PhRvD.106f3520A,   firouzjahi2022cosmological,2022PhRvD.106j3522G, 2022MNRAS.513.3368H,2023arXiv230315523B,2023JHEP...06..012B,2023PhRvD.108h3512D,2023JCAP...12..017F, 2023PhRvD.108j3526G, 2023MNRAS.519.3664H, 2023arXiv230705475J,2023JCAP...02..005K,2023PhRvL.130p1003L,2023PhRvD.107j3523L, 2023arXiv230700418N,2023arXiv230102913R,2023PhRvD.107h3512S,2023PhRvD.107j3507T,2023PDU....3901150T,2023arXiv231116862T,2024Univ...10..122A,2024arXiv240517554B,2024APh...15702925B,2024JCAP...08..052C,2024PDU....4301406D,2024PDU....4501533G,2024PhLB..85138588J, 2024JHEP...07..145J,2024PDU....4601584J, 2024PDU....4501511L,2024PhRvD.109j3531L,  2024FrASS..1113816N, 2024MNRAS.528.1531P, 2024PhyS...99g5043S,2024arXiv240406396S,     2024PDU....4601568T,2024arXiv240701173T,    2024JCAP...01..048V,2024JCAP...03..045W, 2024PhRvD.109f3502Y, 2024PhLB..85538826S, Joseph2021}. In this category, we can also find the running vacuum models \citep{2017PhLB..774..317S, 2021arXiv210212758S,Aich_2022,2023arXiv230713130M} and the models with time-varying dark energy \citep{2017JCAP...02..024K}. The quintessence models and the non-minimally coupled DE models represent a further interesting proposal \citep{2024arXiv240917019W,2024PhRvD.110h3528W}.

\item {\bf Dark Matter and Dark Radiation models.} Discussing other possible solutions, alternative DM models are key to formulating proposals to tackle the $H_0$ tension
\citep{Jodlowski2020,Beltran2021,Blinov2021,  Ghose2021, gutierrezluna2021scalar, Hansen2021, Liu2021a, Parnovsky2021b,  safari2022, 2024arXiv240303484K, 2023arXiv230813617N, 2023ARep...67..115B,2023JHEP...06..012B, 2023JCAP...11..005B,  2023PhRvD.107j3523L,2023arXiv230700418N, 2023arXiv230102913R,  2024arXiv240211947E, 2024FrASS..1113816N, 2024PhRvD.109f3502Y, 2024PDU....4601584J}. The Dark Radiation and Interactive Radiation models constitute another interesting idea to counteract cosmological parameter tensions
\citep{Aloni2021, Ghosh2021, 2022JCAP...12..001S, 2023arXiv230615067G, 2023PhRvD.107f3536Z, 2024arXiv240415220A, 2024arXiv240517554B, 2024EPJC...84..912L}.

\item {\bf Early-Universe proposals.} Other theoretical proposals are based on the modification of early-time cosmology
\citep{2019arXiv190401016A, Galli2021, 2021arXiv210510425V,  Ye2021a,2022JCAP...04..042A, 2022arXiv220801162E, 2022MNRAS.513.3368H,  2023PhRvD.108j3527M, 2023arXiv230705475J, 2024Univ...10..338E, 2024arXiv240718292P, 2023PhRvL.130k1001A, 2023NatCo..14.7523H}.

\item {\bf Late-Universe proposals.} Considering the late-time cosmology instead, alternative formulations can be found in
\citealt{2021MNRAS.504.3956A, 2021arXiv210602963D,  Normann2021,2023PhRvL.131k1002K, 2023EPJC...83..167I, 2023PhRvD.108f3527S, 2022PhRvD.106d3503H}. 

\item {\bf Inhomogeneities and anisotropies.} In addition, further hypotheses have been explored, such as local voids or local under-densities \citep{perivolaropoulos2014large, 2020PhRvD.101l3516A, 2020MNRAS.499.2845H, 2021MNRAS.500.5249A, Castello2021, 2021MNRAS.501.3421K, Martin2021, Wong2022}, anisotropies \citep{2022ChJPh..80..261B}, and local inhomogeneities
\citep{Gasperini_2011, perivolaropoulos2011Tolman,2014PhRvL.112v1301B,Fleury_2017, Adamek2019, Fanizza2021, Rashkovetskyi2021,2021arXiv210503003T,    2024JCAP...05..126M}. 

\item {\bf Exotic particle proposals.} Together with the theoretical proposals that concern the cosmological models, some further formulations involve exotic particles such as the Axi-Higgs model \citep{Fung2021,luu2021axihiggs}, Majoron alternative models \citep{Cuesta2021, GonzalezLopez2021, 2024JHEP...07..145J,fernandez2021}, the Two-Higgs doublet theory \citep{Ghosh2017}, the Mirror Twin Higgs model \citep{Bansal2021}, the Axion \citep{2024arXiv240512268C, Mawas2021} and the Axio-Dilaton \citep{Burgess2021}, the Gravitino mass conjecture \citep{2021arXiv210410181C}, mirror dark sector model \citep{2023PhRvD.107d3529Z} and the alternative neutrino physics \citep{Corona2021,das2021selfinteracting, 2021arXiv210600025D, DiValentinosolution, Gu2021, Khalifeh2021, 2022AstL...48..689C,2022JCAP...08..009G, GomezValent2022,2022ApJ...934..113S, 2023arXiv230805720A, 2023PhRvD.108h3512D,2023PhRvD.107l3030D, 2023arXiv230209958S}.

\item {\bf Cosmography.} The cosmographic approach, which is independent of the scale factor shape, is an interesting alternative tool that can shed more light on the cosmological tensions:
\citet{2022PhRvD.106l3523P, 2022MNRAS.516.2597S, 2022A&A...667A.123S, 2023PhRvD.108d4038S, 2024SSRv..220...48B}.


\item {\bf Other proposals.} Many proposals that go beyond the canonical cosmological models are provided in the literature:
\citet{2021PhRvD.103j3533B, Greene2021,  gutierrezluna2021scalar, Krishnan2021a, Mehrabi2021a,Mercier2021, RuizZapatero2021, cea2022, 2022arXiv221009661G, 2022MNRAS.517.5805L, 2022arXiv220311219W,2023PhRvD.107b3507K, 2023arXiv231003509K,2023arXiv230414465L,2024EPJC...84..167B,  2024JCAP...08..052C, 2024A&A...686A.210F, 2024arXiv240114170H, 2024PhyS...99h5025P}. 
Also, the non-conventional approaches to General Relativity, such as stochastic techniques \citep{Lulli2021, 2023ApJ...959...83L}, covariant formulation \citep{Arjona2021, 2024PDU....4501533G}, and the inclusion of gravitational self-interaction \citep{2024PhyS...99g5043S} may solve $H_0$ tension. Extensions of $\Lambda$CDM models \citep{ADHIKARI2022101005,2023arXiv230710899A} and analytical improvements \citep{2022EPJP..137..819S} have also been suggested as potential solutions to the aforementioned problem.

\item {\bf Machine learning.}The use of novel machine learning approaches is effective in analyzing cosmological probes:
\citet{2017JCAP...03..056C,Dialektopoulos2021, Drees2021, 2021arXiv210514332E,Huber2021,Li2021, Ray2021, sun2021influence, 2022arXiv221004143H, 2023EPJC...83..548B,2023PhRvD.108j3526G,2023arXiv230204324L,2023MNRAS.521.4406L,2024EPJC...84..402D, 2024PhLB..85338699G,   2024MNRAS.528.1232K,2024ApJ...960...61M, 2024PhyS...99k5007P,   2022ApJ...932..131R}.


\item {\bf Link to the $\sigma_8$ tension.} In addition, it is also important to consider the connection of $H_0$ tension with other open problems in cosmology, such as $\sigma_8$ tension and $S_8$ tension. The strength of the clustering of matter in the late Universe is quantified in cosmology by the parameter $S_8$. On the other hand, the related parameter $\sigma_8$ refers to the root mean square of the amplitude of matter perturbations,  which are smoothed over 8 $h^{-1} Mpc$, where $h = H_0/100$ is the reduced $H_0$. $S_8$ and $\sigma_8$ are related by the equation $S_8 = \sigma_8 \sqrt{\Omega_M/0.3}$ \citep{DivalentinointertwinedIV}. The measured values of $S_8$ and $\sigma_8$ parameters from low-$z$ probes such as galaxy clusters and weak lensing are found to be from 2 $\sigma$  to 3 $\sigma$ lower than those measured from the evolution of CMB fluctuations \citep{PhysRevD.107.123538}. The discrepancies in the measured values of the $S_8$ and $\sigma_8$ parameters are known as $S_8$ tension and $\sigma_8$ tension, respectively. These may be related to the $H_0$ tension. 

\end{itemize}

We here acknowledge some review of the change in $H_0$ value over the years, using different probes and methods for its calculation \citep{Schoneberg2021}. The state-of-the-art measurements of  $H_0$ with
the main probes are presented in figure \ref{fig:H0probes}.

\section{Appendix 2: The SH0ES analysis}
\subsection{The SH0ES constraints in the first bins}
\label{sec:SH0ES}

In this Section, we offer a thorough methodology for performing a combined MCMC analysis using two complementary approaches: the SH0ES least-square fitting analysis \citep{2022ApJ...934L...7R} and the first bin analysis of SNe Ia. The approach stresses the systematic integration of different methodologies to improve parameter estimation, leading to reliable and consistent findings. The key features of the approach are the integration of many data sets, the optimization of fitting algorithms, and a uniform framework to evaluate the presence of $H_0$ trend. 
The goal of merging the SH0ES sample with the first SNe Ia bin is to constrain $H_0$ using information from both data sets jointly.

The integrated analysis begins with independent likelihood calculations. The SH0ES analysis computes a matrix-based likelihood using a predetermined covariance matrix, while the first bin analysis computes a likelihood based on $\mu_{\text{obs}}$ obtained from SNe data derived from SNe data.
During MCMC analysis, $H_0$ is left free to vary together with the other parameters of the SH0ES likelihood. Allowing $H_0$ to vary ensures its involvement in both combined likelihoods. 
This method improves the efficiency of the MCMC process while also maintaining the original relationship between $H_0$ and the observational data on which it is based.

As in the analysis discussed above, $H_0$ is then determined throughout the MCMC procedure.\\
A unified log-likelihood function is then created by adding the weighted contributions from the SH0ES and SNe Ia likelihoods, with the weights normalized to ensure equal effect from both data sets. This unified likelihood is passed to the MCMC sampler, which uses the joint parameter space to estimate the needed values of $H_0$.

To validate the reliability of the combined analysis in different data sets, the methodology has been adopted to include P+ with and without duplicates. This approach ensures reliable results across various redshift ranges, where higher-$z$ data are naturally down-weighted by covariance matrices.

The combined study shows that the uncertainties in $H_0$ are of the order of unity, which has no impact on the fitting process due to the weighting system, where the weights are defined as $1/\Delta_{H_0}^2$, with $\Delta_{H_0}$ representing the 1 $\sigma$ uncertainty on $H_0$ in the given bin. As a result, the impact of highly inaccurate data points is significantly minimized.
We also analyze cases where the SNe Ia in SH0ES are weighted according to their number.

\subsection{Formalization of the likelihoods with SH0ES}
In this part, we explicitly define the likelihood functions for the SH0ES analysis and the SNe Ia analysis, together with their combination. The likelihood function, $\mathcal{L}$, is proportional to $\exp(-\chi^2/2)$, where $\chi^2$ quantifies the goodness of fit for a given model relative to the data. 

The combined likelihood $\mathcal{L}_{\text{combined}}$, \citep{Lewis:2002ah, SDSS:2014iwm}, is related to $\chi^2_{\text{combined}}$ as:

\begin{equation}
\mathcal{L}_{\text{combined}} \propto \exp\left(-\frac{1}{2} \chi^2_{\text{combined}}\right).
\end{equation}

To construct the combined likelihood, the individual $\chi^2$ contributions from SH0ES and SNe Ia analyses must be summed:
\begin{equation}
\chi^2_{\text{combined}} = \chi^2_{\text{SH0ES}} + \chi^2_{\text{SNe}},
\end{equation}
where $\chi^2_{\text{SNe}}$ is the one defined in Equation \ref{eq:chi2default} and $\chi^2_{\text{SH0ES}}$ has the same form as $\chi^2_{\text{SNe}}$. Concerning $\chi^2_{\text{SH0ES}}$, it is defined in \citealt{2022ApJ...934L...7R}.

The independence of the SH0ES and the other SNe Ia data sets is necessary to use the combined likelihood. Thus, any duplicated SNe Ia information among the SNe likelihood and the SH0ES likelihood have been removed.  
SNe Ia data sets and SH0ES have a different number of data points and, to ensure balanced contributions, weights can be applied:
\begin{equation}
    \chi^2_{\text{combined}} = w_{\text{SH0ES}} \,  \chi^2_{\text{SH0ES}} + w_{\text{SNe}} \,\chi^2_{\text{SNe}},
\end{equation}
where $w_{\text{SH0ES}}$ and $w_{\text{SNe}}$ are normalization factors inversely proportional to the effective degrees of freedom or uncertainties in each data set. Typically, $w_{\text{SH0ES}}$ and $w_{\text{SNe}}$ are set such that their contributions are comparable:
\begin{equation}
    w_{\text{SH0ES}} \propto \frac{1}{\text{Var}(\mathbf{y})}, \quad w_{\text{SNe}} \propto \frac{1}{\text{Var}(\mu)},
\end{equation}

\noindent $Var(k)$ being the variance on the quantities present in the $k$ vector.

\subsection{The SH0ES results}
\label{sec:sh0es analysis}
\begin{figure}[H]
    \centering
    \includegraphics[width=0.49\linewidth]{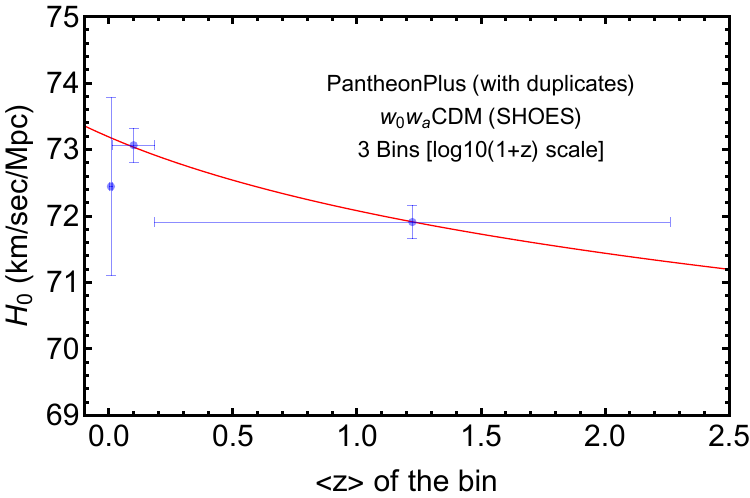}

    \includegraphics[width = 0.49\linewidth]{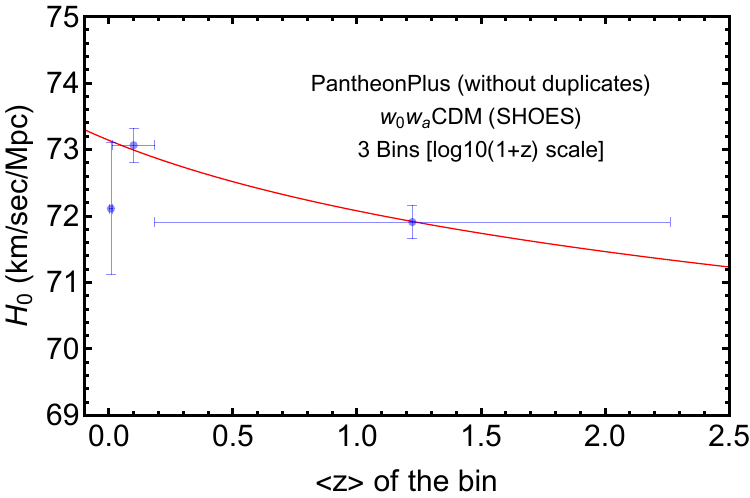}
    \includegraphics[width = 0.49 \linewidth]{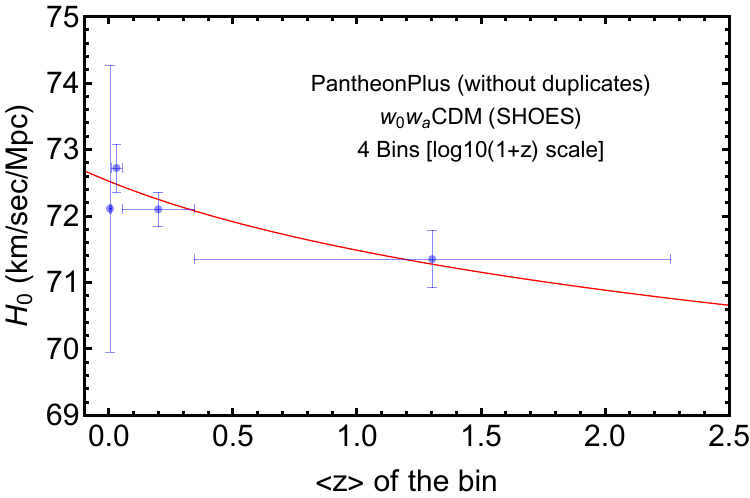}
    
    \caption{The fitting of $H_0$ values as a function of $z$ in the context of equi-spacing binning using the SH0ES constraints on the $\log-z$ for the Diamond samples treated with Marshall's likelihood. The P+ sample with and without duplicates is calibrated with $H_0=73.04$ within the $w_{0}w_{a}$CDM model.  \textbf{Top central:} shows 3 bins with P+ (with duplicates). \textbf{Bottom left:} shows 3 bins with P+ (with duplicates removed). \textbf{Bottom right:} shows 4 bins with P+ (with duplicates removed). The results of these plots are summarized in table \ref{tab:pantheon_fixed3_4} and the corresponding fiducial values are reported in table \ref{tab:Omega}.}
    \label{fig:pantheon_plus_3_bins_sh0es_w}
\end{figure}

We here apply the Marshall's likelihood to the SH0ES sample added to the P+, obtaining Gold cases in both the samples with and without duplicates. 
The Gold result in the P+ with duplicates is reported in the upper part of table \ref{tab:pantheon_fixed3_4} and in the top central panel of figure \ref{fig:pantheon_plus_3_bins_sh0es_w}, In contrast, the Gold results for the P+ without duplicates are reported in the lower part of table \ref{tab:pantheon_fixed3_4} and in the bottom left and bottom right panels of figure \ref{fig:pantheon_plus_3_bins_sh0es_w}.
Considering the P+ sample with duplicates, a Gold case is identified in the 3 bins within the $w_{0}w_{a}$CDM model, where \( \tilde{H}_0 = 73.19 \pm 0.28\) and \(\alpha = 0.022 \pm 0.007\). Consequently, the ratio \(\alpha/\sigma_\alpha = 3.14\), and the \(\mathcal{H}_0(z=1100)\) is \(62.73 \pm 3.08\).

For the P+ sample without duplicates, two Gold cases are identified in the $w_{0}w_{a}$ CDM model. In these cases, the values of \(\tilde{H}_0\) range from \(72.52 \pm 0.26\) to \(73.14 \pm 0.28\), while \(\alpha=0.021\) is the same in both cases with uncertainties of $0.009$ and $0.007$, respectively.
The values of \(\alpha/\sigma_\alpha\) range from $2.33$ to $3.00$, and the values of \(\mathcal{H}_0(z=1100)\) range from \(62.69 \pm 4.04\) to \(63.13 \pm 3.10\). These findings illustrate the consistency in estimating the $\alpha$ parameters when duplicates are included or eliminated. Furthermore, these results show how the SH0ES constraints confirm the findings of our previous analysis.

\begin{table*}[ht!]
\small
\renewcommand{\arraystretch}{1.1} 
\centering
\begin{adjustbox}{width=0.7\textwidth,center}
\begin{tabular}{|c|c|c|c|c|c|c|}
\hline
\multicolumn{5}{|c|}{Equi-spacing on $\text{log}\,z$, Gold cases, P+ (with duplicates),} \\
\multicolumn{5}{|c|}{SH0ES constraints, Marshall's likelihood, $w_{0}w_{a}$CDM model} \\
\hline
Bins  & $\tilde{H}_0$ & $\alpha$ & $\alpha/\sigma_{\alpha}$ & $\mathcal{H}_0(z=1100)$ \\
\hline
3 & 73.19 $\pm$ 0.28  &  0.022 $\pm$ 0.007  & 3.1 & 
$62.73 \pm 3.08$  \\
\hline
\multicolumn{5}{|c|}{Equi-spacing on $\text{log}\,z$, Gold cases, P+ (duplicates removed),} \\
\multicolumn{5}{|c|}{SH0ES constraints, Marshall's likelihood, $w_{0}w_{a}$CDM model} \\
\hline
Bins & $\tilde{H}_0$ & $\alpha$ & $\alpha/\sigma_{\alpha}$ & $\mathcal{H}_0(z=1100)$ \\
\hline
3 & 73.14 $\pm$ 0.28 & $0.021 \pm 0.007$  & 3.0 & $63.13 \pm 3.10$ \\
\hline
4 & $72.52 \pm 0.26$ & $0.021 \pm 0.009$  & 2.3 & $62.69 \pm 4.04$\\
\hline
\end{tabular}
\end{adjustbox}
\caption{Fit parameters for $\mathcal{H}_0(z)$ in the Gold cases considering the equi-space binning of the P+ treated with Marshall's likelihood, assuming the flat $w_{0}w_{a}$CDM models and adding the SH0ES constraints. The first column indicates the number of bins. The second and third columns denote the fit parameters, $\tilde{H}_0$ and $\alpha$.
The fourth column denotes the consistency of the evolutionary parameter $\alpha$ with zero in terms of $1\sigma$, represented by the ratio $\alpha/\sigma_{\alpha}$ and the fifth column denotes the extrapolated $\mathcal{H}_0(z=1100)$, representing the $H_0$ in the LSS.
All the uncertainties are given in $1\sigma$. The fiducial values are the same as table \ref{tab:Omega}.}
\label{tab:pantheon_fixed3_4} 
\end{table*}

\end{document}